\documentclass[aps,prc,twoside,twocolumn,nofootinbib,10pt,showpacs,floatfix]{revtex4-1}
\usepackage{amsmath,amssymb}
\usepackage{graphicx,bm}
\usepackage{slashed}
\usepackage{epstopdf}
\usepackage{ulem} 
\usepackage[usenames]{color}
\usepackage{float}
\usepackage{hyperref}
\usepackage{subfigure}
\usepackage{rotating}
\usepackage{color}
\usepackage{multirow}
\usepackage{dcolumn}
\usepackage{overpic}
\usepackage{booktabs}
\usepackage{makecell}
\usepackage{diagbox}
\usepackage{array}
\newcolumntype{|}{!{\vline}}

\renewcommand\sout{\bgroup \color{red} \ULdepth=-.5ex \ULset}

\newsavebox{\tablebox}
\begin{document}
\title{{Manifestly} exotic pentaquarks with a  single heavy quark}
\author{Hong-Tao An$^{1,2}$}\email{anht14@lzu.edu.cn}
\author{Kan Chen$^{1,2,3,4}$}\email{chenk$_$10@pku.edu.cn}
\author{Xiang Liu$^{1,2,5}$\footnote{Corresponding author}}\email{xiangliu@lzu.edu.cn}
\affiliation{
$^1$School of Physical Science and Technology, Lanzhou University, Lanzhou 730000, China\\
$^2$Research Center for Hadron and CSR Physics, Lanzhou University and Institute of Modern Physics of CAS, Lanzhou 730000, China\\
$^3$Center of High Energy Physics, Peking University, Beijing 100871, China\\
$^4$School of Physics and State Key Laboratory of Nuclear Physics and Technology, Peking University, Beijing 100871, China\\
$^5$Lanzhou Center for Theoretical Physics, Key Laboratory of Theoretical Physics of Gansu Province, and Frontier Science Center for Rare Isotopes, Lanzhou University, Lanzhou, Gansu 730000, China}

\date{\today}
\begin{abstract}
Inspired by the observed $X(2900)$, we study systematically the mass spectra of the ground pentaquark states with the $qqqq\bar{Q}$ ($Q=c,b$; $q=n,s$; $n=u,d$) configuration in the framework of the Chromomagnetic Interaction model.
We present a detailed analysis of their stabilities and decay behaviors.
Our results indicate that there may exist narrow states or even stable states.
We hope that the present study may inspire experimentalist's interest in searching for such a type of the exotic pentaquark state.
\end{abstract}
\maketitle

\section{Introduction}\label{sec1}
Since the 1960s, the observations of different types of hadronic states at least have stimulated three important stages of the development of hadron physics.
Facing more than one hundred light hadrons, the classification of hadrons based on SU(3) symmetry was discovered by Gell-Mann \cite{GellMann:1964nj} and Zweig \cite{Zweig:1981pd,Zweig:1964jf}.
After the observation of $J/\psi$ \cite{Aubert:1974js,Augustin:1974xw}, a dozen of charmonia, which construct the main body of charmonium family collected by Particle Data Group (PDG), were found \cite{Zyla:2020zbs}.
It provides a good chance to construct the Cornell model \cite{Eichten:1979ms} which makes a quantitative depiction of hadron spectroscopy become possible \cite{Isgur:1978xj,Godfrey:1985xj,Capstick:1986bm}. At present, we are experiencing a new stage with the accumulation of these charmoniumlike $XYZ$ states and $P_c$ states \cite{Aaij:2019vzc} announced in experiments. Searching for and identifying exotic hadrons have formed a hot issue (see reviews  \cite{Hosaka:2016pey,Richard:2016eis,Chen:2016qju,Guo:2017jvc,Lebed:2016hpi,Esposito:2016noz,Liu:2019zoy,Ali:2017jda,Brambilla:2019esw} for learning the recent progress). Interestingly, the exotic hadronic states including glueball, hybrid, and multiquark states can provide crucial clues to understanding how the quarks and gluons are bounded together to form different kinds of exotic states, which are involved in the nonperturbative problem of quantum chromodynamics (QCD).

However, identifying a genuine exotic hadronic state is not an easy task which is full of challenge, especially, when mass spectrum of exotic hadronic states overlaps with that of conventional hadrons. A typical example is that the correlation of a neutral charmoniumlike $XYZ$ state and charmonium makes establishing a neutral charmoniumlike $XYZ$ state as an exotic hadronic state become ambiguous. Thus, hunting for manifestly exotic hadronic states accessible in experiment is an optimistic option when constructing an exotic hadron family.

{In 2020, the LHCb Collaboration performed a model-independent analysis of the $B^+\to D^+D^-K^+$ process and they also presented the amplitude analysis in the same decay channel \cite{Aaij:2020hon,LHCb:2020pxc}.}
They found one or more charm-strange resonance structures existing in the $D^-K^+$ invariant mass spectrum, which have masses around 2.9 GeV. Here, the observed charm-strange resonances are referred to the $X(2900)$, which obviously meets the criterion of {a manifestly} exotic hadronic state since the minimal quark content of the $X(2900)$ is $\bar{c}\bar{s}du$ and can be fully distinguished from the conventional meson.

\begin{figure}[htpb]
\includegraphics[width=240pt]{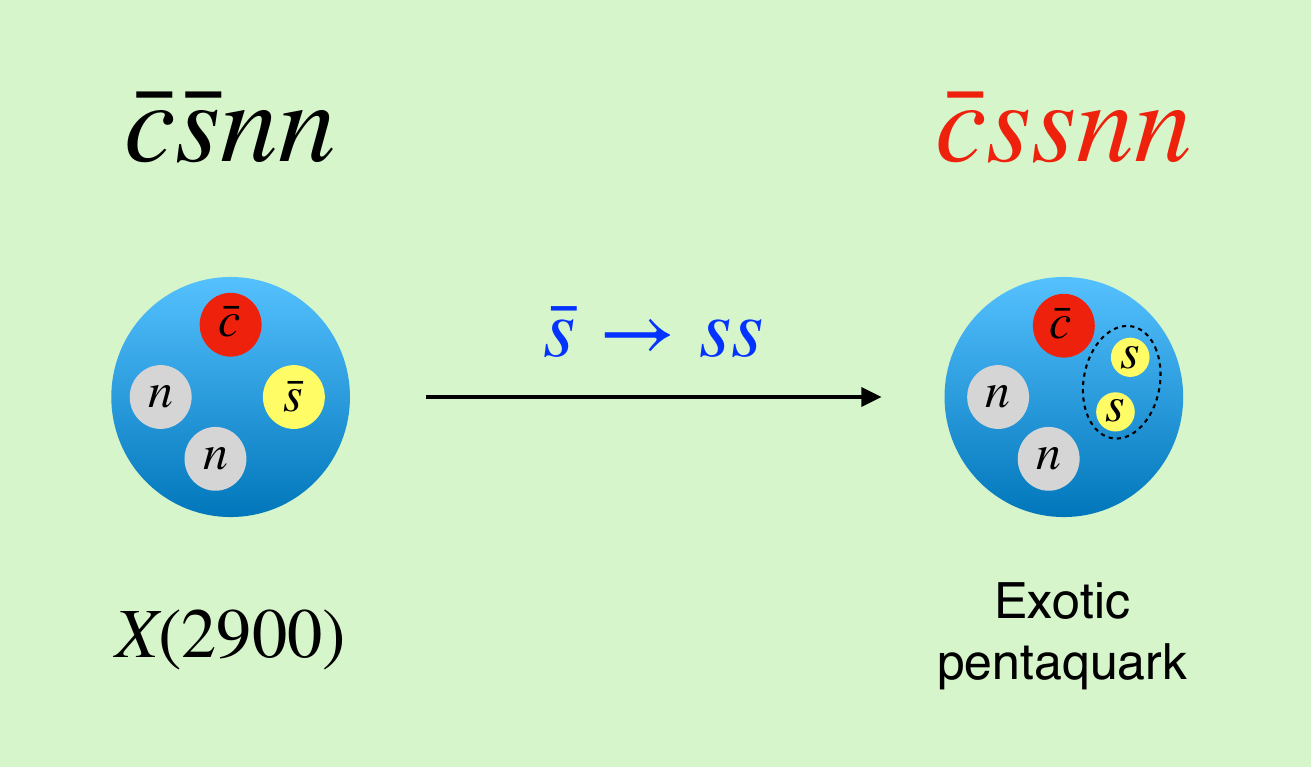}
\caption{Evolution of the $qqqq\bar Q$ $(Q=c,b; q=n,s; n=u,d)$ pentaquark state from the $X(2900)$.}\label{aa}
\end{figure}

Inspired by the observed $X(2900)$, we notice an interesting phenomenon. When replacing the $\bar{s}$ antiquark inside the $X(2900)$ by an $ss$ pair, we obtain a manifestly exotic pentaquark system, which has the $\bar{c}ss nn$ content. Here, we can extend our study to whole $qqqq\bar{Q}$ $(Q=c,b; q=n,s; n=u,d)$ pentaquark state system within the framework of the chromomagnetic interaction model.

We notice that the $qqqq\bar{Q}$ pentaquark states were investigated in some former work.
Genovese {\it et al}.  systemically discussed the stabilities of the $qqqq\bar{Q}$ pentaquark states in the chiral constituent quark model and
found that the $qqqq\bar{Q}$ pentaquark states cannot be bound \cite{Genovese:1997tm}. 
However, assuming that the strength of the chromomagnetic term is the same as that of the conventional baryon, the authors of Ref. \cite{Richard:2018jkw} indicated that the $qqqq\bar{Q}$ pentaquark state can exist and lie about 150 MeV below the $\bar{Q}q + qqq$ meson-baryon threshold in the $m_Q\rightarrow \infty$ limit. The weak decay properties for the stable states of the $qqqq\bar{Q}$ pentaquark system were discussed in Ref. \cite{Stewart:2004pd}.
In addition, Sarac {\it et al}. presented a QCD sum rule analysis of the anti-charmed pentaquark state ($\Theta_{c}$) \cite{Sarac:2005fn}.
Lee {\it et al}. explored the possibility of observing the anticharmed pentaquark state in the $B^+\rightarrow\Theta_{c}\bar{n}\pi^{+}$ process \cite{Lee:2005pn}.
Experimentally,
this possible pentaquark state was studied in the Fermilab experiment \cite{Aitala:1997ja,Aitala:1999ij}, but no evidence was found.
The signal for the $\Theta^{0}_{c}$ $(uudd\bar{c})$  was only observed in the DIS experiment by the H1 Collaboration \cite{Aktas:2004qf}.
In the distribution of $M_{D^{*}p}= M_{K\pi\pi p}-M_{K\pi\pi}+M_{D^{*}}$ with opposite-charge combinations, a peak was observed at $3099\pm3\pm5$ MeV with a Gaussian width of $12\pm3$ MeV.
However, this resonance was not confirmed by other experiments \cite{Chekanov:2004qm,Link:2005ti,Aubert:2006qu,Aubert:2006qx}.
Moreover, the LHCb Collaboration tried to find the pentaquark signal in the $P^{+}_{B^{0}p}(uudd\bar{b})\rightarrow J/\psi K^{+}\pi^{-} p$ weak decay mode via the $b\rightarrow c\bar{c}s$ transition, while no evidence for such state was found \cite{Aaij:2017jgf}.
Recently, the LHCb Collaboration \cite{LHCb:2021ohr} reported the observation of the $\Lambda_{b}^{0}\rightarrow DpK^{-}$ channel, where the invariant mass spectrum of $Dp$ was measured, however, more detailed analysis is still needed to identify the structures existing in the $Dp$ invariant mass spectrum.
In conclusion, whether there exists the  $qqqq\bar{Q}$ pentaquark state is still an open question.

This paper is organized as follows.
After the Introduction, in Sec. \ref{sec2}, we introduce the chromomagnetic interaction model and construct the $flavor \otimes color \otimes spin$ wave functions of the $qqqq\bar{Q}$ pentaquark states.
In Sec. \ref{sec4}, we present the mass spectra, possible strong decay channels, relative partial decay widths, and discuss the stabilities of the discussed states.
The discussion and conclusion are given in Sec. \ref{sec5}. 
Finally, we will present some useful expressions in appendix in Sec. \ref{sec8}.

\section{The chromomagnetic Interaction Model and determination of parameters}\label{sec2}
We adopt an extended chromomagnetic interaction (CMI) model \cite{Weng:2019ynva,Weng:2018mmf,Weng:2020jao,Weng:2021ngd,Weng:2021hje,Hogaasen:2013nca,Karliner:2016zzc} to describe the mass of the ground hadron state. The effective Hamiltonian is
\begin{eqnarray}\label{Eq3}
H&=&\sum_im^{0}_i+H_{\rm CEI}+H_{\rm CMI}\nonumber\\
&=&\sum_im^{0}_i-\sum_{i<j}A_{ij} \vec\lambda_i\cdot \vec\lambda_j-\sum_{i<j}v_{ij} \vec\lambda_i\cdot \vec\lambda_j \vec\sigma_i\cdot\vec\sigma_j, \nonumber\\
&=&-\frac{3}{4}\sum_{i<j}m_{ij}V^{\rm C}_{ij}-\sum_{i<j}v_{ij}V^{\rm CMI}_{ij},
\end{eqnarray}
where $V^{\rm C}_{ij}=\vec\lambda_i\cdot \vec\lambda_j$ and $ V^{\rm CMI}_{ij}=\vec\lambda_i\cdot \vec\lambda_j \vec\sigma_i\cdot\vec\sigma_j$
are the chromoelectric and chromomagnetic interaction between quarks, respectively.
$\sigma_{i}$ denotes the Pauli matrices and $\lambda_i$ is the Gell-Mann matrices.
For the antiquark, $\lambda_i$ is replaced by $-\lambda_i^{*}$.
The $v_{ij}$ is the effective coupling constant of the interaction between the $i$-th quark and $j$-th quark, which depends on the quark masses and the spatial wave function of the ground state.
Meanwhile, $m_{ij}$ is the mass parameter of quark pair, i.e.,
\begin{eqnarray}
m_{ij}=\frac{1}{4}(m^{0}_{i}+m^{0}_{j})+\frac{4}{3}A_{ij},
\end{eqnarray}
which contains the effective quark mass $m_{i}$ ($m_j$) and the color interaction strength $A_{ij}$.
The parameters $m_{ij}$ and $v_{ij}$ are determined by the observed hadron masses \cite{ParticleDataGroup:2020ssz}.
Here, we collect these adopted coupling parameters in Table \ref{parameter2}. The interested
readers may further refer to Refs. \cite{Weng:2019ynva,Weng:2018mmf,Liu:2019zoy,Weng:2020jao,Weng:2021ngd,Weng:2021hje} for more details.

In principle, the values of $m_{ij}$ and $v_{ij}$ should be different for different systems.
However, it is difficult to take this way to carry out a realistic study. Thus, we take an approximation, i.e., 
we extract these coupling parameters by reproducing the masses of these conventional hadrons if assuming that the quark-(anti)quark interactions are the same for all the hadron systems.
Of course, this treatment results in the uncertainty on the mass estimate of the multiquark state.
Note that the size of a multiquark state is expected to be larger than that of a conventional hadron. Correspondingly, the distance between different quark (anti-quark) components in a multiquark state should be larger than that in a conventional hadron. Thus, the attractive forces of different quark (anti-quark) components in a multiquark state are expected to be weaker than that in a conventional hadron. Thus, if such multiquak state exists, its mass calculated by our model should be slightly smaller than its realistic masses.

\begin{table}[t]
\centering \caption{Coupling parameters of the $qq$ and $q\bar{q}$ pairs (in units of MeV).
}\label{parameter2}
\renewcommand\arraystretch{1.25}
\begin{tabular}{ccc|cccc}
\bottomrule[1.5pt]
\bottomrule[0.5pt]
$m_{nn}$&$m_{ns}$&$m_{ss}$&$m_{n\bar{c}}$&$m_{s\bar{c}}$&$m_{n\bar{b}}$&$m_{s\bar{b}}$\\
181.2&226.7&262.3&493.3&519.0&1328.3&1350.8\\
\bottomrule[0.7pt]
$v_{nn}$&$v_{ns}$&$v_{ss}$&$v_{n\bar{c}}$&$v_{s\bar{c}}$&$v_{n\bar{b}}$&$v_{s\bar{b}}$\\
19.1&13.3&12.2&6.6&6.7&2.1&2.3\\
\bottomrule[0.5pt]
\bottomrule[1.5pt]
\end{tabular}
\end{table}

In order to calculate the mass spectrum of the discussed $qqqq\bar{Q}$ pentaquark state, we need to construct the corresponding total wave functions, which are the direct product of the spatial, flavor, color, and spin wave functions
\begin{eqnarray}
\psi_{\textrm{tot}}=\psi_{\textrm{space}}\otimes\psi_{\textrm{flavor}}\otimes\psi_{\textrm{color}}\otimes\psi_{\textrm{spin}}.
\end{eqnarray}
Since we only consider these low-lying $S$-wave pentaquark states, the constraint from the symmetry to the spatial wave functions of pentaquark becomes trivial.
In detail, the $\psi_{\textrm{flavor}}\otimes\psi_{\textrm{color}}\otimes\psi_{\textrm{spin}}$ wave functions of the discussed pentaquark system should be fully antisymmetric when exchanging identical quarks.
In Table \ref{flavor1}, we list these possible flavor combinations for the $qqqq\bar{Q}$ pentaquark system.
According to their symmetry properties, the $qqqq\bar{Q}$ pentaquark systems can be categorized into three groups which are shown in Table \ref{flavor1}. Thus, we need to construct the corresponding $\psi_{\textrm{flavor}}\otimes\psi_{\textrm{color}}\otimes\psi_{\textrm{spin}}$ wave functions with the $\{12\}\{34\}$, $\{1234\}$, and $\{123\}$ symmetries. Here, we use the notation $\{1234\}$ to label that the quarks 1, 2, 3, and 4 have the antisymmetry property.

\begin{table}[t]
\centering \caption{All possible flavor combinations for the $qqqq\bar{Q}$ pentaquark system. Here, $q=n$, $s$ ($n=u$, $d$) and $Q=c$, $b$.
}\label{flavor1}
\renewcommand\arraystretch{1.4}
\begin{tabular}{l|cccc}
\bottomrule[1.5pt]
\bottomrule[0.5pt]
System&\multicolumn{4}{c}{Flavor combinations}\\
\bottomrule[0.5pt]
\multirow{3}*{$qqqq\bar{Q}\quad$}&$nnss\bar{c}\quad$&$nnss\bar{b}\quad$\\
&$nnnn\bar{c}\quad$&$nnnn\bar{b}\quad$&$ssss\bar{c}\quad$&$ssss\bar{b}$\\
&$nnns\bar{c}\quad$&$nnns\bar{b}\quad$&$sssn\bar{c}\quad$&$sssn\bar{b}$\\
\bottomrule[0.5pt]
\bottomrule[1.5pt]
\end{tabular}
\end{table}

Additionally, the Young tableaus, which represent the irreducible bases of the permutation group,
enable us to easily identify the pentaquark configuration with the concrete symmetry.
Thus, we may use the Young tableaus and the Young-Yamanouchi bases to describe the wave functions of these discussed pentaquark states.
The procedure of constructing the $qqqq\bar{Q}$ pentaquark wave functions has been illustrated in Refs. \cite{An:2020jix,An:2021vwi}. Here, 
we only list the values of the multiplicities (the numbers of physical allowed $\psi_{\textrm{flavor}}\otimes\psi_{\textrm{color}}\otimes\psi_{\textrm{spin}}$ bases) for the $qqqq\bar{Q}$ pentaquark subsystems with the different light quark components in Table \ref{qqqqQ}.

In Table \ref{CMI1}, we present the explicit expressions of the CMI Hamiltonian for the $nnnn\bar{Q}$ ($I = 2,1,0$) states. Besides, the expressions of the CMI Hamiltonian for the $nnns\bar{Q}$ ($I=3/2,1/2$), $nnss\bar{Q}$ ($I=1,0$) states are listed in Table \ref{CMI} of the Appendix.

The explicit forms of the CMI Hamiltonian for the $ssss\bar{Q}$ and $sssn\bar{Q}$ pentaquark subsystems are the same as those of the $nnnn\bar{Q}$ ($I=2$) and $nnns\bar{Q}$ ($I = 3/2$) pentaquark subsystems, respectively, after appropriately replacing the corresponding $v_{ij}$ constants.
For example, to obtain the expressions of the CMI Hamiltonian for the $ssss\bar{Q}$ pentaquark subsystem,
we should replace $v_{nn}$ and $v_{n\bar{Q}}$ in the explicit form of the CMI Hamiltonian for the $nnnn\bar{Q}$ ($I = 2$) pentaquark
subsystem with the effective constants $v_{ss}$ and $v_{s\bar{Q}}$, respectively.
Similar treatment is also applied to the calculation of the $sssn\bar{Q}$ subsystem.

\begin{table}[htp]
\caption{
The multiplicity for the studied $qqqq\bar{Q}$ pentaquark system.
$M$ denotes the multiplicity of the pentaquark state with the  $\psi_{flavor}\otimes\psi_{color}\otimes\psi_{spin}$ wave function. }\label{qqqqQ}
\begin{center}
\begin{tabular}{|c|c|c|c|c|c|c|c|}
\toprule[1pt] \toprule[0.5pt]
Flavor state&Isospin&Spin&$M$&Flavor state&Isospin&Spin&$M$ \\
\hline
\multirow{9}*{$nnnn\bar{Q}$}&\multirow{3}*{2}&$5/2$&0&\multirow{6}*{$nnns\bar{Q}$}&\multirow{3}*{$3/2$}&$5/2$&1\\
&&$3/2$&1&&&$3/2$&3\\
&&$1/2$&1&&&$1/2$&3\\ \cline{2-4} \cline{6-8}
&\multirow{3}*{1}&$5/2$&1&&\multirow{3}*{$1/2$}&$5/2$&1\\
&&$3/2$&2&&&$3/2$&4\\
&&$1/2$&2&&&$1/2$&5\\ \cline{2-8}
&\multirow{3}*{0}&$5/2$&0&\multirow{6}*{$nnss\bar{Q}$}&\multirow{3}*{1}&$5/2$&1\\
&&$3/2$&1&&&$3/2$&4\\
&&$1/2$&1&&&$1/2$&4\\ \cline{1-4} \cline{6-8}
\multirow{3}*{$sssn\bar{Q}$}&\multirow{3}*{1/2}&$5/2$&1&&\multirow{3}*{$0$}&$5/2$&1\\
&&$3/2$&3&&&$3/2$&3\\
&&$1/2$&3&&&$1/2$&4\\
\cline{1-8}
\multirow{3}*{$ssss\bar{Q}$}&\multirow{3}*{0}&$5/2$&0&\multicolumn{4}{l|}{}\\
&&$3/2$&1&\multicolumn{4}{l|}{}\\
&&$1/2$&1&\multicolumn{4}{l|}{}\\
\cline{1-4}  \cline{4-8}
\toprule[0.5pt] \toprule[1pt]
\end{tabular}
\end{center}
\end{table}

\begin{table}[t]
\centering \caption{
The CMI Hamiltonian of $nnnn\bar{Q}$ with $I=2,1,0$ $(n= u, d$; $Q =c, d$). Here, $I(J)$ represents the isospin(spin) of the pentaquark states.
}\label{CMI1}
\begin{lrbox}{\tablebox}
\renewcommand\arraystretch{1.43}
\renewcommand\tabcolsep{3.8pt}
\begin{tabular}{c|c}
\bottomrule[1.5pt]
\bottomrule[0.5pt]
$I(J^{P})$&The CMI Hamiltonian \\
\hline
$2(\frac32^{-})$&$\frac{56}{3}v_{nn}-\frac{16}{3}v_{n\bar{Q}}$\\
\cline{2-2}
$2(\frac12^{-})$&$\frac{56}{3}v_{nn}+\frac{32}{3}v_{n\bar{Q}}$\\ \cline{1-2}
$1(\frac52^{-})$&$8v_{nn}+\frac{16}{3}v_{n\bar{Q}}$\\
\cline{2-2}
$1(\frac32^{-})$&
$\begin{pmatrix}
8v_{nn}-8v_{n\bar{Q}}&4\sqrt{10}v_{n\bar{Q}}\\
4\sqrt{10}v_{n\bar{Q}}&\frac{8}{3}v_{nn}-\frac{4}{3}v_{n\bar{Q}}
\end{pmatrix}$
\\ \cline{2-2}
$1(\frac12^{-})$&
$\begin{pmatrix}
\frac{8}{3}v_{nn}+\frac{8}{3}v_{n\bar{Q}}&8v_{n\bar{Q}}\\
8v_{n\bar{Q}}&0
\end{pmatrix}$
\\ \cline{1-2}
$0(\frac32^{-})$&$-\frac{16}{3}v_{nn}+\frac{20}{3}v_{n\bar{Q}}$\\
\cline{2-2}
$0(\frac12^{-})$&$-\frac{16}{3}v_{nn}-\frac{40}{3}v_{n\bar{Q}}$\\
\bottomrule[0.5pt]
\bottomrule[1.5pt]
\end{tabular}
\end{lrbox}\scalebox{1.05}{\usebox{\tablebox}}
\end{table}
\section{Mass Spectra, stabilities, and decay behaviors}\label{sec4}
\subsubsection{The mass spectrum of pentaquark}

\begin{figure*}[t]
\begin{tabular}{ccc}
\includegraphics[width=240pt]{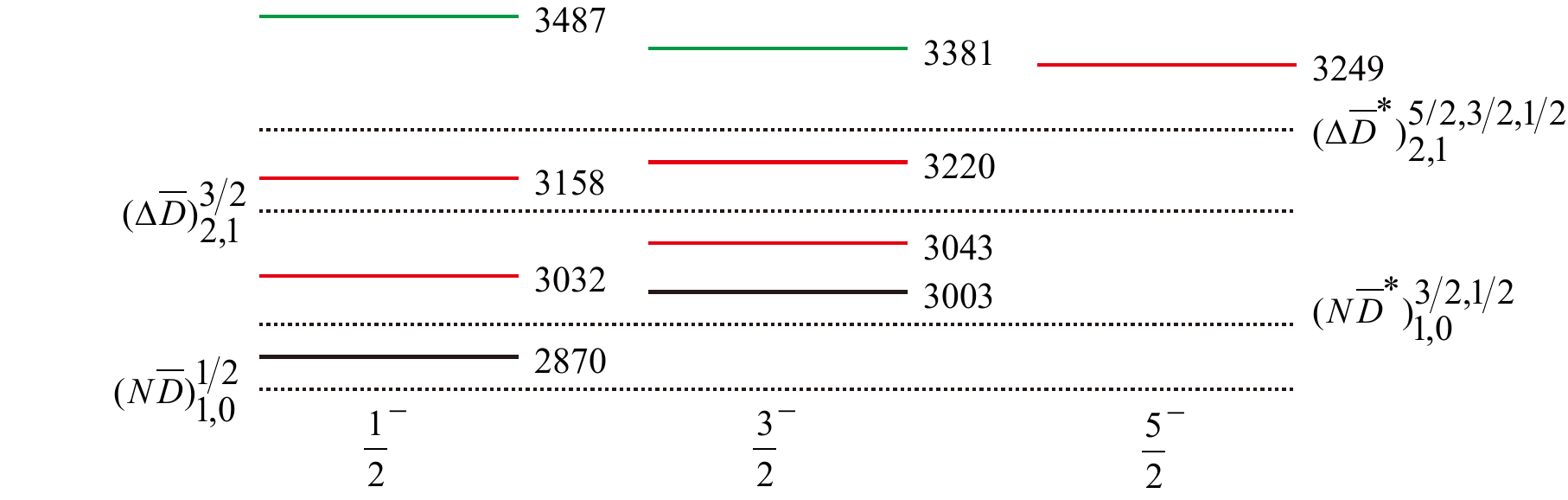}&$\qquad$& \includegraphics[width=240pt]{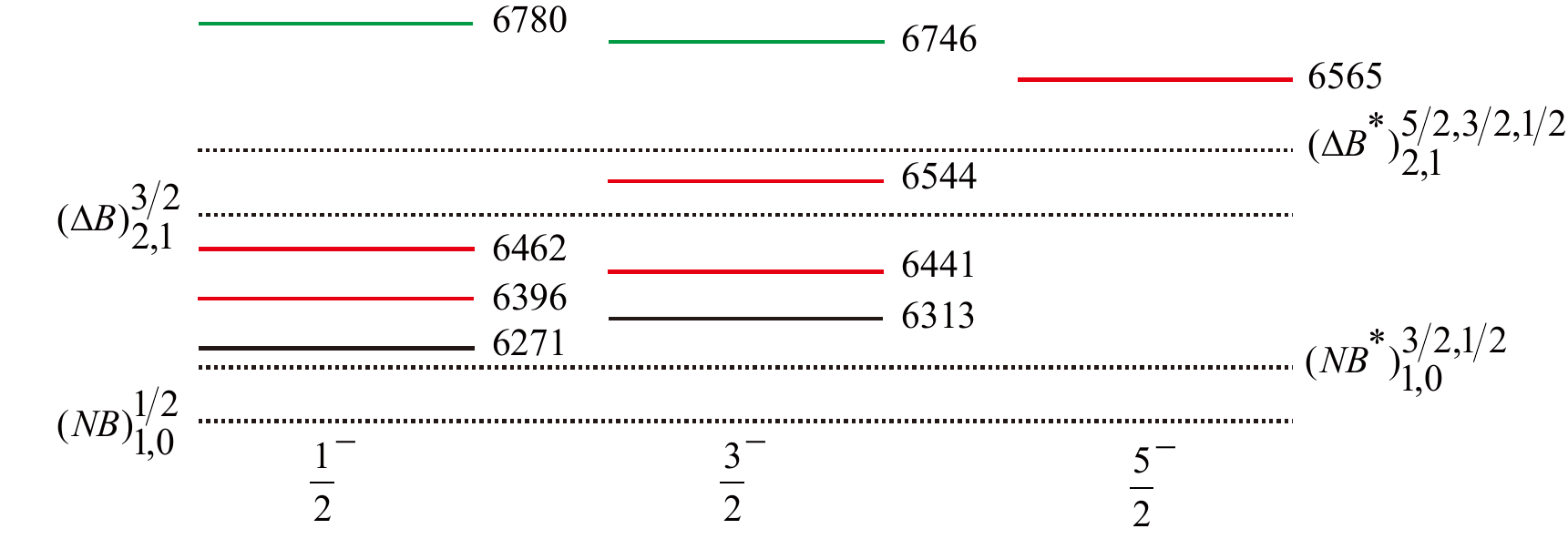}\\
(a) $nnnn\bar{c}$ states&$\qquad$& (b)  $nnnn\bar{b}$ states\\
\includegraphics[width=240pt]{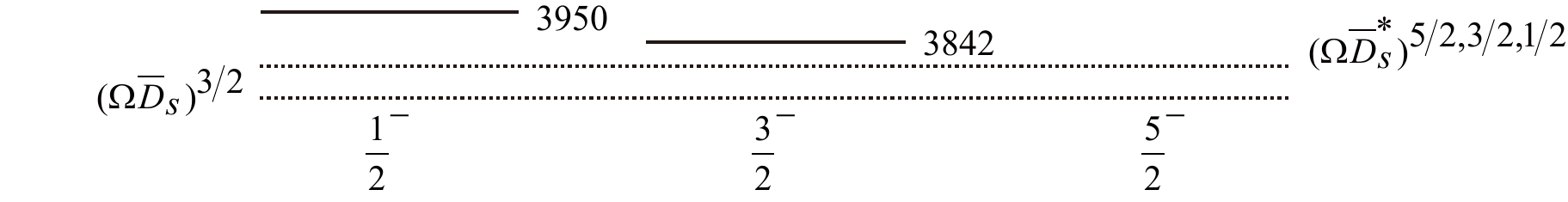}&$\qquad$&\includegraphics[width=240pt]{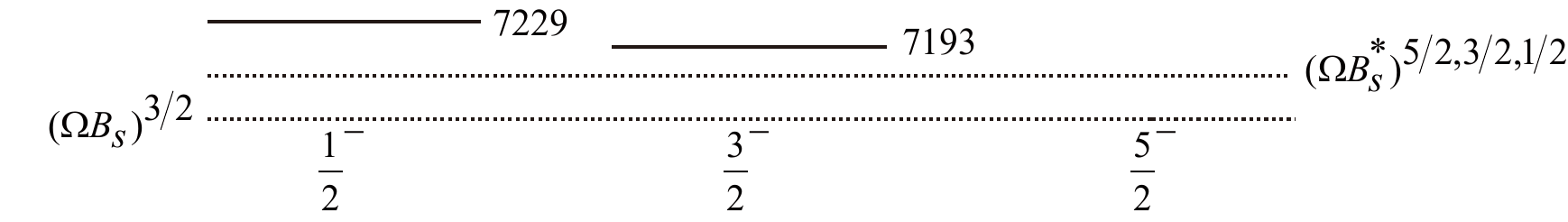}\\
(c)  $ssss\bar{c}$ states&$\qquad$& (d)  $ssss\bar{b}$ states\\
\includegraphics[width=240pt]{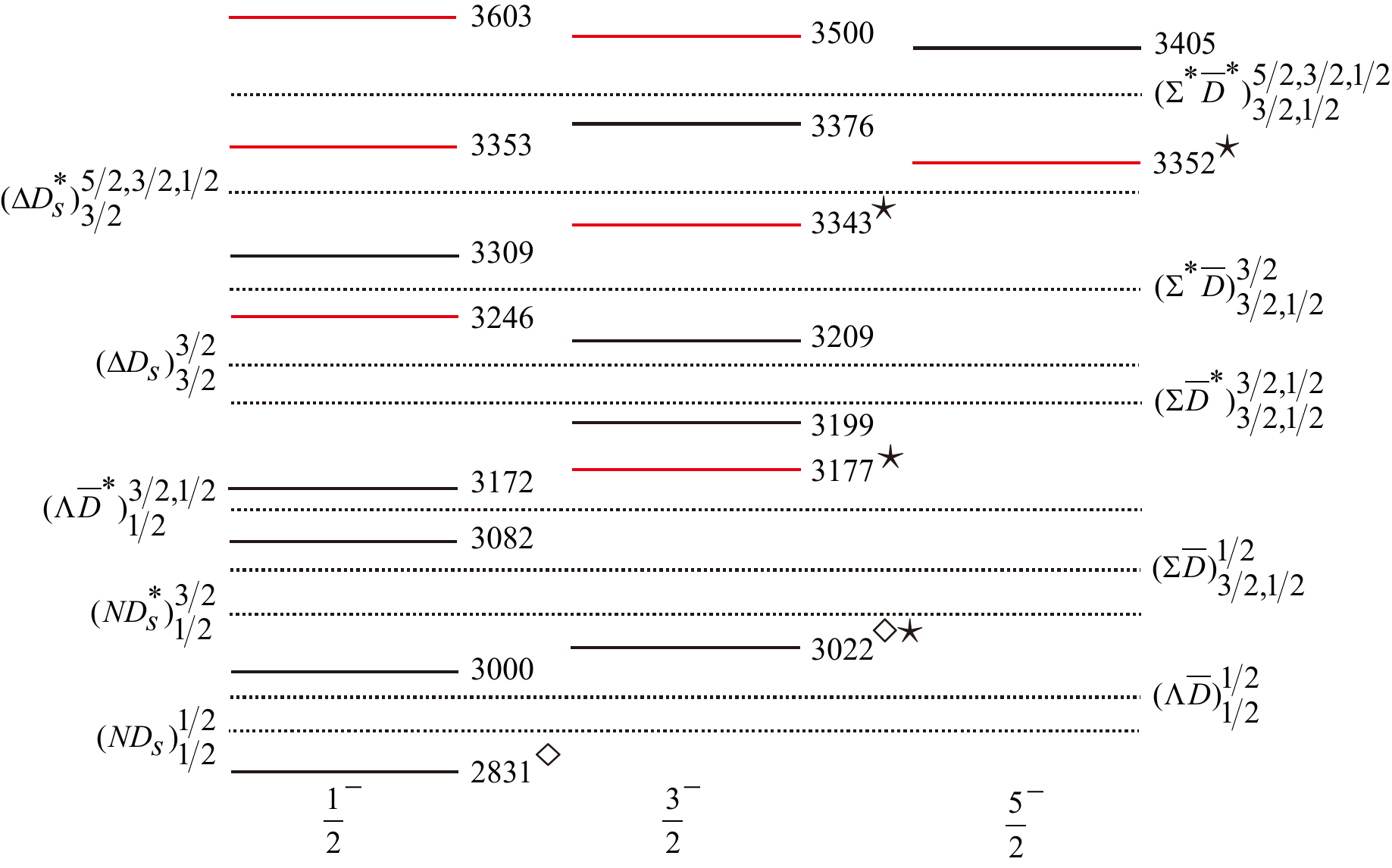}&$\qquad$&
\includegraphics[width=240pt]{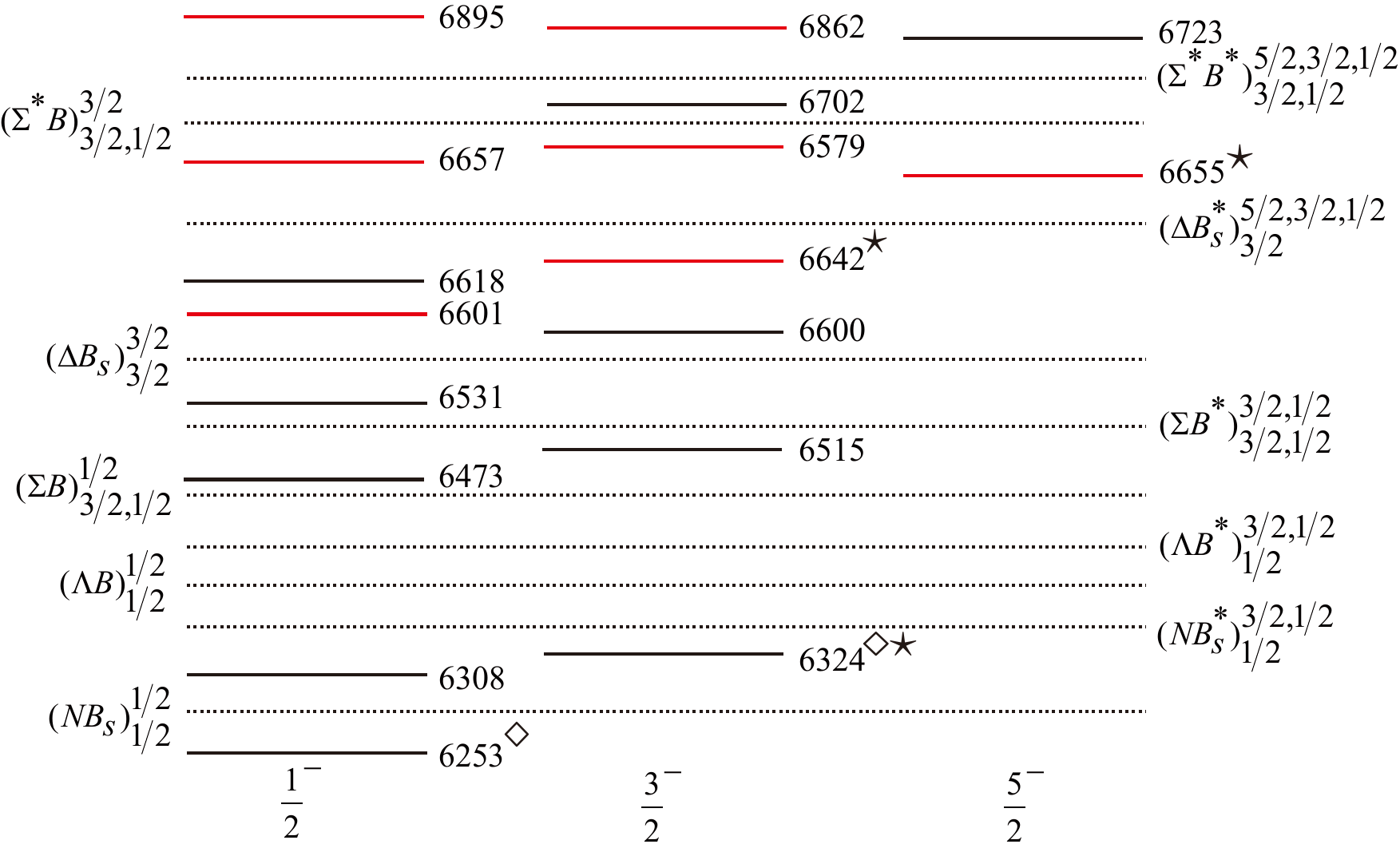}\\
(e) \begin{tabular}{c}  $nnns\bar{c}$ states\end{tabular} &&(f)  $nnns\bar{b}$ states\\
\includegraphics[width=240pt]{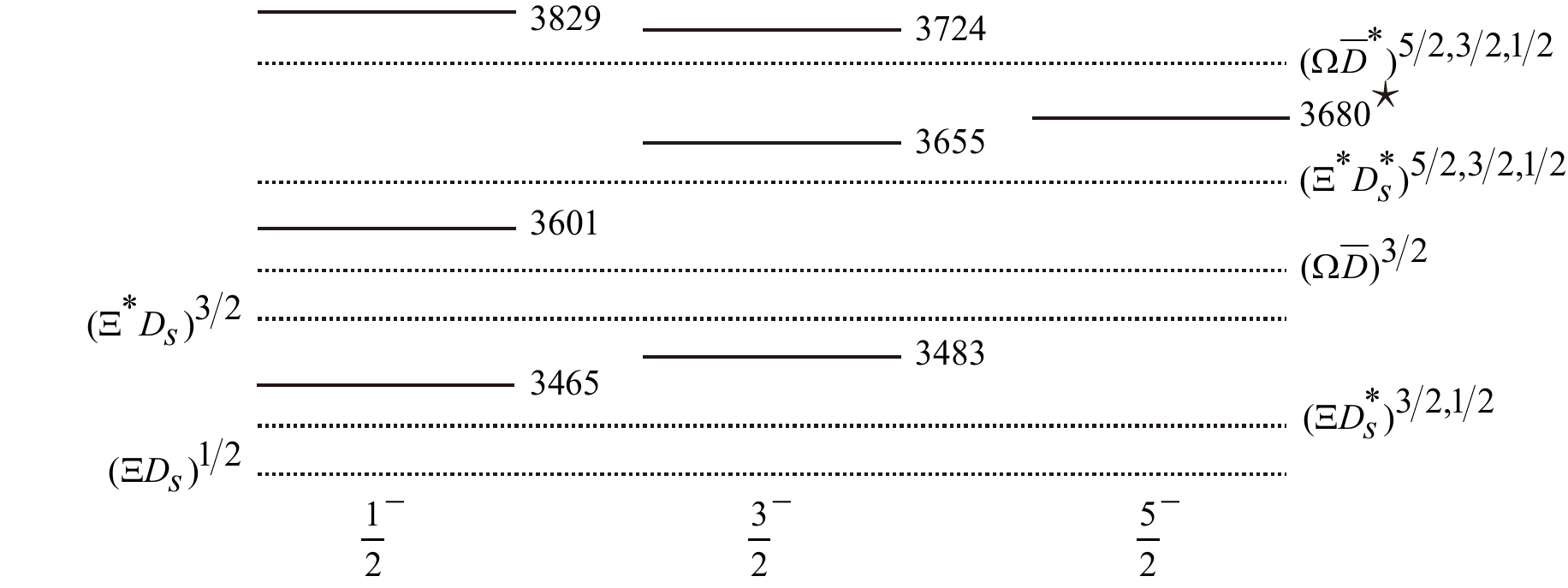}&$\qquad$&
\includegraphics[width=240pt]{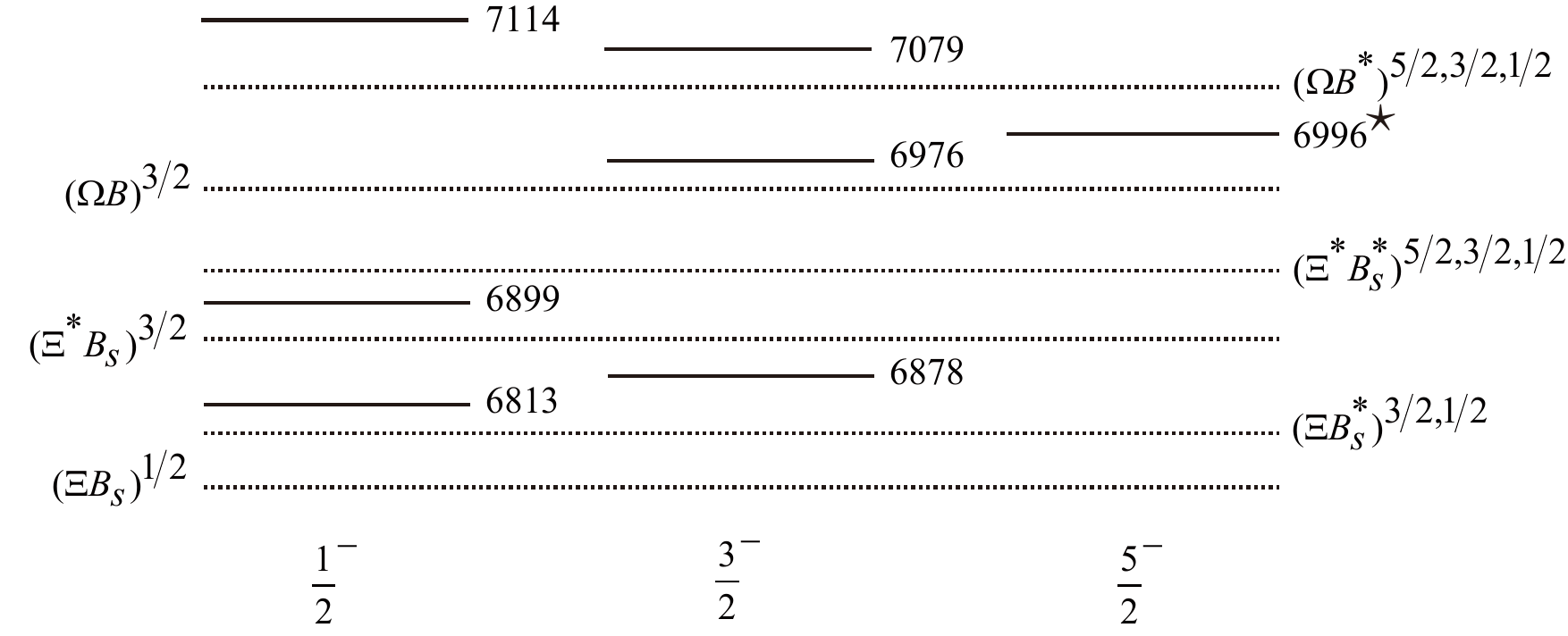}\\
(g) \begin{tabular}{c}  $sssn\bar{c}$ states\end{tabular} &&(h)  $sssn\bar{b}$ states\\
\includegraphics[width=240pt]{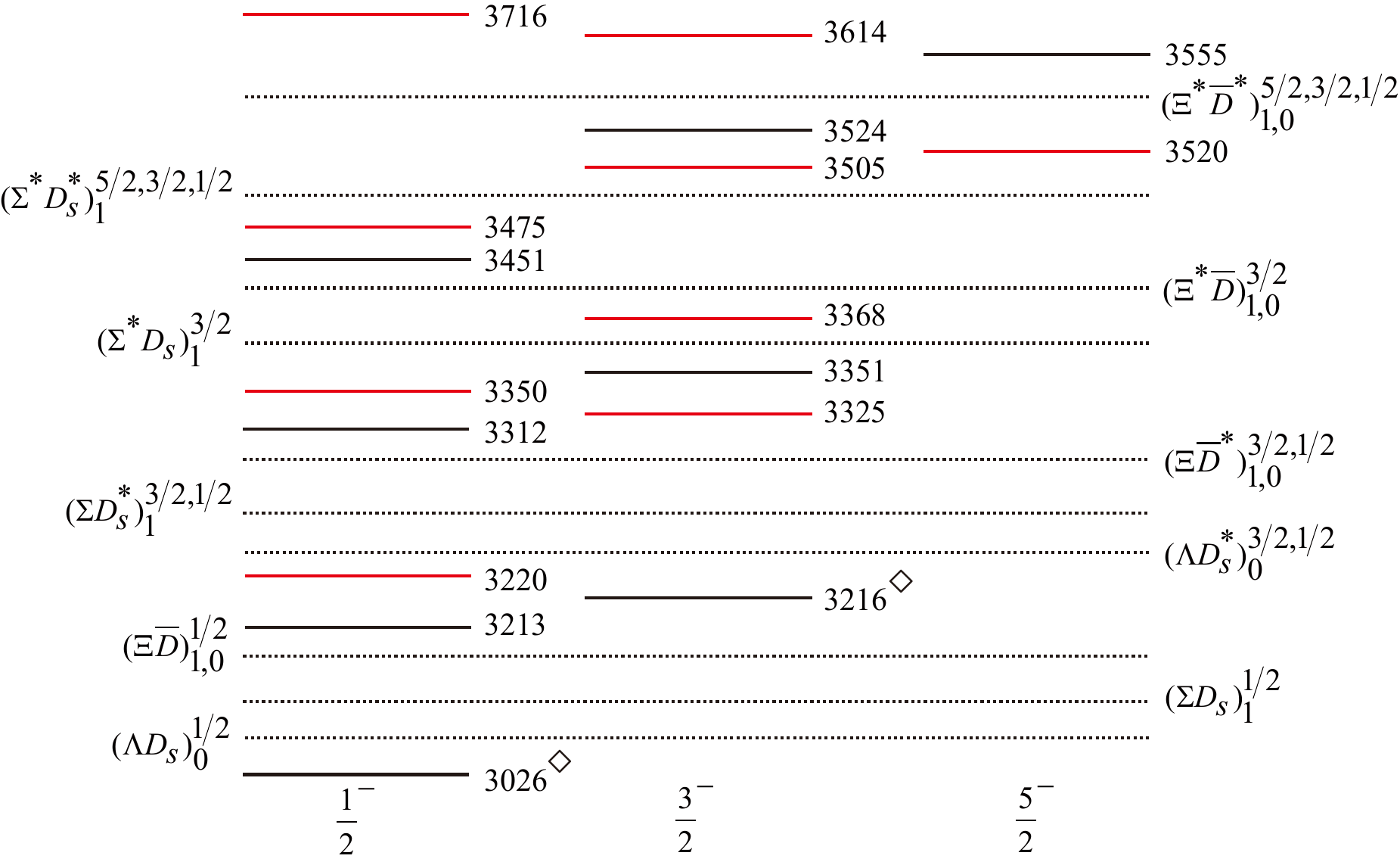}&$\qquad$&
\includegraphics[width=240pt]{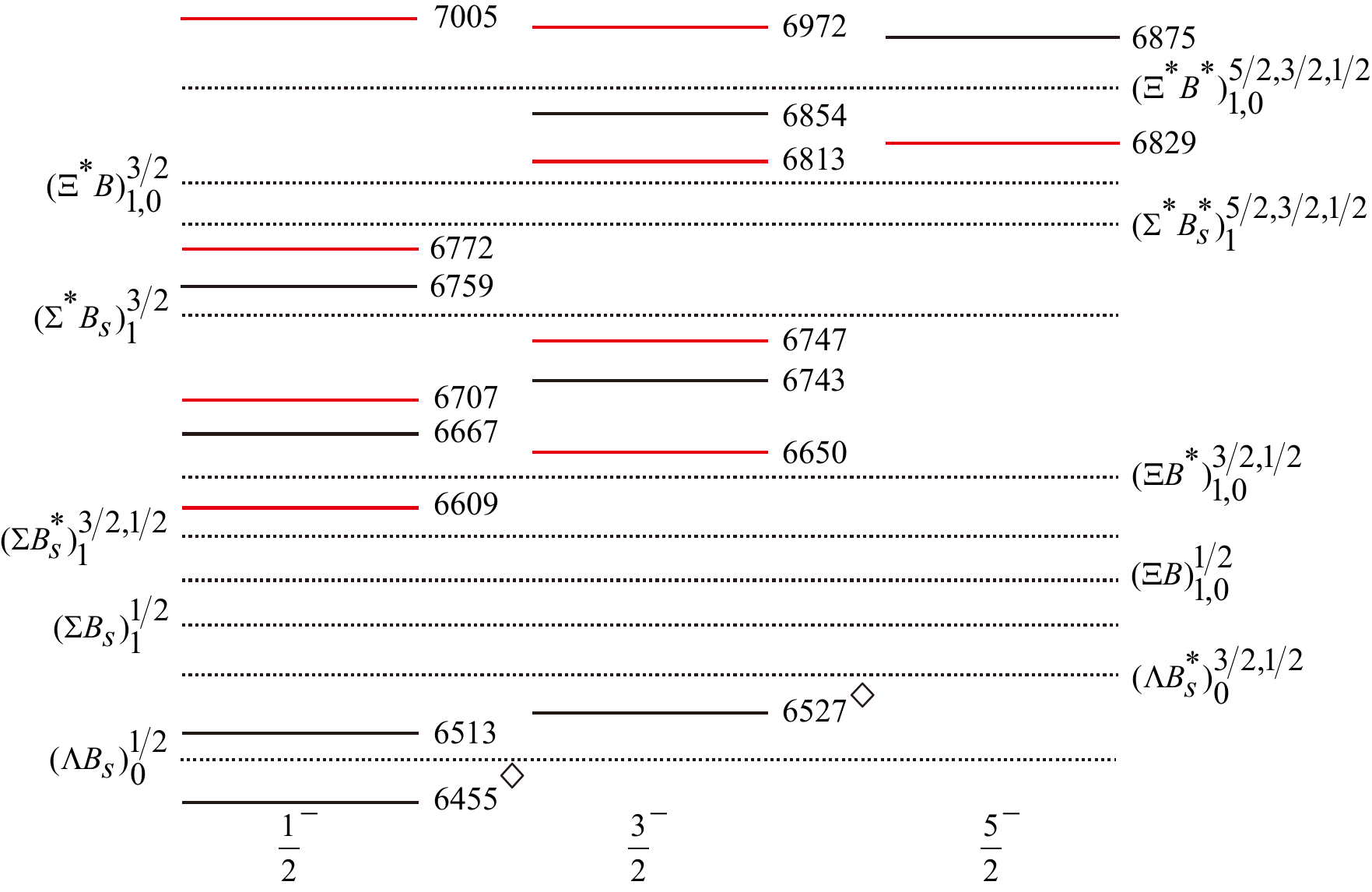}\\
(i) \begin{tabular}{c}  $nnss\bar{c}$ states\end{tabular} &&(j)  $nnss\bar{b}$ states\\
\end{tabular}
\caption{
Relative positions (units: MeV) for the $nnnn\bar{c}$, $nnnn\bar{b}$, $ssss\bar{c}$,
$ssss\bar{b}$, $nnns\bar{c}$, $nnns\bar{b}$, $sssn\bar{c}$, $sssn\bar{b}$, $nnss\bar{c}$, and $nnss\bar{b}$ pentaquark states labeled with solid lines.
In the $nnnn\bar{c}$ ($nnnn\bar{b}$) and $nnss\bar{c}$ ($nnss\bar{b}$) subsystems, the green, red, and black lines represent the pentaquark states with $I=2$, $I=1$, and $I=0$, respectively. In the $nnns\bar{c}$ ($nnns\bar{b}$) subsystem, the red and black lines denote the pentaquark states with $I=3/2$ and $I=1/2$, respectively.
The dotted lines denote various S-wave baryon-meson thresholds, and the superscripts (subscript) of the labels, e.g. $(\Delta\bar{D}^{*})^{5/2,3/2,1/2}_{2,1}$, represent the possible total angular momenta (isospin) of the channels.
Since the $ssss\bar{c}$ ($ssss\bar{b}$) and $sssn\bar{c}$ ($sssn\bar{b}$) pentaquark states have the same isospin quantum number, we do not label isospin quantum number of their baryon-meson thresholds.
We mark these ``stable" pentaquarks, which cannot decay into baryon-meson final states via the S-wave interaction, with ``$\diamond$" after their masses.
We mark the pentaquark whose wave function overlaps with that of one special baryon-meson state more than 90\% with
``$\star$" after their masses.
}
\label{fig-nnnsQ}
\end{figure*}

Assigning the value to these parameters in the expressions of the CMI Hamiltonian for these discussed $qqqq\bar{Q}$ pentaquark subsystems, 
we obtain the corresponding mass spectrum as shown in Fig. \ref{fig-nnnsQ}.
Meanwhile, we also list the baryon-meson thresholds relevant to the allowed decay channels of the corresponding pentaquark states. 
For these rearranged decay channels, we label the spin (isospin) of the baryon-meson states with superscript (subscript).
When the spin (isospin) of an initial pentaquark state is equal to the number in the superscript (subscript) of a baryon-meson state,
this pentaquark may decay into the corresponding baryon-meson channel via the $S$-wave interaction.
In addition, we define the ``stable" pentaquark state if the state is below the lowest
baryon-meson threshold, which is marked by ``$\diamond$'' in Tables \ref{Kij}, \ref{eigenvector-nnnsC}, \ref{value-nnnsC}, and Figs. \ref{fig-nnnsQ}, \ref{nnSC}.
For simplicity, we use $\rm P_{content}$(Mass, $I$, $J^{P}$) [$\rm T_{content}$(Mass, $I$, $J^{P}$)] to label a particular pentaquark [tetraquark] state.

\subsubsection{Stability of pentaquark}

We further explore the stabilities of the pentaquark states in the $qqqq\bar{Q}$ system.
The authors in Refs. \cite{Wu:2016vtq,Li:2018vhp,Wu:2018xdi,Cheng:2019obk,Cheng:2020nho} proposed a method to check the evolution of effective interaction between two quarks by varying the corresponding effective coupling strengths, by which we can roughly test whether the involved multiquark states are stable. In this work, we also adopt the same approach to discussing the stabilities of the $qqqq\bar{Q}$ pentaquark states.

Due to the complicated couplings among different color-spin structures, the properties of the interaction between quark pairs become ambiguous.
Here, we need to further introduce a new quantity, by which we can determine whether the effective interaction is attractive or repulsive.
To find it, we may study the effects induced by the artificial change of the coupling strengths in the Hamiltonian.
The mass may increase or decrease when  reducing the coupling strength. If the effective interaction between the considered components is attractive (repulsive), the mass would be shifted upward (downward) and vice versa \cite{Wu:2016vtq}.
To illustrate this effect, we define a dimensionless variable
\begin{eqnarray}\label{Eq19}
K_{ij}=\frac{\Delta m}{\Delta v_{ij}}\rightarrow\frac{\partial M}{\partial v_{ij}},
\end{eqnarray}
where $\Delta v_{ij}$ ($\Delta m$) is the variation of the coupling strength (the eigenvalue of the CMI Hamiltonian) in Eq. (\ref{Eq3}).
When $\Delta v_{ij}$ is small enough, $K_{ij}$ tends to be a constant $\partial M/\partial v_{ij}$.
{In fact, the value of the $K_{ij}$ mainly depends on the matrix element of $\vec\lambda_i\cdot \vec\lambda_j \vec\sigma_i\cdot\vec\sigma_j$.
Besides, in a $n\times n$ CMI Hamiltonian, the contributions from other diagonal components and off-diagonal
components would give small corrections to the $K_{ij}$ values,
but these corrections can hardly affect the sign of $K_{ij}$.}


We decrease one relevant coupling strength $v_{ij}$ to 99\% of its original value and keep other coupling parameters unchanged, then the corresponding $K_{ij}$ value can be determined.
For the discussed $qqqq\bar{Q}$ pentaquark states, the $K_{ij}$ values are presented in Table \ref{Kij}.
For a tetraquark state ($qq\bar{q}\bar{q}$), one can easily understand that a relatively stable state is favored if the effective $qq$ and $\bar{q}\bar{q}$ interactions are all attractive.

However, the situation becomes complicated for a pentaquark state since more types of quark pairs and interactions may exist.
{In fact, it is hard to judge whether a pentaquark state is stable or not just from the signs of $K_{ij}$. At present, the defined $K_{ij}$ is expected to be a characteristic quantity to describe the stability of the multiquark state before an actual dynamical calculation is performed.}

Later, we will give some further discussions on the stabilities of the $nn\bar{s}\bar{c}$ and $qqqq\bar{Q}$ pentaquark states.

\subsubsection{Relative decay widths of pentaquarks}

Besides the studies of the mass spectra and stabilities, we also discuss the two-body strong decays of the $qqqq\bar{Q}$ pentaquark states based on the obtained eigenvectors \cite{An:2020jix,An:2021vwi,Jaffe:1976ig,Strottman:1979qu,Weng:2019ynva,Weng:2020jao,Zhao:2014qva,Wang:2015epa,Weng:2021ngd,Weng:2021hje}.
The overlap between the pentaquark and a specific baryon $\otimes$ meson state can be calculated by transforming the eigenvectors of the pentaquark state into the baryon $\otimes$ meson configuration.
The baryon and meson components inside the pentaquark can be either the 1 $\otimes$ 1 component or 8 $\otimes$ 8 component.
The 1 $\otimes$ 1 component can be easily dissociated into an $S$-wave baryon and an $S$-wave meson, which is denoted as the OZI-superallowed decay process, while the 8 $\otimes$ 8 component cannot fall apart without the gluon exchange force.
In our work, we only focus on the OZI-superallowed decay process.
Thus, we present these possible overlaps between a pentaquark state and its meson $\otimes$ baryon component corresponding to the 1 $\otimes$ 1 dissociation in Table \ref{eigenvector-nnnsC}.
Here, the overlaps are proportional to the $1\otimes 1$ components in the pentaquark states.
Although the relative signs may affect the shapes of the wave function for the corresponding pentaquark states, the relative decay widths will not be affected because they depend on the square of the overlaps.

As shown in Table \ref{eigenvector-nnnsC}, the $\rm P_{n^{3}s\bar{c}}(3352, 3/2, 5/2^{-})$ state  completely couples to the $\Delta\bar{D}_{s}^{*}$ system, which can be written as the direct product of a $\Delta$ baryon and a $\bar{D}_{s}^{*}$ meson in the $nnns\bar{c}$ pentaquark subsystem.
Meanwhile, the $\rm P_{n^{3}s\bar{c}}(3343, 3/2, 3/2^{-})$, $\rm P_{n^{3}s\bar{c}}(3177, 3/2, 3/2^{-})$, and $\rm P_{n^{3}s\bar{c}}(3022, 1/2, 3/2^{-})$ states couple almost completely to the $\Delta\bar{D}^{*}_{s}$, $\Delta\bar{D}_{s}$, and $N\bar{D}^{*}_{s}$ baryon-meson systems, respectively.
This kind of pentaquark state behaves similarly to the ordinary scattering state that is composed of a baryon and a meson if its inner interaction is not strong enough. However, we still cannot exclude the possibility of it as the resonance or the bound state dynamically generated from the strong interaction.
These kinds of pentaquarks deserve a more careful study with some hadron-hadron interaction models in the future. We label these states with $``\star"$ in Tables \ref{Kij}, \ref{eigenvector-nnnsC}, \ref{value-nnnsC}, and Fig. \ref{fig-nnnsQ}.
{
For the $\rm P_{n^{3}s\bar{c}}(3022,1/2,3/2^{-})$ state, its mass mainly comes from the $97.8 \%$ $N D_{s}^{*}$ ($1\otimes 1$) component, while other $8\otimes8$ components and the contributions from the off-diagonal matrix elements in the Hamiltonian give small corrections to the mass of the 
$\rm P_{n^{3}s\bar{c}}(3022,1/2,3/2^{-})$ state. It results in that this state should lie below the $N \bar{D}^{*}_{s}$ threshold.
Since the $\rm P_{n^{3}s\bar{c}}(3022,1/2,3/2^{-})$ state is below the $N \bar{D}^{*}_{s}$ threshold and its strong decay channels are kinematically forbidden, $\rm P_{n^{3}s\bar{c}}(3022,1/2,3/2^{-})$ could be a good candidate of the molecular state.
Besides, $D^{*}_{s}$ can decay into the $D_{s}\gamma$ channel or the isospin breaking channel $D_{s}\pi$; the
$\rm P_{n^{3}s\bar{c}}(3022,1/2,3/2^{-})$ state is expected to be a narrow state.}

Moreover, we find that the $\rm P_{n^{3}s\bar{c}}(3000, 1/2, 3/2^{-})$ state has $87.9 \%$ of the $N\bar{D}_{s}^{*}$ component, and the $\rm P_{n^{3}s\bar{c}}(2831, 1/2, 1/2^{-})$ state also has more than $85 \%$ of the $N\bar{D}_{s}$ component.
For such states, we still cannot rule out the possibilities of assigning them as genuine pentaquark states.
Thus, except for the states labeled with ``$\star$'', other states are safely regarded as the genuine pentaquark states in the $qqqq\bar{Q}$ systems.

Note that the $qqqq\bar{Q}$ pentaquark states have no constituent light antiquarks and have four valance light quarks. If such a  state could be observed in its two-body strong decay pattern, this state must be a $qqqq\bar{Q}$ pentaquark state.
{
However, since we do not consider any kinetic effects in the CMI model, it is still difficult to estimate the total width and line shape of a pentaquark state.
Thus, based on the CMI model, we mainly focus on the relative decay widths of the discussed pentaquark states, and provide an effective approach to identifying the configurations of multiquark states via their relative decay widths.}

For the $L$-wave two-body decay, its partial wave decay width reads as \cite{Weng:2019ynva,Weng:2020jao,An:2020jix,An:2021vwi,Weng:2021ngd,Weng:2021hje}
\begin{eqnarray}\label{Eq20}
\Gamma_{i}=\gamma_{i}\alpha\frac{k^{2L+1}}{m^{2L}}\,|c_{i}|^{2}.
\end{eqnarray}
Here, $m$ is the mass of initial state and
$k$ is the momentum of the final state in the rest frame of the initial state.
Since $(k/m)^{2}$ is of the $\mathcal{O}(10^{-2})$ order or even smaller, the contribution from higher partial wave decays would be suppressed. Thus, we only need to consider the $S$-wave decays.
$\alpha$ is an effective coupling constant and $c_{i}$ is the overlap between the pentaquark and a specific baryon $\otimes$ meson state, which is given in Tables \ref{eigenvector-nnnnQ}, and \ref{eigenvector-nnnsC}.
\begin{table}[htp]
\caption{The approximate relation about $\gamma_{i}$ for the $qqqq\bar{Q}$ system.
}\label{gamma}
\begin{center}
\renewcommand\arraystretch{1.4}
\begin{tabular}{r|llr}
\toprule[1pt]
\toprule[0.5pt]
Subsystem&\multicolumn{3}{c}{$\gamma_{i}$}\\
\hline
$nnnn\bar{c}\quad$&$\gamma_{\Delta\bar{D}}=\gamma_{\Delta\bar{D}^{*}}$&$\gamma_{N\bar{D}}=\gamma_{N\bar{D}^{*}}$\\
$nnnn\bar{b}\quad$&$\gamma_{\Delta B}=\gamma_{\Delta B^{*}}$&$\gamma_{N B}=\gamma_{N B^{*}}$\\
$ssss\bar{c}\quad$&$\gamma_{\Omega \bar{D}_{s}}=\gamma_{\Omega\bar{D}_{s}^{*}}$\\
$ssss\bar{b}\quad$&$\gamma_{\Omega B_{s}}=\gamma_{\Omega B_{s}^{*}}$\\
\toprule[0.5pt]
\multirow{2}*{$nnns\bar{c}\quad$}&$\gamma_{\Delta\bar{D}_{s}}=\gamma_{\Delta\bar{D}^{*}_{s}}$&$\gamma_{N\bar{D_{s}}}=\gamma_{N\bar{D}^{*}_{s}}$\\
&\multicolumn{2}{l}{$\gamma_{\Sigma^{*} \bar{D}}=\gamma_{\Sigma^{*} \bar{D}^{*}}\approx\gamma_{\Sigma \bar{D}^{*}}=\gamma_{\Sigma \bar{D}}$}&$\gamma_{\Lambda \bar{D}}=\gamma_{\Lambda \bar{D}^{*}}$\\
\multirow{2}*{$nnns\bar{b}\quad$}&$\gamma_{\Delta B_{s}}=\gamma_{\Delta B^{*}_{s}}$&$\gamma_{N B_{s}}=\gamma_{N B_{s}}$\\
&\multicolumn{2}{l}{$\gamma_{\Sigma^{*} B}=\gamma_{\Sigma^{*} B^{*}}\approx\gamma_{\Sigma B^{*}}=\gamma_{\Sigma B}$}&$\gamma_{\Lambda B^{*}}=\gamma_{\Lambda B}$\\
$sssn\bar{c}\quad$&\multicolumn{2}{l}{$\gamma_{\Xi^{*} \bar{D}_{s}}=\gamma_{\Xi^{*} \bar{D}^{*}_{s}}\approx\gamma_{\Xi \bar{D}^{*}_{s}}=\gamma_{\Xi \bar{D}_{s}}$}&$\gamma_{\Omega \bar{D}}=\gamma_{\Omega \bar{D}^{*}}$\\
$sssn\bar{b}\quad$&\multicolumn{2}{l}{$\gamma_{\Xi^{*} B_{s}}=\gamma_{\Xi^{*} B^{*}_{s}}\approx\gamma_{\Xi B^{*}_{s}}=\gamma_{\Xi B_{s}}$}&$
\gamma_{\Omega B}=\gamma_{\Omega B^{*}}$\\
\toprule[0.5pt]
\multirow{2}*{$nnss\bar{c}\quad$}&\multicolumn{2}{l}{$\gamma_{\Sigma^{*} \bar{D}^{*}_{s}}=\gamma_{\Sigma^{*} \bar{D}_{s}}\approx\gamma_{\Sigma \bar{D}^{*}_{s}}=\gamma_{\Sigma \bar{D}_{s}}$}&$\gamma_{\Lambda \bar{D}_{s}}=\gamma_{\Lambda \bar{D}^{*}_{s}}$\\
&\multicolumn{2}{l}{$\gamma_{\Xi^{*} \bar{D}^{*}}=\gamma_{\Xi^{*} \bar{D}}\approx\gamma_{\Xi \bar{D}^{*}}=\gamma_{\Xi \bar{D}}$}\\
\multirow{2}*{$nnss\bar{b}\quad$}&\multicolumn{2}{l}{$\gamma_{\Sigma^*B_{s}^*}=\gamma_{\Sigma^*B_s}\approx \gamma_{\Sigma B^*_s}=\gamma_{\Sigma B_s}$}&$\gamma_{\Lambda B_s}=\gamma_{\Lambda B^*_s}$\\
&\multicolumn{2}{l}{$\gamma_{\Xi^{*} B^{*}}=\gamma_{\Xi^{*} B}\approx\gamma_{\Xi B^{*}}=\gamma_{\Xi B}$}\\
\toprule[0.5pt]
\toprule[1pt]
\end{tabular}
\end{center}
\end{table}

The parameter $\gamma_{i}$ depends on the spatial wave functions of the initial and final states. 
In the quark model, the spatial wave functions of the ground scalar mesons are the same as those of the ground vector mesons \cite{Weng:2019ynva}. Thus, we adopt the approximation to ignore the differences of spatial wave functions of the $\Sigma^{*}$ ($\Xi^{*}$) and $\Sigma$ ($\Xi$), where the approximated relations for $\gamma_i$ are collected  in Table \ref{gamma}. 
Note that the spatial wave function of the $\Lambda$ baryon is different from those of the $\Sigma$ and $\Sigma^{*}$ baryons, and thus we can not directly calculate the ratios of relative partial decay widths between the decay channels with $\Lambda$ and $\Sigma$ ($\Sigma^*$) in the final states.


Based on Eq. (\ref{Eq20}), we present the value of $k\,|c_{i}|^{2}$ for each decay process in Tables \ref{value-nnnnQ} and \ref{value-nnnsC}. From Tables \ref{value-nnnnQ} and \ref{value-nnnsC}, one can roughly estimate the ratios of decay widths for related decay channels if neglecting the differences of $\gamma_i$ parameters. Such an approximation was also adopted in Refs. \cite{Cheng:2019obk,Cheng:2020nho}. 
In the following, we  specifically discuss the mass spectra, stability, and strong decay properties according to the results in Table \ref{flavor1}.

\subsection{The $X_{0}(2900)$ and its partner states}

\begin{table*}[htbp]
\centering \caption{The mass spectra, $K_{ij}$, overlaps, and relative widths of the $nn\bar{s}\bar{c}$ states.
 The masses are all in units of MeV.
}\label{mass1}
\renewcommand\arraystretch{1.5}
\renewcommand\tabcolsep{2.3pt}
\scalebox{1}{
\begin{tabular}{cl|cccc|cccc|cccc|c}
\bottomrule[1.5pt]
\bottomrule[0.5pt]
\multicolumn{2}{c}{$nn\bar{s}\bar{c}$}&\multicolumn{4}{c}{$K_{ij}$}&\multicolumn{4}{c}{Overlaps}&\multicolumn{4}{c}{Relative widths}&\multicolumn{1}{c}{Uncertainties}\\
$I(J^{P})$&\multicolumn{1}{c|}{Mass}&$nn$&$\bar{s}\bar{c}$&$n\bar{c}$&$n\bar{s}$&$K^{*}D^{*}$&$K^{*}D$&$KD^{*}$&$KD$&$K^{*}D^{*}$&$K^{*}D$&$KD^{*}$&$KD$&$+2.6\%m_{ij}$\\
\bottomrule[0.5pt]
$1(2^{+})$&2968.1&2.67&2.67&2.67&2.67&0.577&&&&96&&&&3041.9\\
$1(1^{+})$&2980.0&3.29&-4.29&3.82&6.54&0.677&-0.176&-0.016&&144&16&0.2&&3052.5\\
          &2879.4&2.71&0.09&-8.62&2.67&0.088&0.564&0.136&&$\times$&122&11&&2953.1\\
          &2630.1&3.33&-2.47&2.13&-11.88&0.185&0.259&0.631&&$\times$&$\times$&131&&2702.5\\
$1(0^{+})$&3045.6&3.41&3.41&7.38&7.38&0.721&&&-0.026&223&&&0.6&3117.8\\
          &2546.4&3.25&3.25&-12.71&-12.71&0.253&&&0.645&$\times$&&&166&2619.0\\
\bottomrule[0.5pt]
$0(2^{+})$&2869.4&-1.33&-1.33&6.67&6.67&-0.816&&&&$\times$&&&&2940.5\\
$0(1^{+})$&2855.9&-4.08&3.42&5.17&5.82&0.703&0.171&-0.082&&$\times$&10&4&&2928.1\\
          &2697.2&-3.79&1.19&-17.86&5.97&0.008&0.737&0.024&&$\times$&$\times$&0.2&&2769.3\\
          &2400.6&-2.80&0.72&6.02&-18.46&0.079&-0.105&-0.759&&$\times$&$\times$&$\times$&&2472.3\\
$0(0^{+})$&2793.0&-6.34&-6.33&5.15&5.15&-0.637&&&0.103&$\times$&&&7&\bf{2866.1}\\
          &2220.7$\star$&-3.00&-3.00&-18.49&-18.49&-0.104&&&0.757&$\times$&&&$\times$&2293.4\\
\bottomrule[0.5pt]
\bottomrule[1.5pt]
\end{tabular}}
\end{table*}

\begin{figure}[t]
\begin{tabular}{ccc}
\includegraphics[width=250pt]{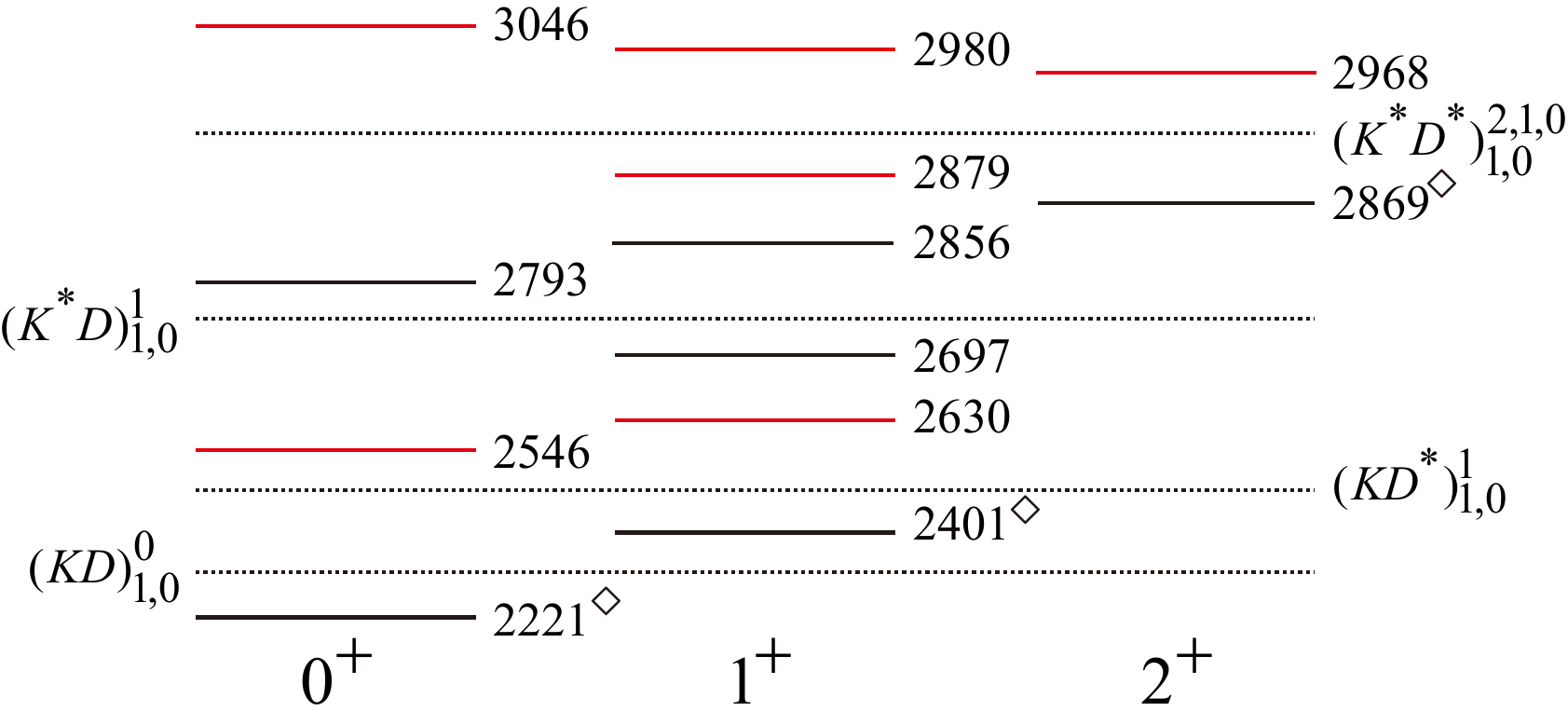}\\
$nn\bar{s}\bar{c}$ states\\
\end{tabular}
\caption{
Relative positions (units: MeV) for the $nn\bar{s}\bar{c}$ states.
Here, the red (black) lines represent the $nn\bar{s}\bar{c}$ states with $I=1 (0)$.
The dotted lines denote various S-wave meson-meson thresholds, and the superscripts (subscript) of the labels, e.g. $(K^{*}D^{*})^{2,1,0}_{2,1}$, represent the possible total angular momenta (isospin) of the channels.
Moreover, we mark these ``stable" states, which cannot decay into meson-meson final states via S-wave interaction, with ``$\diamond$" after their masses.
}
\label{nnSC}
\end{figure}

As mentioned in the Introduction, there are some similarities between the $nn\bar{s}\bar{c}$ and $nnss\bar{c}$ systems. Before discussing the properties of 
these involved pentaquarks, we should firstly study the $nn\bar{s}\bar{c}$ tetraquark relevant to the 
$X(2900)$ \cite{Aaij:2020hon,LHCb:2020pxc}.

In 2020, the LHCb Collaboration reported an enhancement on the $D^{-}K^{+}$ invariant mass distribution in the decay channel of $B^+\to D^+D^-K^+$  \cite{Aaij:2020hon,LHCb:2020pxc}.
The best fit requires spin-0 and spin-1 states, and their resonance parameters were determined to be
\begin{eqnarray}
X_0(2900):M&=&2.866\pm0.007\pm0.002\ \rm{GeV},\nonumber\\
\Gamma&=&57\pm12\pm4\ \rm{MeV},\nonumber\\
X_1(2900):M&=&2.904\pm0.005\pm0.001\ \rm{GeV},\nonumber\\
\Gamma&=&110\pm11\pm4\ \rm{MeV}.\nonumber
\end{eqnarray}
The $D^-K^+$ decay channel indicates that their quark components should be $nn\bar{s}\bar{c}$.
In the following, we present a brief discussion on the possible tetraquark spectrum with the $nn\bar{s}\bar{c}$ configuration. The $nn\bar{s}\bar{c}$ tetraquarks can be grouped into isoscalar and isovector systems since the isospin of two $nn$ quarks can couple to either $I=1$ or $I= 0$.
The results for the $nn\bar{s}\bar{c}$ tetraquark states are listed in Table \ref{mass1}, including their masses, $K_{ij}$ values, overlaps, and relative decay widths.
According to the overlaps shown in Table \ref{mass1}, all of $nn\bar{s}\bar{c}$ states could be safely regarded as the genuine tetraquark states, and we also plot the mass spectrum and the corresponding meson-meson thresholds  relevant to the allowed decay channels in Fig. \ref{nnSC}. These $nn\bar{s}\bar{c}$ states only have the $n\bar{s}-n\bar{c}$ rearrangement decay mode.
Thus, their rearrangement decay channels include $K^{*}D^{*}$, $K^{*}D$, $KD^{*}$, and $KD$.

From Fig. \ref{nnSC}, we notice that these  lowest isoscalar tetraquark states with $I(J^{P})=0(2^{+}), 0(1^{+}),$ and $0(0^{+})$ are below the thresholds of all allowed strong decay channels. Thus, they are considered as the stable tetraquark states.


Next, we specify the $nn\bar{s}\bar{c}$ subsystem with other $I(J^{P})$ quantum numbers. There is only one tetraquark state $\rm T_{n^{2}\bar{s}\bar{c}}(2968,1,2^{+})$ with quantum number $I(J^{P})=1(2^{+})$.
We notice these values of $K_{nn}$, $K_{\bar{s}\bar{c}}$, $K_{n\bar{s}}$ and $K_{n\bar{c}}$ are all positive. Thus, it is difficult to form a bound tetraquark state since the interactions between different quark components are all repulsive.
There are three tetraquark states $\rm T_{n^{2}\bar{s}\bar{c}}(2980,1,1^{+})$,
$\rm T_{n^{2}\bar{s}\bar{c}}(2879,1,1^{+})$, and $\rm T_{n^{2}\bar{s}\bar{c}}(2630,1,1^{+})$ with the quantum number $I(J^{P})=1(1^{+})$.
The lowest $I(J^P)=1(1^{+})$ state $\rm T_{n^{2}\bar{s}\bar{c}}(2630,1,1^{+})$ can only decay into the $KD^{*}$ final states.
Meanwhile, the $\rm T_{n^{2}\bar{s}\bar{c}}(2630,1,1^{+})$ state can decay into the $KD^{*}$ and $K^{*}D$ final states.
For the $\rm T_{n^{2}\bar{s}\bar{c}}(2980,1,1^{+})$ state, its mass is larger than the corresponding allowed decay channels, and we have
\begin{equation}
\Gamma_{K^{*}D^{*}}:\Gamma_{K^{*}D}=9:1,
\end{equation}
which shows that its dominant decay mode is $K^{*}D^{*}$.

Corresponding to the quantum number $I(J^{P})=1(0^{+})$, there are two tetraquark states: $\rm T_{n^{2}\bar{s}\bar{c}}(3046,1,0^{+})$ and $\rm T_{n^{2}\bar{s}\bar{c}}(2546,1,0^{+})$.
For the $\rm T_{n^{2}\bar{s}\bar{c}}(3046,1,0^{+})$ state, it can only decay into the $KD$ final states. This state is expected to be broad due to  large phase space of this decay mode. Similarly, the $\rm T_{n^{2}\bar{s}\bar{c}}(2546,1,0^{+})$ state can also decay into the $KD$ final states but with a relatively small phase space. 

We can use the similar way to analyze the property of these higher tetraquark states with $I=0$.
Focusing on the $X_0(2900)$, we find that it is suitable to assign the $X_{0}(2900)$ as an $S$-wave tetraquark state with $I(J^{P})=0(0^{+})$ since the mass of the $X_0(2900)$ can be reproduced, where we can 
obtain the measured mass of the  $T_{n^{2}\bar{s}\bar{c}}(2793,0,0^{+})$ state
corresponding to the $X_{0}(2900)$ when the value of the parameters $m_{ij}$ is increased by $2.6\%$.

We further use the mass of the $X_{0}(2900)$ as input to recalculate the masses of $nn\bar{s}\bar{c}$ subsystems and present them in the last row of Table \ref{mass1}.
In this case, there is only one stable state $T_{n^{2}\bar{s}\bar{c}}(2293.4,0,0^{+})$, which is still below all allowed thresholds.
However, the lowest $0(2^{+})$ and $0(1^{+})$ $nn\bar{s}\bar{c}$ states become unstable, which can decay into $K^{*}D^{*}$ and $KD^{*}$, respectively.

\subsection{The $nnss\bar{Q}$ pentaquark states}

In the following, we discuss the $nnss\bar{c}$ and $nnss\bar{b}$ pentaquark subsystems.
For the $nnss\bar{c}$ $(nnss\bar{b})$ pentaquark subsystem, the isospin of the first two light quarks can couple to $I=0,1$. When checking Fig. \ref{fig-nnnsQ} (i)-(j),
we find that the lowest $I(J^P)=0(1/2^{-})$ and $I(J^P)=0(3/2^{-})$ $nnss\bar{Q}$ pentaquark states, i.e., the $\rm P_{n^2s^2\bar{c}}(3026,0,1/2^-)$, $\rm P_{n^2s^2\bar{c}}(3216,0,3/2^-)$, $\rm P_{n^2s^2\bar{b}}(6527,0,3/2^-)$, and $\rm P_{n^2s^2\bar{b}}(6455,0,1/2^-)$ states are below
the corresponding thresholds of the lowest strong decay channels in the $nnss\bar{Q}$ pentaquark subsystem.
Thus, they can be considered as the stable pentaquark states.

On the contrary, for the $I(J^P)=1(5/2^{-})$ $nnss\bar{Q}$ state, the values of $K_{nn}$, $K_{ns}$, $K_{n\bar{c}}$, and $K_{s\bar{c}}$ are all positive as shown in Table \ref{Kij}.
In our opinion, it seems difficult to form a bound state for the $I(J^P)=1(5/2^{-})$ $nnss\bar{Q}$ system since the interactions between different quark components are all repulsive. One can perform similar analysis to the other $nnss\bar{Q}$ states based on the information given in Table \ref{Kij}.

Next, we discuss their decay behaviors. For convenience, we mainly focus on the $nnss\bar{c}$ pentaquark states according to Table \ref{value-nnnsC}. One can perform similar discussion on the $nnss\bar{b}$ pentaquark system correspondingly.
For the $nnss\bar{c}$ pentaquark states with $I=1$,
if the obtained bound state $\rm P_{n^{2}s^{2}\bar{c}}(3520, 1, 5/2^{-})$ exists, this state should lie below the $\Xi^{*}\bar{D}_{s}^{*}$ threshold, and can only decay into the $\Sigma^{*}\bar{D}_{s}^{*}$ final states due to the requirement of the angular momentum conservation.

The lowest $I(J^P)=1(3/2^{-})$ state $\rm P_{n^{2}s^{2}\bar{c}}(3325, 1, 3/2^{-})$ can only decay into the $\Sigma\bar{D}_{s}^{*}$ final states.
Since its negative values of $K_{ns}$, $K_{n\bar{c}}$, and $K_{s\bar{c}}$ lead to a small decay phase space, this pentaquark is expected to be a narrow state.
Moreover, note that $\rm P_{n^{2}s^{2}\bar{c}}(3325, 1, 3/2^{-})$ is slightly above the threshold of the $\Sigma\bar{D}_{s}^{*}$ channel.
Considering the uncertainty of parameters introduced in the CMI model, we still cannot rule out the possibility of the $\rm P_{n^{2}s^{2}\bar{c}}(3325, 1, 3/2^{-})$ as a stable state.

For the other $I(J^P)=1(3/2^{-})$ $nnss\bar{c}$ pentaquark states, the $\rm P_{n^{2}s^{2}\bar{c}}(3614, 1, 3/2^{-})$ has
\begin{equation}
\Gamma_{\Sigma^{*}\bar{D}^{*}_{s}}:\Gamma_{\Sigma^{*}\bar{D}_{s}}:\Gamma_{\Sigma\bar{D}^{*}_{s}}=20.9:35.0:1,
\end{equation}
and
\begin{equation}
\Gamma_{\Xi^{*}\bar{D}^{*}}:\Gamma_{\Xi^{*}\bar{D}}:\Gamma_{\Xi\bar{D}^{*}}=11.4:7.6:1.
\end{equation}
Thus, the dominant decay channels for the $\rm P_{n^{2}s^{2}\bar{c}}(3614, 1, 3/2^{-})$ are the $\Sigma^{*}\bar{D}_{s}$ and $\Sigma^{*}\bar{D}^{*}_{s}$ in the $nns$-$s\bar{c}$ decay mode. Similarly, in the $nss$-$n\bar{c}$ decay mode, the dominant decay channels are the $\Xi^{*}\bar{D}^{*}$ and $\Xi^{*}\bar{D}$ channels. In addition, the $\rm P_{n^{2}s^{2}\bar{c}}(3505, 1, 3/2^{-})$ and $\rm P_{n^{2}s^{2}\bar{c}}(3367, 1, 3/2^{-})$ have various two-body strong decay channels, and they are expected to be broad states.

For the $I(J^P)=1(1/2^{-})$ $nnss\bar{c}$ pentaquark states, they all have several two-body strong decay channels. Specifically, the $\rm P_{n^2s^2\bar{c}}(3716,1,1/2^-)$ has two dominant decay channels, i.e., the $\Sigma^*\bar{D}^*_s$ and $\Xi^*\bar{D}^*$ decays, which have partial decay widths much larger than those of the $\Sigma\bar{D}_s$, $\Sigma \bar{D}_s$, $\Xi\bar{D}^*$, and $\Xi\bar{D}$ channels.

For the $nnss\bar{c}$ pentaquark states with $I=0$,  the $\rm P_{n^{2}s^{2}\bar{c}}(3555, 0, 5/2^{-})$ can only decay into the $\Xi^{*}\bar{D}_{s}^{*}$ channel via $S$-wave.
The angular momentum conservation results in the suppression of the decay rate of the $\rm P_{n^{2}s^{2}\bar{c}}(3555, 0, 5/2^{-})$ state via higher partial waves. This state only has the $nns$-$s\bar{c}$ decay mode.
For the states with $I(J^P)=0(3/2^{-})$, they can only decay into the $\Lambda\bar{D}_{s}$ channel in the $nns$-$s\bar{c}$ mode.
In the $ssn-n\bar{c}$ decay mode, we have
\begin{equation}
\Gamma_{\Xi^{*}\bar{D}^{*}}:\Gamma_{\Xi^{*}\bar{D}}:\Gamma_{\Xi\bar{D}^{*}}=0:7.5:1,
\end{equation}
for the $\rm P_{n^{2}s^{2}\bar{c}}(3524, 0, 3/2^{-})$ state. It suggests that the relative partial decay width of the $\Xi^{*}\bar{D}$ channel is much larger than that of the $\Xi\bar{D}^{*}$ channel.
The $\rm P_{n^{2}s^{2}\bar{c}}(3351, 0, 3/2^{-})$ state can decay into the $\Sigma^*\bar{D}_s^{*}$ and $\Xi\bar{D}^*$ channels for the $ssn-n\bar{c}$ and $snn-s\bar{c}$ decay modes, respectively.
For the $\rm P_{n^{2}s^{2}\bar{c}}(3312, 0, 1/2^{-})$ state,
we find that
\begin{equation}
\Gamma_{\Lambda\bar{D}^{*}_{s}}:\Gamma_{\Lambda\bar{D}_{s}}=1:3.2,
\end{equation}
and for the $\rm P_{n^{2}s^{2}\bar{c}}(3451, 0, 1/2^{-})$ state, we have
\begin{eqnarray}
\Gamma_{\Lambda\bar{D}^{*}_{s}}:\Gamma_{\Lambda\bar{D}_{s}}=1:0.2,\quad\Gamma_{\Xi\bar{D}^{*}}:\Gamma_{\Xi\bar{D}}=1:0.2.
\end{eqnarray}

\begin{table*}[t]
\centering \caption{
The values of $K_{ij}$ when reducing the coupling strength $v_{ij}$ between the $i$-th and $j$-th quark components by 1\% for the $qqqq\bar{Q}$ pentaquark states.
The positive (negative) value represents that the corresponding color-spin interaction is repulsive (attractive).
The masses are all in units of MeV. Here, the reader can refer to the caption of Fig. \ref{fig-nnnsQ} for the meanings of ``$\diamond$" and ``$\star$".
}\label{Kij}
\begin{lrbox}{\tablebox}
\renewcommand\arraystretch{1.31}
\renewcommand\tabcolsep{1.562pt}
\begin{tabular}{cl|rrrrr|cl|rrrr|cl|rrrrr}
\bottomrule[1.5pt]
\bottomrule[0.5pt]
\multicolumn{7}{l|}{$nnnn\bar{c}$/$ssss\bar{c}$/$sssn\bar{c}$}&\multicolumn{6}{l|}{$nnns\bar{c}$}&\multicolumn{7}{l}{$nnss\bar{c}$}\\
$I(J^{P})$&Mass&\multicolumn{1}{c}{$nn$}&\multicolumn{1}{c}{$n\bar{c}$}&\multicolumn{1}{c}{$ss$}&\multicolumn{1}{c}{$s\bar{c}$}&\multicolumn{1}{c|}{$ns$}&
$I(J^{P})$&Mass&\multicolumn{1}{c}{$nn$}&\multicolumn{1}{c}{$ns$}&\multicolumn{1}{c}{$n\bar{c}$}&\multicolumn{1}{c|}{$s\bar{c}$}&
$I(J^{P})$&Mass&\multicolumn{1}{c}{$nn$}&\multicolumn{1}{c}{$ss$}&\multicolumn{1}{c}{$ns$}&\multicolumn{1}{c}{$n\bar{c}$}&\multicolumn{1}{c}{$s\bar{c}$}\\
\bottomrule[0.5pt]
$2(\frac{3}{2}^{-})$&3381&18.67&-5.33&&&&$\frac32(\frac52^{-})$&$3352\star$&8.00&0.00&0.00&5.33&$1(\frac52^{-})$&3520&2.67&2.67&2.67&2.67&2.67\\
$2(\frac{1}{2}^{-})$&3487&18.67&10.67&&&&$\frac32(\frac32^{-})$&$3500$&9.40&9.24&-3.91&-1.44&$1(\frac32^{-})$&3614&3.14&3.09&12.42&-2.63&-2.74\\
$1(\frac{5}{2}^{-})$&3249&8.00&5.33&&&&&$3343\star$&8.25&-2.56&4.21&3.22&                                    &3505&2.94&2.97&-0.21&3.56&3.86\\
$1(\frac{3}{2}^{-})$&3220&6.19&6.19&&&&&$3177\star\diamond$&8.35&-3.56&-3.64&-13.12&                         &3368&2.93&2.83&-11.01&3.42&3.01\\
                    &3043&4.47&-15.52&&&&$\frac32(\frac12^{-})$&$3603$&9.37&9.29&8.08&2.58&                  &3325&2.98&3.11&-1.20&-8.35&-8.13\\
$1(\frac{1}{2}^{-})$&3158&2.06&8.77&&&&&3353&9.04&-7.18&6.94&2.26&            $1(\frac12^{-})$&3716&3.13&3.10&12.44&5.39&5.27\\
                    &3032&0.61&-6.10&&&&&3246&9.59&-8.78&-8.35&1.82&                          &3475&3.15&3.16&-4.35&4.42&4.61\\
$0(\frac{3}{2}^{-})$&3003&-5.33&6.67&&&&$\frac12(\frac52^{-})$&3405&2.00&6.00&6.00&-0.67&     &3350&2.85&2.84&-4.98&-3.47&-2.95\\
$0(\frac{1}{2}^{-})$&2870&-5.33&-13.33&&&&           $\frac12(\frac32^{-})$&3376&0.58&5.50&5.14&1.31&  &3220&2.87&2.90&-11.10&-6.34&-6.93\\
\Xcline{1-7}{0.5pt}
$0(\frac{3}{2}^{-})$&3842&&&18.67&-5.33&&                                  &3208&-1.02&-4.68&5.62&0.93&          $0(\frac52^{-})$&3555&-1.33&2.67&6.67&6.67&-1.33\\
$0(\frac{1}{2}^{-})$&3950&&&18.67&10.67&&                                  &3199&-0.33&4.63&-15.14&-0.26&     $0(\frac32^{-})$&3524&-2.40&2.67&5.71&4.05&2.63\\
\Xcline{1-7}{0.5pt}
$\frac12(\frac{5}{2}^{-})$&$3680\star$&&5.33&8.00&0.00&0.00&&$3022\star\diamond$&-5.22&-7.44&2.38&4.02&                &$3351$&-2.66&2.67&4.22&-15.38&-0.22\\
$\frac12(\frac{3}{2}^{-})$&3724&&-1.38&9.32&-3.99&9.34   &$\frac12(\frac12^{-})$&3309&-2.58&4.53&6.84&1.94&    &$3216\diamond$&-5.60&2.67&-9.93&4.66&1.59\\
                          &3655&&2.83&8.28&4.47&8.28&                           &3172&-3.95&4.44&-3.42&-2.74&   $0(\frac12^{-})$&3451&-4.59&2.88&3.60&4.81&4.06\\
                          &3483&&-12.78&8.39&-3.82&8.39&                        &3082&-0.95&-9.70&-1.16&-0.54&                &3312&-5.44&3.33&2.63&-0.35&5.90\\
$\frac12(\frac{1}{2}^{-})$&3829&&2.68&9.33&7.99&9.34&                           &3000&-5.79&-3.84&-4.47&1.07&                       &3213&-4.14&2.98&-13.75&4.29&4.28\\
                          &3601&&2.29&-7.11&6.77&-7.11&                  &$2831\diamond$&-4.73&-9.44&-11.79&-9.73&    &$3026\diamond$&-4.49&2.81&-12.48&-15.41&6.43\\
                          &3465&&1.70&-8.90&-8.09&-8.90&&&&&&&&&&\\
\bottomrule[1.0pt]
\multicolumn{7}{l|}{$nnnn\bar{b}$/$ssss\bar{b}$/$sssn\bar{b}$}&\multicolumn{6}{l|}{$nnns\bar{b}$}&\multicolumn{7}{l}{$nnss\bar{b}$}\\
$I(J^{P})$&Mass&\multicolumn{1}{c}{$nn$}&\multicolumn{1}{c}{$n\bar{b}$}&\multicolumn{1}{c}{$ss$}&\multicolumn{1}{c}{$s\bar{b}$}&\multicolumn{1}{c|}{$ns$}&
$I(J^{P})$&Mass&\multicolumn{1}{c}{$nn$}&\multicolumn{1}{c}{$ns$}&\multicolumn{1}{c}{$n\bar{b}$}&\multicolumn{1}{c|}{$s\bar{b}$}&
$I(J^{P})$&Mass&\multicolumn{1}{c}{$nn$}&\multicolumn{1}{c}{$ss$}&\multicolumn{1}{c}{$ns$}&\multicolumn{1}{c}{$n\bar{b}$}&\multicolumn{1}{c}{$s\bar{b}$}\\
\bottomrule[0.5pt]
$2(\frac{3}{2}^{-})$&6746&18.67&-5.33&&&&$\frac32(\frac52^{-})$&$6655\star$&8.00&0.00&0.00&5.33&        $1(\frac52^{-})$&6829&2.67&2.67&2.67&2.67&2.67\\
$2(\frac{1}{2}^{-})$&6780&18.67&10.67&&&&$\frac32(\frac32^{-})$&6862       &9.38&9.27&-4.07&-1.27&      $1(\frac32^{-})$&6972&3.13&3.09&12.43&-2.77&-2.58\\
$1(\frac{5}{2}^{-})$&6565&8.00&5.33  &&&&                      &$6642\star$&8.10&-0.96&3.10&-0.77&                      &6813&2.77&2.77&1.63&1.06&1.11\\
$1(\frac{3}{2}^{-})$&6544&7.60&-0.95 &&&&                      &6579       &8.52&-4.98&-2.37&-9.29&                     &6747&3.23&3.23&-2.96&-5.49&-5.98\\
                    &6441&3.06&-8.39 &&&&$\frac32(\frac12^{-})$&6895       &9.37&9.28&8.08&2.58&                        &6650&2.87&2.91&-11.10&3.20&3.46\\
$1(\frac{1}{2}^{-})$&6462&2.48&6.58  &&&&                      &6657       &8.82&-6.54&4.86&3.04&       $1(\frac12^{-})$&7705&3.13&3.10&12.44&5.40&5.27\\
                    &6396&0.19&-3.91 &&&&                      &6601       &9.80&-9.41&-6.28&1.05&                      &6772&3.26&3.26&-4.14&3.16&2.86\\
$0(\frac{3}{2}^{-})$&6313&-5.33&6.67 &&&&$\frac12(\frac52^{-})$&6723       &2.00&6.00&6.00&-0.67&                       &6707&2.74&2.73&-5.19&-2.63&-2.06\\
$0(\frac{1}{2}^{-})$&6271&-5.33&-13.33&&&&$\frac12(\frac32^{-})$&6702      &1.66&5.88&-1.74&1.26&                       &6609&2.87&2.91&-11.01&-6.23&-7.07\\
\Xcline{1-7}{0.5pt}
$0(\frac{3}{2}^{-})$&7193&&&18.67&-5.33&&                      &6600       &-1.53&4.53&-8.78&-0.01&                       $0(\frac52^{-})$&6875&-1.33&2.67&6.67&6.67&-1.33\\
$0(\frac{1}{2}^{-})$&7229&&&18.67&10.67&&                      &6515       &-1.10&-4.61&5.78&0.88&                        $0(\frac52^{-})$&6854&-1.61&2.67&6.42&-2.55&2.54\\
\Xcline{1-7}{0.5pt}
$\frac12(\frac{5}{2}^{-})$&$6996\star$&&5.33&8.00&0.00&0.00&                &$6324\star\diamond$&-5.03&-7.80&2.73&3.87&                   &6743&-3.76&2.67&4.10&-9.34&0.12\\
$\frac12(\frac{3}{2}^{-})$&7079&&-1.42&9.32&-3.93&9.34&$\frac12(\frac12^{-})$&6618              &-2.14&4.52&5.67&0.96&          &$6527\diamond$&-5.30&2.67&-10.5&5.23&1.34\\
                          &6976&&-1.32&8.09&2.97&-0.84&                      &6531              &-4.33&4.42&-2.53&-1.52&  $0(\frac52^{-})$&6759&-4.29&2.75&3.84&4.83&2.09\\
                          &6878&&-8.60&8.59&-2.38&-5.17&                     &6473              &-0.97&-4.81&-11.03&-2.01&                &6667&-5.66&3.48&2.28&-0.77&-3.60\\
$\frac12(\frac{1}{2}^{-})$&7114&&2.68&9.33&7.99&9.34&                        &6308              &-5.39&-8.53&2.87&1.08&                   &6513&-4.34&2.90&-13.15&3.37&4.08\\
                          &6899&&3.06&8.83&4.49&-6.50&                       &$6253\diamond$    &-5.16&-9.59&-9.97&-8.51&       &$6455\diamond$&-4.38&2.88&-12.97&-14.10&-6.58\\
                          &6813&&0.93&9.84&-5.81&-9.51&&&&&&&&&\\
\bottomrule[0.5pt]
\bottomrule[1.5pt]
\end{tabular}
\end{lrbox}\scalebox{0.975}{\usebox{\tablebox}}
\end{table*}

\begin{table*}[t]
\centering \caption{
The overlaps of wave functions between a $nnnn\bar{Q}$ ($ssss\bar{Q}$) pentaquark state and a particular baryon $\otimes$ meson state. The masses are all in units of MeV.
See the caption of Fig. \ref{fig-nnnsQ} for the meanings of ``$\diamond$" and ``$\star$".
}\label{eigenvector-nnnnQ}
\renewcommand\arraystretch{1.15}
\begin{tabular}{c|c|cccc|c|cccc}
\bottomrule[1.5pt]
\bottomrule[0.5pt]
Subsystem&$nnnn\bar{c}$&\multicolumn{4}{c|}{$nnn\bigotimes n\bar{c}$}&$nnnn\bar{b}$&\multicolumn{4}{c}{$nnn\bigotimes n\bar{b}$}\\
$I(J^P)$&Mass&$\Delta \bar{D}^{*}$&$\Delta \bar{D}$&$N\bar{D}^{*}$&$N\bar{D}$&Mass&$\Delta B^{*}$&$\Delta B$&$NB^{*}$&$NB$\\
\bottomrule[0.7pt]
$2(\frac{3}{2}^{-})$&3381.4&0.456&-0.354&&&6745.6&0.456&-0.354\\
$2(\frac{1}{2}^{-})$&3487.4&-0.577&&&&6779.5&-0.577\\
\bottomrule[0.5pt]
$1(\frac{5}{2}^{-})$&3248.5&0.707&&&&6564.6&0.707\\
$1(\frac{3}{2}^{-})$&3212.0&-0.618&-0.450&0.168&&6543.9&-0.540&-0.442&-0.078\\
&3042.7&-0.099&0.613&0.235&&6441.2&-0.322&0.492&0.278\\
$1(\frac{1}{2}^{-})$&3157.9&0.507&&0.334&0.100&6461.9&0.556&&0.255&0.174\\
&3031.7&0.276&&0.311&-0.339&6395.9&0.154&&0.379&-0.308\\
\bottomrule[0.5pt]
$0(\frac{3}{2}^{-})$&3002.9&&&0.577&&6313.0&&&0.577\\
$0(\frac{1}{2}^{-})$&2870.4&&&-0.289&-0.500&6270.6&&&-0.289&-0.500\\
\bottomrule[1.0pt]
Subsystem&$ssss\bar{c}$&\multicolumn{4}{c|}{$sss\bigotimes s\bar{c}$}&$ssss\bar{b}$&\multicolumn{4}{c}{$sss\bigotimes s\bar{b}$}\\
&Mass&$\Omega \bar{D}_{s}^{*}$&$\Omega \bar{D}_{s}$&&&Mass&$\Delta B_{s}^{*}$&$\Delta B_s$\\
\bottomrule[0.7pt]
$0(\frac{3}{2}^{-})$&3842.1&0.456&-0.354&&&7193.0&0.456&-0.354\\
$0(\frac{1}{2}^{-})$&3949.9&-0.577&&&&7229.4&-0.577\\
\bottomrule[0.5pt]
\bottomrule[1.5pt]
\end{tabular}
\end{table*}

\begin{table}[t]
\centering \caption{
The values of $k\cdot |c_{i}|^{2}$ for the $nnnn\bar{Q}$ and $ssss\bar{Q}$ pentaquark states. The masses are all in units of MeV.
The kinetically forbidden decay channel is marked with ``$\times$''.
See the caption of Fig. \ref{fig-nnnsQ} for the meanings of ``$\diamond$" and ``$\star$".
One can roughly estimate the relative decay widths between different decay processes of different initial pentaquark states with this table if neglecting the $\gamma_i$ differences.
}\label{value-nnnnQ}
\begin{lrbox}{\tablebox}
\renewcommand\arraystretch{1.15}
\begin{tabular}{c|c|cccc|c|cccc}
\bottomrule[1.5pt]
\bottomrule[0.5pt]
&$nnnn\bar{c}$&\multicolumn{4}{c|}{$nnn\bigotimes n\bar{c}$}&$nnnn\bar{b}$&\multicolumn{4}{c}{$nnn\bigotimes n\bar{b}$}\\
$I(J^P)$&Mass&$\Delta \bar{D}^{*}$&$\Delta \bar{D}$&$N \bar{D}^{*}$&$N\bar{D}$&Mass&$\Delta B^{*}$&$\Delta B$&$N B^{*}$&$N B$\\
\bottomrule[0.7pt]
$2(\frac{3}{2}^{-})$&3381&97&83&&&6746&131&88\\
$2(\frac{1}{2}^{-})$&3487&208&&&&6780&229\\
\bottomrule[0.5pt]
$1(\frac{5}{2}^{-})$&3249&49&&&&6565&63\\
$1(\frac{3}{2}^{-})$&3212&$\times$&83&17&&6544&$\times$&50&4\\
&3043&$\times$&$\times$&19&&6441&$\times$&$\times$&43\\
$1(\frac{1}{2}^{-})$&3158&$\times$&&59&7&6462&$\times$&&31&20\\
&3032&$\times$&&32&63&6396&$\times$&&68&52\\
\bottomrule[0.5pt]
$0(\frac{3}{2}^{-})$&3003&&&88&&6313&&&95\\
$0(\frac{1}{2}^{-})$&2870&&&$\times$&70&6271&&&9&73\\
\bottomrule[1.0pt]
&$ssss\bar{c}$&\multicolumn{4}{c|}{$sss\bigotimes s\bar{c}$}&$ssss\bar{b}$&\multicolumn{4}{c}{$sss\bigotimes s\bar{b}$}\\
&Mass&$\Omega \bar{D}_{s}^{*}$&$\Omega \bar{D}_{s}$&&&Mass&$\Delta B_{s}^{*}$&$\Delta B_s$\\
\bottomrule[0.7pt]
$0(\frac{3}{2}^{-})$&3842&68&77&&&7193&109&80\\
$0(\frac{1}{2}^{-})$&3950&187&&&&7229&203\\
\bottomrule[0.5pt]
\bottomrule[1.5pt]
\end{tabular}
\end{lrbox}\scalebox{0.89}{\usebox{\tablebox}}
\end{table}

\subsection{The $nnnn\bar{Q}$ and $ssss\bar{Q}$ pentaquark states}\label{secw}

For the $nnnn\bar{c}$ $(nnnn\bar{b})$ pentaquark subsystem, the first four quarks inside this system
can be described by the SU(2) isospin group. The isospin quantum numbers for such pentaquark subsystem are $I=2$, $1$, and $0$. Because of the constraint from the Pauli Principle, the ground $nnnn\bar{Q}$ pentaquark states with quantum number $I(J^P)=2(5/2^-)$, $0(5/2^-)$ do not exist.
Finally, there exist six ground $nnnn\bar{c}$ $(nnnn\bar{b})$ pentaquark states.
Meanwhile, for the $ssss\bar{c}$ $(ssss\bar{b})$ subsystem, the first four strange quarks inside this system are identical, and are regarded as the flavor singlet. Similarly, there are only two ground $ssss\bar{c}$ $(ssss\bar{b})$ pentaquark states, while the states with quantum number $J^P=5/2^{-}$ do not exist.

From Fig. \ref{fig-nnnsQ} (a)-(b), we can see that in the $nnnn\bar{c}$ $(nnnn\bar{b})$ subsystem, the pentaquark states with the smallest and largest masses both have the assignment $J^P={1/2}^-$.
Besides, 
we can easily find that the $I=0$ states have lower masses than those of the $I=1$ $nnnn\bar{c}$ $(nnnn\bar{b})$ pentaquark states. Meanwhile, the masses of $I=1$ pentaquark states are lower than those of the $I=2$ pentaquark states.
Our results indicate that the $nnnn\bar{c}$ $(nnnn\bar{b})$ states with a lower isospin quantum number are expected to form more compact $nnnn\bar{c}$ $(nnnn\bar{b})$ pentaquarks and thus have lower masses.

Now we discuss the possible decay patterns for the $nnnn\bar{c}$ $(nnnn\bar{b})$ and $ssss\bar{c}$ $(ssss\bar{b})$ pentaquark states.
Possible reference meson-baryon systems for the $nnnn\bar{c}$ $(nnnn\bar{b})$ and $ssss\bar{c}$ $(ssss\bar{b})$ pentaquark states can be obtained by rearranging their constituent quarks and regrouping them into meson-baryon systems. As shown in Fig. \ref{fig-nnnsQ} (a)-(d), the reference meson-baryon systems for the $nnnn\bar{c}$ $(nnnn\bar{b})$ pentaquark states are the $\Delta \bar{D}^{*}$ $(\Delta B^*)$, $\Delta \bar{D}$ $(\Delta B)$, $N \bar{D}^{*}$ $(N B^*)$, and $N \bar{D}$ $(N B)$, while the reference meson-baryon systems for the $ssss\bar{c}$ $(ssss\bar{b})$ pentaquark states are the $\Omega\bar{D}_s^*$ $(\Omega B^*_s)$ and $\Omega \bar{D}_s$ $(\Omega B_s)$.

If we only consider the pentaquark decay through these S-wave strong decay channels, we can see that all the $nnnn\bar{Q}$ and $ssss\bar{Q}$ pentaquark states are higher than the lowest thresholds of the corresponding strong decay channels,
which suggests that there exists no stable pentaquark state with the $nnnn\bar{Q}$ and $ssss\bar{Q}$ configurations.
According to Table \ref{Kij}, we notice that many states have a repulsive $K_{n\bar{c}}$ ($K_{n\bar{b}}$) interaction, and thus these states could hardly exist. However, for the $I(J^{P})=0(1/2^{-})$ $nnnn\bar{c}$ ($nnnn\bar{b}$) state, the $K_{nn}$ and $K_{n\bar{c}}$ interactions are both attractive and its width should be narrower compared to that of the other $nnnn\bar{c}$ $(nnnn\bar{b})$ pentaquark states.

Indeed, the stabilities of the $nnnn\bar{Q}$ pentaquark states have been discussed for a long time. Especially, in Ref. \cite{Jaffe:2003sg}, Jaffe and Wilczek found that the $\Theta^{+}$ \cite{Nakano:2003qx} could be a bound state with two spin-0 $ud$ diquarks in $P$-wave attached with an $\bar{s}$ antiquark.
Thus, they made a simple mass estimate and suggested that the states analog to the $\Theta^{+}(1540)$, in which the $\bar{s}$ is replaced by a heavy antiquark, may also be bound.
They denoted the states with flavour structures $(ud)(ud)\bar{c}$ and $(ud)(ud)\bar{b}$ as $\Theta_{c}$ and $\Theta_{b}$ states, respectively. They predicted their masses as $m_{\Theta_{c}}\backsimeq$ 2710 MeV and $m_{\Theta_{b}}\backsimeq$ 6050 MeV, lying 100 MeV and 165 MeV below the strong decay thresholds of $pD^{-}$ and $nB^{+}$, respectively.
Based on the conclusion of Ref. \cite{Jaffe:2003sg}, Leibovich {\it et al}. suggested that the $P$-wave $I(J^{P})=0(3/2^{+})$ $\Theta^{*}_{Q}$ could also be stable with respect to strong interaction and can decay into the $\Theta_{Q}\gamma$ final state \cite{Leibovich:2003tw}.
Similarly, Oh {\it et al}. investigated the pentaquark (P) exotic baryons as soliton-antiflavored heavy
mesons bound states by considering the chiral symmetry and heavy quark symmetry.
Their results support the existence of the loosely bound non-strange P-baryon(s) ($nnnn\bar{c}$ and $nnnn\bar{b}$) \cite{Oh:1994np}.
Moreover, for these subsystems, Park {\it et al}. presented systemically the results of the corresponding binding
energies (defined as the difference between the hyperfine interaction of the pentaquark against its lowest threshold
values) in Table IV of Ref. \cite{Park:2018oib}.
Until now, this topic is still an open issue. In the following, we discuss the possible decay behaviors of the $nnnn\bar{Q}$ and $ssss\bar{Q}$ pentaquark states in the framework of the modified CMI model.
Here, we mainly discuss the decay behaviors of the $nnnn\bar{c}$ pentaquark states,
one can perform very similar discussions on the decay behaviors of the $nnnn\bar{b}$, $ssss\bar{c}$, and $ssss\bar{b}$ pentaquark states according to Tables \ref{eigenvector-nnnnQ} and \ref{value-nnnnQ}.

From Table \ref{value-nnnnQ} and Fig. \ref{fig-nnnsQ} (a), we find that the $\rm P_{n^{4}\bar{c}}(3487,2,1/2^{-})$ state can only decay into the $\Delta\bar{D}^{*}$ final states, while the $\rm P_{n^{4}\bar{c}}(3381,2,3/2^{-})$ state has two decay channels, i.e., decaying into the $\Delta\bar{D}^{*}$ and $\Delta\bar{D}$ final states.
The ratio of relative decay widths between the $\Delta\bar{D}^{*}$ and $\Delta\bar{D}$ mode is
\begin{equation}
\Gamma_{\Delta\bar{D}^{*}}:\Gamma_{\Delta\bar{D}}=1:0.9,
\end{equation}
where both the $\Delta\bar{D}^{*}$ and $\Delta\bar{D}$ channels are the dominant decay modes for the $\rm P_{n^{4}\bar{c}}(3381,2,3/2^{-})$ pentaquark state.

Due to the conservation of angular momentum, the $\rm P_{n^{4}\bar{c}}(3249,1,5/2^{-})$ state can decay into the $\Delta\bar{D}^{*}$ channel via $S$-wave.
The $\rm P_{n^{4}\bar{c}}(3220,1,3/2^{-})$ state can decay into the $\Delta\bar{D}$ and $N\bar{D}^{*}$ final states.
As presented in Tables  \ref{eigenvector-nnnnQ} and \ref{value-nnnnQ}, although the $\Delta\bar{D}^{*}$ has the largest eigenvector component, this mode is kinematically forbidden.
The $\rm P_{n^{4}\bar{c}}(3043,1,3/2^{-})$ state can only decay into the $N\bar{D}^{*}$ channel. Due to small eigenvector component, this state is expected to be a narrow state.

For the two $I(J^{P})=1({1/2}^{-})$ states: $\rm P_{n^{4}\bar{c}}(3158,1,1/2^{-})$ and $\rm P_{n^{4}\bar{c}}(3032,1,1/2^{-})$, we obtain the following relative ratios of decay widths:
\begin{equation}
\Gamma_{N\bar{D}^{*}}:\Gamma_{N\bar{D}}=1:0.1,
\end{equation}
and
\begin{equation}
\Gamma_{N\bar{D}^{*}}:\Gamma_{N\bar{D}}=1:2.0,
\end{equation}
respectively. The dominant decay mode for the $\rm P_{n^{4}\bar{c}}(3003,1,1/2^{-})$ state is the $N\bar{D}^{*}$. Besides, the $I(J^{P})=0({3/2}^{-})$ states $\rm P_{n^{4}\bar{c}}(3003,0,3/2^{-})$ and the $I(J^{P})=0({1/2}^{-})$ state $\rm P_{n^{4}\bar{c}}(2870,0,1/2^{-})$ can only decay into the $N\bar{D}^{*}$ and $N\bar{D}$ channels, respectively.

In addition, for the $nnnn\bar{c}$ subsystem, the H1 Collaboration find the $\Theta^{0}_{c}$ signal at $3099$ MeV in the $ep\rightarrow eD^{*-}pX$ reaction \cite{Aktas:2004qf}.
However, this resonance was not observed in any other experiment including ZEUS \cite{Chekanov:2004qm}, FOCUS \cite{Link:2005ti}, BaBar \cite{Aubert:2006qu}, ALEPH \cite{Schael:2004nm}, and CDF \cite{Litvintsev:2004yw}.
According to Fig. \ref{fig-nnnsQ} (a), we suggest that the future experiment could check the pentaquark signal existing in the $2800-3050$ MeV mass range. For the $nnnn\bar{b}$ subsystem, the LHCb Collaboration tried to find the pentaquark signal in the $P^{+}_{B^{0}p}(uudd\bar{b})\rightarrow J/\psi K^{+}\pi^{-} p$ weak decay mode via the $b\rightarrow c\bar{c}s$ transition. They search for the  $nnnn\bar{b}$ pentaquark in the energy range $4668-6220$ MeV. However, no evidence for such a state is found \cite{Aaij:2017jgf}. According to Fig. \ref{fig-nnnsQ} (b), our results suggest that the LHCb Collaboration may check the $nnnn\bar{b}$ pentaquark signal in the $6200-6900$ MeV energy window.

\subsection{The $nnns\bar{Q}$ and $sssn\bar{Q}$ pentaquark states}
Lastly, we discuss the $nnns\bar{c}$ $(nnns\bar{b})$ and $sssn\bar{c}$ $(sssn\bar{b})$ pentaquark subsystems.
With the less constraint from the Pauli principle,
the corresponding mass spectra are more complicated.
For the $nnns\bar{c}$ $(nnns\bar{b})$ pentaquark subsystem, the isospin of the first three light quarks can couple to $I=3/2$, $1/2$.

Similar to the previous discussion, in the following, we firstly distinguish the scattering states from the calculated $nnns\bar{Q}$ and $sssn\bar{Q}$ subsystems, the remaining states can be regarded as the genuine pentaquark states. Then we discuss the strong decay properties of these genuine pentaquarks.


In Fig. \ref{fig-nnnsQ}, we find that the lowest $I(J^{P})=1/2(1/2^{-})$ and $I(J^{P})=1/2(3/2^{-})$ $nnns\bar{Q}$ states are all below the lowest allowed strong decay channels.
However, from Table \ref{eigenvector-nnnsC}, we find that the lowest $I(J^{P})=1/2(3/2^{-})$ $nnns\bar{c}$ $(nnns\bar{b})$ state has quite large fraction of the $N\bar{D}_s^*$ ($NB_s^*$) component.
Thus, it is more reasonable to take this state as a scattering state.
For the lowest $I(J^{P})=1/2(1/2^{-})$ $nnns\bar{c}$ $(nnns\bar{b})$ state, we consider it as a stable state, although it also has relatively large fraction of the meson-baryon color-singlet component.
From Table \ref{Kij}, we also find that the $K_{nn}$, $K_{ns}$, $K_{n\bar{c}}$ ($K_{n\bar{b}}$), and $K_{s\bar{c}}$ ($K_{s\bar{b}}$) interactions for the lowest $I(J^{P})=1/2(1/2^{-})$ $nnns\bar{c}$ $(nnns\bar{b})$ state are all attractive, thus the width of this state is suppressed by its small decay phase space. Extending to the entire $nnns\bar{c}$ ($nnns\bar{b}$) subsystem, our results also suggest that the states with the lowest isospin quantum number can form bound states easily due to the attractive $K_{ij}$ interactions from their quark pairs.

There are some theoretical discussions on the existence of the $qqqs\bar{Q}$ pentaquark states. Gignoux {\it et al}. found that the states $P^{0}=\bar{c}uuds$ and $P^{-}=\bar{c}ddus$ with spin 1/2 and their beauty analogs are very likely to be stable multiquarks \cite{Gignoux:1987cn}.
Similarly, possible stable pentaquark configurations $\bar{Q}sqqq$ were also proposed in Ref. \cite{Lipkin:1987sk}. In addition, the mass of $T_{s} (nnns\bar{c}, I=1/2)$ was estimated to be $m_{T_{s}}\simeq 2580$ MeV in Ref. \cite{Stewart:2004pd}.
For the $T_{s}\rightarrow D_{s}p$ decay process, the sum of the masses of $D_s$ and proton is 2910 MeV, i.e., the state is below the lowest meson-baryon threshold about 330 MeV. For the $R_{s} (nnns\bar{b},I=1/2)$, they have the prediction $m_{R_{s}}\simeq 5920$ MeV, which is 390 MeV less than the threshold of $B_{s}p$. Meanwhile, they find that there is no stable pentaquark in the $sssn\bar{Q}$ pentaquark subsystem. Moreover, the $\bar{K}\bar{D}N$ three-body system
with $I = 1/2$ has the minimal quark component with $uuds\bar{c}$ or $udds\bar{c}$.
In Ref. \cite{Yamagata-Sekihara:2018gah}, they found that such three-body system may form a bound state and acts like an explicit ``$uuds\bar{c}$" pentaquark.

Next, we focus on the decay behaviors of the $nnns\bar{c}$ pentaquark states.
And one can perform very similar discussions on the decay behaviors of the $nnns\bar{b}$, $sssn\bar{c}$, and $sssn\bar{b}$ pentaquark subsystems according to Tables  \ref{eigenvector-nnnsC} and \ref{value-nnnsC}.

For the $I(J^{P})=1/2(5/2^{-})$ state, the $\rm P_{n^{3}s\bar{c}}(3405, 1/2, 5/2^{-})$ can dominantly decay into $\Sigma^{*}_{c}\bar{D}^{*}$ final states via $S$-wave. The decay widths for the $\rm P_{n^{3}s\bar{c}}(3405, 1/2, 5/2^{-})$ into other higher partial wave channels are suppressed.

The other $P_{n^3s\bar{c}}$ pentaquark states all have two types of decay mode, i.e., the $nnn$-$s\bar{c}$ and $nns$-$n\bar{c}$ modes. As the only genuine pentaquark state with the quantum number $I(J^{P})=3/2(3/2^{-})$, the $\rm P_{n^{3}s\bar{c}}(3500, 3/2, 3/2^{-})$ has two $nnn$-$s\bar{c}$ decay channels, namely the $\Delta\bar{D}_{s}$ and $\Delta\bar{D}^{*}_{s}$. The corresponding ratio of partial decay widths is
\begin{equation}
\Gamma_{\Delta\bar{D}_{s}^{*}}:\Gamma_{\Delta\bar{D}_{s}}=1:1.9.
\end{equation}
On the other hand, as shown in Table \ref{value-nnnsC}, the $\rm P_{n^{3}s\bar{c}}(3500, 3/2, 3/2^{-})$ state also has three $nns$-$n\bar{c}$ decay modes.
The ratio of their partial decay widths is
\begin{equation}
\Gamma_{\Sigma^{*}\bar{D}^{*}}:\Gamma_{\Sigma^{*}\bar{D}}:\Gamma_{\Sigma\bar{D}^{*}}=20.5:15.6:1.
\end{equation}
Our results suggest that the $\Sigma^{*}\bar{D}$ and $\Sigma^{*}\bar{D}^{*}$ channels are the dominant decay modes for the $\rm P_{n^{3}s\bar{c}}(3500, 3/2, 3/2^{-})$ state.

Moreover, for the $nnn-s\bar{c}$ decay mode, three genuine $I(J^{P})=1/2(3/2^{-})$ pentaquark states have the only one allowed decay channel $N\bar{D}^{*}_{s}$.
While for the $nns-n\bar{c}$ decay mode, the three $I(J^{P})=1/2(3/2^{-})$ pentaquark states can decay freely to the $\Lambda\bar{D}^{*}$ final states.
For the $\rm P_{n^{3}s\bar{c}}(3209, 1/2, 3/2^{-})$ and $\rm P_{n^{3}s\bar{c}}(3199, 1/2, 3/2^{-})$ states, they have the same quantum numbers and similar masses, but we can distinguish them from their decay behaviors. As presented in Table \ref{value-nnnsC}, the $\rm P_{n^{3}s\bar{c}}(3209, 1/2, 3/2^{-})$ can decay into the $\Sigma\bar{D}^{*}$, while this channel is forbidden for the $\rm P_{n^{3}s\bar{c}}(3199, 1/2, 3/2^{-})$.
Besides, the relative partial decay width ratio of the $\Sigma_c^*\bar{D}$ and $\Sigma_c\bar{D}^*$ channels for the $\rm P_{n^{3}s\bar{c}}(3376, 1/2, 3/2^{-})$ is
\begin{equation}
\Gamma_{\Sigma^{*}\bar{D}}:\Gamma_{\Sigma\bar{D}^{*}}=10.1:1.
\end{equation}
Thus, the dominant decay channel for the $\rm P_{n^{3}s\bar{c}}(3376, 1/2, 3/2^{-})$ is the $\Sigma^{*}\bar{D}$ channel in $nns$-$n\bar{c}$ decay mode.

For the five $I(J^{P})=1/2(1/2^{-})$ pentaquark states, all of them can be considered as genuine pentaquark states.
The lowest state $\rm P_{n^{3}s\bar{c}}(2831, 1/2, 1/2^{-})$ is expected to be a stable pentaquark state.
For the $\rm P_{n^{3}s\bar{c}}(3309, 1/2, 1/2^{-})$ state, we find
\begin{equation}
\Gamma_{N \bar{D}^{*}_{s}}:\Gamma_{N \bar{D}_{s}}=1:0.1,
\end{equation}
and
\begin{equation}
\Gamma_{\Sigma\bar{D}^{*}}:\Gamma_{\Sigma\bar{D}}=1:0.1,\quad\Gamma_{\Lambda\bar{D}^{*}}:\Gamma_{\Lambda\bar{D}}=1:0.2.
\end{equation}

Similarly, for the $\rm P_{n^{3}s\bar{c}}(3172, 1/2, 1/2^{-})$ and $\rm P_{n^{3}s\bar{c}}(3081.9, 1/2, 1/2^{-})$  states, we have:
\begin{equation}
\Gamma_{N \bar{D}^{*}_{s}}:\Gamma_{N \bar{D}_{s}}=1:2.3, \quad\Gamma_{\Lambda\bar{D}^{*}}:\Gamma_{\Lambda\bar{D}}=1:2.0,
\end{equation}
and
\begin{equation}
\Gamma_{N \bar{D}^{*}_{s}}:\Gamma_{N \bar{D}_{s}}=1:4.3,
\end{equation}
respectively.
Meanwhile, they can decay into the $\Sigma\bar{D}$ channel in the $nns-n\bar{c}$ decay mode.
These two pentaquark states may have broad widths since they can decay freely to many strong decay channels.

{
\subsection{The uncertainties from CMI model}
In this subsection, we take the $I(J^{P})=1/2(1/2^{-})$ $nnns\bar{c}$ states and the obtained six stable states to discuss the uncertainties of the CMI model. 

The uncertainties we encountered are mainly from the parameters $m_{ij}$ and $v_{ij}$, and their uncertainties will mainly affect the position of whole pentaquark mass spectra and the mass gaps between the pentaquark states in the same multiplet, respectively.

Firstly, we have discussed the uncertainties about $m_{ij}$ in the $nn\bar{s}\bar{c}$ subsystem, and we obtained an uncertainty of $2.6 \%$ for the parameters $m_{ij}$ based on the mass of the $X_{0}(2900)$ by assuming that $X_0(2900)$ is an $I(J^{P})=0(0^+)$ $S$-wave tetraquark state.
To further discuss the uncertainties of the $I(J^{P})=1/2(1/2^{-})$ $nnns\bar{c}$ states, we assume that the $m_{ij}$ and $v_{ij}$ have at most $5\%$ and $10\% $ deviations from their physical values, respectively. The corresponding results are shown in Table \ref{aaa1}.

According to Table \ref{aaa1}, the whole $nnns\bar{c}$ pentaquark mass spectra moves up (down) relative to the baryon-meson thresholds as $m_{ij}$ increases (decreases). On the other hand, since the parameters $v_{ij}$ are suppressed by $1/m_Q$, thus, they mainly affect the mass gaps between different $nnns\bar{c}$ pentaquark states in the same multiplet.

Moreover, if the $I(J^{P})=1/2(1/2^{-})$ $nnns\bar{c}$ states also have $+2.6\%$ correction as that of the $nn\bar{s}\bar{c}$ subsystem,
then the whole $nnns\bar{c}$ mass spectra would shift up by about 80 MeV.
In this case, the lowest state lies slightly above the lowest threshold and is no longer a stable state. 




\begin{table*}[t]
\centering \caption{
The discussion of uncertainty of the $m_{ij}$ and $v_{ij}$ values for the $nnns\bar{c}$ states with $I(J^{P})=1/2(1/2^{-})$.
}\label{aaa1}
\renewcommand\arraystretch{1.45}
\renewcommand\tabcolsep{1.75pt}
\begin{tabular}{c|ccc|c|ccc}
\bottomrule[1.5pt]
\bottomrule[0.5pt]
\multicolumn{6}{l}{The $I(J^{P})=1/2(1/2^{-})$ $nnns\bar{c}$ states.  }\\
$m_{ij}$&$+5.0\%m_{ij}$&$+3.0\%m_{ij}$&$+2.6\%m_{ij}$&$m_{ij}$&$-2.6\%m_{ij}$&$-3.0\%m_{ij}$&$-5.0\%m_{ij}$\\
\bottomrule[0.7pt]
$+10.0\%v_{ij}$&
$\begin{pmatrix}3478\\3328\\3228\\3141\\2955\end{pmatrix}$&
$\begin{pmatrix}3413\\3263\\3163\\3077\\2891\end{pmatrix}$&
$\begin{pmatrix}3400\\3250\\3150\\3064\\2878\end{pmatrix}$&
$\begin{pmatrix}3315\\3166\\3066\\2982\\2795\end{pmatrix}$&
$\begin{pmatrix}3231\\3082\\2982\\2899\\2713\end{pmatrix}$&
$\begin{pmatrix}3218\\3069\\2969\\2886\\2700\end{pmatrix}$&
$\begin{pmatrix}3153\\3004\\2904\\2823\\2636\end{pmatrix}$
\\
$v_{ij}$&
$\begin{pmatrix}3471\\3334\\3244\\3159\\2990\end{pmatrix}$&
$\begin{pmatrix}3406\\3269\\3179\\3095\\2926\end{pmatrix}$&
$\begin{pmatrix}3393\\3256\\3166\\3082\\2914\end{pmatrix}$&
$\begin{pmatrix}\bf{3309}\\ \bf{3172}\\ \bf{3082}\\ \bf{3000}\\ \bf{2831}\end{pmatrix}$&
$\begin{pmatrix}3224\\3088\\2998\\2917\\2748\end{pmatrix}$&
$\begin{pmatrix}3211\\3075\\2985\\2904\\2735\end{pmatrix}$&
$\begin{pmatrix}3146\\3010\\2920\\2841\\2672\end{pmatrix}$
\\
$-10.0\%v_{ij}$&
$\begin{pmatrix}3464\\3340\\3260\\3177\\3025\end{pmatrix}$&
$\begin{pmatrix}3399\\3276\\3195\\3113\\2962\end{pmatrix}$&
$\begin{pmatrix}3386\\3263\\3182\\3100\\2949\end{pmatrix}$&
$\begin{pmatrix}3302\\3179\\3098\\3017\\2866\end{pmatrix}$&
$\begin{pmatrix}3218\\3094\\3013\\2935\\2783\end{pmatrix}$&
$\begin{pmatrix}3205\\3081\\3001\\2922\\2771\end{pmatrix}$&
$\begin{pmatrix}3140\\3017\\2936\\2859\\2707\end{pmatrix}$
\\
\bottomrule[0.5pt]
\bottomrule[1.5pt]
\end{tabular}
\end{table*}

Next, we discuss the uncertainties of the obtained six stable pentaquark states. 
We also assume that the $m_{ij}$ and $v_{ij}$ have at most 5\% and 10\% deviations from their physical values, respectively.
Then we present how their masses vary with the coupling parameters $m_{ij}$ and $v_{ij}$ in Table \ref{aaa2}.

From Table \ref{aaa2}, we find that when we set $m_{ij}$ at $0.95$ and $v_{ij}$ at $1.1v_{ij}$, respectively, the obtained pentaquark states are deeply bound. On the contrary, as we increase the $m_{ij}$ and decrease $v_{ij}$, the absolute values of binding energies become small, and some of the stable states disappear.
Thus, further exploration on such type of pentaquark states are crucial to narrow the uncertainties encountered in our model.

\begin{table*}[htbp]
\caption{The change of the six stable states by varying the $m_{ij}$ and $v_{ij}$ couplings.
Here, the binding energy is the difference between the mass of the pentaquark state and the lowest threshold.
The masses of pentaquark states, the masses of lowest threshold, and the binding energies are all in units of MeV.}\label{aaa2}
\renewcommand\arraystretch{1.45}
\scalebox{0.95}{
\begin{tabular}{c|c|c|ccc|c|c|ccc}
\bottomrule[1.5pt]
\bottomrule[0.5pt]
\multirow{2}*{States}&\multirow{2}*{$I(J^{P})$}&\multicolumn{4}{c|}{Mass}&\multirow{2}*{Lowest threshold}&\multicolumn{4}{c}{Binding energy}\\ \cline{3-6}\cline{8-11}
&&&$1.1v_{ij}$&$v_{ij}$&$0.9v_{ij}$&&&$1.1v_{ij}$&$v_{ij}$&$0.9v_{ij}$\\
\bottomrule[0.7pt]
\multirow{4}*{$nnns\bar{c}$}&\multirow{4}*{$1/2(1/2^{-})$}&$1.05m_{ij}$&2955&2990&3025&\multirow{4}*{$N\bar{D}_{s}$ (2907)}&$1.05m_{ij}$&48&83&118\\
&&$1.026m_{ij}$&2878&2914&2949&&$1.026m_{ij}$&-28&7&42\\
&&$m_{ij}$&2795&\bf{2831}&2866&&$m_{ij}$&-112&\bf{-76}&-41\\
&&$0.95m_{ij}$&2636&2672&2707&&$0.95m_{ij}$&-271&-235&-200\\
\bottomrule[0.7pt]
\multirow{4}*{$nnns\bar{b}$}&\multirow{4}*{$1/2(1/2^{-})$}&$1.05m_{ij}$&6552&6578&6604&\multirow{4}*{$NB_{s}$ (6305)}&$1.05m_{ij}$&247&273&299\\
&&$1.01m_{ij}$&6292&6318&6344&&$1.01m_{ij}$&-13&13&39\\
&&$m_{ij}$&6227&\bf{6253}&6278&&$m_{ij}$&-78&\bf{-52}&-27\\
&&$0.95m_{ij}$&5901&5927&5953&&$0.95m_{ij}$&-404&-378&-353\\
\bottomrule[0.7pt]
\multirow{7}*{$nnss\bar{c}$}&\multirow{3}*{$0(3/2^{-})$}&$1.05m_{ij}$&3368&3384&3401&\multirow{3}*{$\Lambda \bar{D}^{*}_{s}$ (3228)}&$1.1m_{ij}$&140&156&173\\
&&$m_{ij}$&3199&\bf{3216}&3232&&$m_{ij}$&-29&\bf{-12}&4\\
&&$0.95m_{ij}$&3031&3047&3063&&$0.95m_{ij}$&-197&-181&-165\\ \cline{2-11}
&\multirow{4}*{$0(1/2^{-})$}&$1.05m_{ij}$&3159&3195&3231&\multirow{4}*{$\Lambda \bar{D}_{s}$ (3084)}&$1.05m_{ij}$&75&111&147\\
&&$1.02m_{ij}$&3058&3093&3129&&$1.02m_{ij}$&-26&9&45\\
&&$m_{ij}$&2990&\bf{3026}&3062&&$m_{ij}$&-94&\bf{-58}&-22\\
&&$0.95m_{ij}$&2821&2857&2892&&$0.95m_{ij}$&-263&-227&-192\\
\bottomrule[0.7pt]
\multirow{6}*{$nnss\bar{b}$}&\multirow{3}*{$0(3/2^{-})$}&$1.05m_{ij}$&6843&6862&6881&\multirow{3}*{$\Lambda B^{*}_{s}$ (6531)}&$1.1m_{ij}$&312&331&350\\
&&$m_{ij}$&6507&\bf{6526}&6545&&$m_{ij}$&-24&\bf{-5}&14\\
&&$0.95m_{ij}$&6172&6191&6210&&$0.9m_{ij}$&-359&-340&-321\\ \cline{2-11}
&\multirow{3}*{$0(1/2^{-})$}&$1.05m_{ij}$&6765&6790&6816&\multirow{3}*{$\Lambda B_{s}$ (6483)}&$1.05m_{ij}$&282&307&333\\
&&$m_{ij}$&6429&\bf{6455}&6480&&$m_{ij}$&-54&\bf{-28}&-3\\
&&$0.95m_{ij}$&6093&6119&6145&&$0.95m_{ij}$&-390&-364&-338\\
\bottomrule[0.5pt]
\midrule[1.5pt]
\end{tabular}}
\end{table*}

}

\section{Discussion and conclusion}\label{sec5}
Exotic multiquark candidates are constantly discovered experimentally. 
The lessons from the study of tetraquark candidates $X(2900)$ \cite{Aaij:2020hon,LHCb:2020pxc} and the observation of the $P_c(4312)$, $P_c(4440)$, and $P_c(4457)$ states achieved by the LHCb Collaboration \cite{Aaij:2019vzc} give us strong confidence to explore the $qqqq\bar{Q}$ pentaquark system.

In this work, we firstly construct the $\psi_{\textrm{flavor}}\otimes\psi_{\textrm{color}}\otimes\psi_{\textrm{spin}}$ wave functions of the $qqqq\bar{Q}$ pentaquark states and extract the effective coupling constants from the conventional hadrons. Then we systematically calculate the chromomagnetic Hamiltonian matrices and obtain the corresponding mass spectra.
Besides the mass spectra, we also provide the eigenvectors to extract useful information about the decay properties from the possible quark rearrangement decay channels, and calculate the $K_{ij}$ values to discuss the stabilities and decay phase spaces of the obtained $qqqq\bar{Q}$ pentaquark states.

For the $qqqq\bar{Q}$ pentaquark system, due to the constraint from symmetry, there are no ground $I(J^{P})=0(5/2^{-})$, $I(J^{P})=2(5/2^{-})$ $nnnn\bar{Q}$ states and $I(J^{P})=0(5/2^{-})$ $ssss\bar{Q}$ state. 
Meanwhile, for the $I(J^{P})=3/2(5/2^{-})$ $nnns\bar{Q}$ and $I(J^{P})=1/2(5/2^{-})$ $sssn\bar{Q}$ states, all of them are scattering state since they only have the 1 $\otimes$ 1 component.
Besides, in the framework of CMI model, our results suggest that there exist no stable $nnnn\bar{Q}$, $ssss\bar{Q}$, and $sssn\bar{Q}$ pentaquark states. This conclusion is consistent with that in Ref. \cite{Park:2018oib}. Moreover, our results indicate that the pentaquark states with a lower isospin quantum number are expected to form more compact pentaquark structures and thus have smaller masses.

{According to our results for the $nnnn\bar{b}$ subsystem,
we suggest that the LHCb Collaboration could change the search window from $4600-6220$ MeV to $6200-6800$ MeV to search for the $nnnn\bar{b}$ $(I=0)$ pentaquark states in the $NB^{*}$ final states.
As for $nnns\bar{b}$ subsystem, the lowest $I(J^{P})=1/2(1/2^{-})$ state is stable, and thus we suggest that the LHCb Collaboration could change the search window from $4600-6220$ MeV to $6200-6900$ MeV to search for the $nnns\bar{b}$ $(I=1/2)$ via the $b\rightarrow c\bar{c}s$ transition in the $J/\psi \phi p$ \cite{Aaij:2017jgf} final states.}
In the $nnns\bar{Q}$ subsystem, we find that the lowest $I(J^P)=1/2(1/2^{-})$ $nnns\bar{Q}$ pentaquark state is below all the allowed strong decay channels and are good stable pentaquark candidates.
This conclusion has already been proposed in Refs. \cite{Gignoux:1987cn,Lipkin:1987sk,Park:2018oib}.
In the $nnss\bar{Q}$ subsystem, although our results are larger than the predictions from Ref. \cite{Stewart:2004pd}, our results still suggest that the lowest $I(J^P)=0(1/2^{-})$ and $I(J^P)=0(3/2^{-})$ $nnss\bar{Q}$ states are stable states. In addition, the  $K_{nn}$, $K_{ss}$, $K_{ns}$, $K_{n\bar{b}}$, and $K_{s\bar{b}}$ values are all negative for the lowest $I(J^{P})=0(1/2^{-})$ $nnss\bar{b}$ state $\rm P_{n^{2}s^{2}\bar{b}}(6455, 0, 1/2^{-})$.
Thus, its inner interactions between quarks are all attractive, and its width is suppressed by the small decay phase space.

We collect the obtained six stable candidates in Table \ref{aaa2}. However, due to the uncertainty of the CMI model, further dynamical calculations are still needed to clarify their natures.
Specifically, some stable states are close to the meson-baryon thresholds of the lowest strong decay channels, if the mass deviations in the CMI model are larger than the difference between the pentaquark states and the corresponding meson-baryon thresholds, these states can no longer be considered as stable pentaquark states.
On the contrary, some unstable states, which are a little higher than the meson-baryon thresholds of lowest strong decay channels, also have possibilities to becoming stable states.
Meanwhile, the whole mass spectra has a slight shift or down due to the mass deviations of constituent quarks.
While the mass gaps between different pentaquark states are relatively stable, if one pentaquark state is observed in experiment, we can use these mass gaps to predict their corresponding multiplets.

Among the studied $qqqq\bar{Q}$ pentaquark states, all of them are explicit exotic states. If such pentaquark states are observed, their exotic nature can be easily identified. However, up to now, none of them was found. Our systematical study may provide theorists and experimentalists some preliminary hints toward these pentaquark systems. More detailed dynamical investigations on these pentaquark systems are still needed. Besides,
we hope that the present study may inspire the LHCb, BESIII, Belle II, JLAB,
PANDA, EIC and other relevant experiments to search for these exotic states.

\section*{ACKNOWLEDGMENTS}
We would like to thank Dr. Zhan-Wei Liu for reading the manuscript. 
This work is supported by the China National Funds for Distinguished Young Scientists under Grant No. 11825503, the National Key Research and Development Program of China under Contract No. 2020YFA0406400, the 111 Project under Grant No. B20063, the National Natural Science Foundation of China under Grant No. 12047501, and the Fundamental Research Funds for the Central Universities under Grant No. lzujbky-2021-sp24.

\section{Appendix}\label{sec8}
The CMI Hamiltonian expressions of the $nnns\bar{Q}$ ($I=3/2, 1/2$) and $nnss\bar{Q}$ ($I=1, 0$) pentaquark states are shown in Table \ref{CMI}.

The overlaps of the $nnns\bar{Q}$ ($I=3/2, 1/2$) and $nnss\bar{Q}$ ($I=1, 0$) pentaquark states are shown in Table \ref{eigenvector-nnnsC}.

The The values of $k\cdot |c_{i}|^{2}$ for the $nnns\bar{Q}$ ($I=3/2, 1/2$) and $nnss\bar{Q}$ ($I=1, 0$) pentaquark states are shown in Table \ref{value-nnnsC}.
\begin{table*}[t]
\centering \caption{
The CMI Hamiltonian of the $nnns\bar{Q}$ and $nnss\bar{Q}$ states ($n= u, d$; $Q =c, d$).
}\label{CMI}
\begin{lrbox}{\tablebox}
\renewcommand\arraystretch{1.2}
\renewcommand\tabcolsep{1.85pt}
\begin{tabular}{|c|c|c|c|c|c|}
\bottomrule[1.5pt]
\bottomrule[0.5pt]
$I(J^{P})$&$nnns\bar{Q}$\\
\hline
$\frac32(\frac52^{-})$&$8v_{nn}+\frac{16}{3}v_{s\bar{Q}}$\\
$\frac32(\frac32^{-})$&$\begin{pmatrix}
\begin{pmatrix}\frac{28}{3}v_{nn}+\frac{28}{3}v_{ns}-4v_{n\bar{Q}}-\frac43v_{s\bar{Q}}\end{pmatrix}&
\begin{pmatrix}-\frac{2\sqrt{2}}{3}v_{nn}+\frac{2\sqrt{2}}{3}v_{ns}+\frac{2\sqrt{2}}{3}v_{n\bar{Q}}-\frac{2\sqrt{2}}{3}v_{s\bar{Q}}\end{pmatrix}&
-\frac{8\sqrt{5}}{3}v_{n\bar{Q}}+\frac{8\sqrt{5}}{3}v_{s\bar{Q}}\\

\begin{pmatrix}-\frac{2\sqrt{2}}{3}v_{nn}+\frac{2\sqrt{2}}{3}v_{ns}+\frac{2\sqrt{2}}{3}v_{n\bar{Q}}-\frac{2\sqrt{2}}{3}v_{s\bar{Q}}\end{pmatrix}&
\begin{pmatrix}\frac{26}{3}v_{nn}-6v_{ns}+\frac23v_{n\bar{Q}}-2v_{s\bar{Q}}\end{pmatrix}&
\frac{4\sqrt{10}}{3}v_{s\bar{Q}}+\frac{8\sqrt{10}}{3}v_{s\bar{Q}}\\

-\frac{8\sqrt{5}}{3}v_{n\bar{Q}}+\frac{8\sqrt{5}}{3}v_{s\bar{Q}}&
\frac{4\sqrt{10}}{3}v_{n\bar{Q}}+\frac{8\sqrt{10}}{3}v_{s\bar{Q}}&
8v_{nn}-8v_{s\bar{Q}}
\end{pmatrix}$\\
$\frac32(\frac12^{-})$&$\begin{pmatrix}
\begin{pmatrix}\frac{28}{3}v_{nn}+\frac{28}{3}v_{ns}+8v_{n\bar{Q}}+\frac83v_{s\bar{Q}}\end{pmatrix}&
\begin{pmatrix}-\frac{2\sqrt{2}}{3}v_{nn}+\frac{2\sqrt{2}}{3}v_{ns}-\frac{4\sqrt{2}}{3}v_{n\bar{Q}}+\frac{4\sqrt{2}}{3}v_{s\bar{Q}}\end{pmatrix}&
\frac{2\sqrt{2}}{3}v_{n\bar{Q}}-\frac{2\sqrt{2}}{3}v_{s\bar{Q}}\\

\begin{pmatrix}-\frac{2\sqrt{2}}{3}v_{nn}+\frac{2\sqrt{2}}{3}v_{ns}-\frac{4\sqrt{2}}{3}v_{n\bar{Q}}+\frac{4\sqrt{2}}{3}v_{s\bar{Q}}\end{pmatrix}&
\begin{pmatrix}\frac{26}{3}v_{nn}-6v_{ns}-\frac43v_{n\bar{Q}}+4v_{s\bar{Q}}\end{pmatrix}&
-\frac{26}{3}v_{n\bar{Q}}+\frac{2}{3}v_{s\bar{Q}}\\

\frac{2\sqrt{2}}{3}v_{n\bar{Q}}-\frac{2\sqrt{2}}{3}v_{s\bar{Q}}&
-\frac{26}{3}v_{n\bar{Q}}+\frac{2}{3}v_{s\bar{Q}}&
10v_{nn}-10v_{ns}
\end{pmatrix}$\\
\cline{1-2}
$\frac12(\frac52^{-})$&$8v_{nn}+\frac{16}{3}v_{s\bar{Q}}$\\
$\frac12(\frac32^{-})$&$\begin{pmatrix}
\begin{pmatrix}-\frac{7}{3}v_{nn}+5v_{ns}\\-\frac{11}{6}v_{n\bar{Q}}+\frac12v_{s\bar{Q}}\end{pmatrix}&
\begin{pmatrix}-\frac{\sqrt{2}}{3}v_{nn}+\frac{\sqrt{2}}{3}v_{ns}\\-\frac{7}{3\sqrt2}v_{n\bar{Q}}+\frac{7}{3\sqrt2}v_{s\bar{Q}}\end{pmatrix}&
\begin{pmatrix}\frac{5}{\sqrt{3}}v_{nn}-\frac{5}{\sqrt{3}}v_{ns}\\+\frac{5}{2\sqrt{3}}v_{n\bar{Q}}-\frac{5}{2\sqrt{3}}v_{s\bar{Q}}\end{pmatrix}&
-\frac{23\sqrt{5}}{3\sqrt{2}}v_{n\bar{Q}}+\frac{\sqrt{5}}{3\sqrt{2}}v_{s\bar{Q}}\\

\begin{pmatrix}-\frac{\sqrt{2}}{3}v_{nn}+\frac{\sqrt{2}}{3}v_{ns}\\-\frac{7}{3\sqrt2}v_{n\bar{Q}}+\frac{7}{3\sqrt2}v_{s\bar{Q}}\end{pmatrix}&
\begin{pmatrix}-\frac{8}{3}v_{nn}-\frac{8}{3}v_{ns}\\+5v_{n\bar{Q}}+\frac53v_{s\bar{Q}}\end{pmatrix}&
\begin{pmatrix}\frac{5\sqrt{2}}{\sqrt{3}}v_{nn}-\frac{5\sqrt{2}}{\sqrt{3}}v_{ns}\\+\frac{5}{\sqrt{6}}v_{n\bar{Q}}-\frac{5}{\sqrt{6}}v_{s\bar{Q}}\end{pmatrix}&
-\frac{\sqrt{5}}{3}v_{n\bar{Q}}+\frac{\sqrt{5}}{3}v_{s\bar{Q}}\\

\begin{pmatrix}\frac{5}{\sqrt{3}}v_{nn}-\frac{5}{\sqrt{3}}v_{ns}\\+\frac{5}{2\sqrt{3}}v_{n\bar{Q}}-\frac{5}{2\sqrt{3}}v_{s\bar{Q}}\end{pmatrix}&
\begin{pmatrix}\frac{5\sqrt{2}}{\sqrt{3}}v_{nn}-\frac{5\sqrt{2}}{\sqrt{3}}v_{ns}\\+\frac{5}{\sqrt{6}}v_{n\bar{Q}}-\frac{5}{\sqrt{6}}v_{s\bar{Q}}\end{pmatrix}&
\begin{pmatrix}-3v_{nn}-\frac{31}{3}v_{ns}\\+\frac{23}{6}v_{n\bar{Q}}+\frac{17}{6}v_{s\bar{Q}}\end{pmatrix}&
-\frac{\sqrt{5}}{\sqrt{6}}v_{n\bar{Q}}+\frac{\sqrt{5}}{\sqrt{6}}v_{s\bar{Q}}\\

-\frac{23\sqrt{5}}{3\sqrt{2}}v_{n\bar{Q}}+\frac{\sqrt{5}}{3\sqrt{2}}v_{s\bar{Q}}&
-\frac{\sqrt{5}}{3}v_{n\bar{Q}}+\frac{\sqrt{5}}{3}v_{s\bar{Q}}&
-\frac{\sqrt{5}}{\sqrt{6}}v_{n\bar{Q}}+\frac{\sqrt{5}}{\sqrt{6}}v_{s\bar{Q}}&
\begin{pmatrix}2v_{nn}+6v_{ns}-9v_{n\bar{Q}}+v_{s\bar{Q}}\end{pmatrix}
\end{pmatrix}$
\\
$\frac12(\frac12^{-})$&$\begin{pmatrix}
\begin{pmatrix}-\frac{7}{3}v_{nn}+5v_{ns}\\+\frac{11}{3}v_{n\bar{Q}}-v_{s\bar{Q}}\end{pmatrix}&
\begin{pmatrix}-\frac{\sqrt{2}}{3}v_{nn}+\frac{\sqrt{2}}{3}v_{ns}\\+\frac{7}{3\sqrt2}v_{n\bar{Q}}-\frac{7}{3\sqrt2}v_{s\bar{Q}}\end{pmatrix}&
\begin{pmatrix}\frac{5}{\sqrt{3}}v_{nn}-\frac{5}{\sqrt{3}}v_{ns}\\-\frac{5}{\sqrt{3}}v_{n\bar{Q}}+\frac{5}{\sqrt{3}}v_{s\bar{Q}}\end{pmatrix}&
-\frac{14}{3}v_{n\bar{Q}}-\frac{10}{3}v_{s\bar{Q}}&
2v_{n\bar{Q}}-2v_{s\bar{Q}}\\

\begin{pmatrix}-\frac{\sqrt{2}}{3}v_{nn}+\frac{\sqrt{2}}{3}v_{ns}\\+\frac{7}{3\sqrt2}v_{n\bar{Q}}-\frac{7}{3\sqrt2}v_{s\bar{Q}}\end{pmatrix}&
\begin{pmatrix}-\frac{8}{3}v_{nn}-\frac{8}{3}v_{ns}\\-10v_{n\bar{Q}}-\frac{10}{3}v_{s\bar{Q}}\end{pmatrix}&
\begin{pmatrix}\frac{5\sqrt{2}}{\sqrt{3}}v_{nn}-\frac{5\sqrt{2}}{\sqrt{3}}v_{ns}\\-\frac{5\sqrt{2}}{\sqrt{3}}v_{n\bar{Q}}+\frac{5\sqrt{2}}{\sqrt{3}}v_{s\bar{Q}}\end{pmatrix}&
\frac{7\sqrt{2}}{3}(v_{n\bar{Q}}-v_{s\bar{Q}})&
3\sqrt{3}(v_{n\bar{Q}}-v_{s\bar{Q}})\\

\begin{pmatrix}\frac{5}{\sqrt{3}}v_{nn}-\frac{5}{\sqrt{3}}v_{ns}\\-\frac{5}{\sqrt{3}}v_{n\bar{Q}}+\frac{5}{\sqrt{3}}v_{s\bar{Q}}\end{pmatrix}&
\begin{pmatrix}\frac{5\sqrt{2}}{\sqrt{3}}v_{nn}-\frac{5\sqrt{2}}{\sqrt{3}}v_{ns}\\-\frac{5\sqrt{2}}{\sqrt{3}}v_{n\bar{Q}}+\frac{5\sqrt{2}}{\sqrt{3}}v_{s\bar{Q}}\end{pmatrix}&
\begin{pmatrix}-3v_{nn}-\frac{31}{3}v_{ns}\\-\frac{23}{3}v_{n\bar{Q}}-\frac{17}{3}v_{s\bar{Q}}\end{pmatrix}&
\frac{-8}{\sqrt{3}}(v_{n\bar{Q}}-v_{s\bar{Q}})&
\frac{8}{\sqrt{3}}(2v_{n\bar{Q}}+v_{s\bar{Q}})\\

-\frac{14}{3}v_{n\bar{Q}}-\frac{10}{3}v_{s\bar{Q}}&
\frac{7\sqrt{2}}{3}v_{n\bar{Q}}-\frac{7\sqrt{2}}{3}v_{s\bar{Q}}&
-\frac{8}{\sqrt{3}}v_{n\bar{Q}}+\frac{8}{\sqrt{3}}v_{s\bar{Q}}&
-5v_{nn}+5v_{ns}&
-3v_{nn}+3v_{ns}\\

2v_{n\bar{Q}}-2v_{c\bar{Q}}&
3\sqrt{3}v_{n\bar{Q}}-3\sqrt{3}v_{s\bar{Q}}&
\frac{16}{\sqrt{3}}v_{n\bar{Q}}+\frac{8}{\sqrt{3}}v_{s\bar{Q}}&
-3v_{nn}+3v_{ns}&
-5v_{nn}-11v_{ns}
\end{pmatrix}$
\\
\bottomrule[0.8pt]
&$nnss\bar{Q}$\\
\cline{2-2}
$1(\frac52^{-})$&$\frac83v_{nn}+\frac83v_{ss}+\frac83v_{ns}+\frac83v_{n\bar{Q}}+\frac83v_{s\bar{Q}}$\\
$1(\frac32^{-})$&$\begin{pmatrix}
\begin{pmatrix}\frac{28}{9}v_{nn}+\frac{28}{9}v_{ss}+\frac{112}{9}v_{ns}\\-\frac83v_{n\bar{Q}}-\frac83v_{s\bar{Q}}\end{pmatrix}&
\begin{pmatrix}\frac{2}{3}\sqrt{\frac23}v_{nn}-\frac{2}{3}\sqrt{\frac23}v_{ss}\\-\frac{4}{3}\sqrt{\frac23}v_{n\bar{Q}}+\frac{4}{3}\sqrt{\frac23}v_{s\bar{Q}}\end{pmatrix}&
\begin{pmatrix}-\frac{2}{9}\sqrt{2}v_{nn}-\frac{2}{9}\sqrt{2}v_{ss}\\-\frac{4}{9}\sqrt{2}v_{ns}\end{pmatrix}&
\frac{16}{3}\sqrt{\frac53}(v_{n\bar{Q}}-v_{s\bar{Q}})\\

\begin{pmatrix}\frac{2}{3}\sqrt{\frac23}v_{nn}+\frac{2}{3}\sqrt{\frac23}v_{ss}\\-\frac{4}{3}\sqrt{\frac23}v_{n\bar{Q}}+\frac{4}{3}\sqrt{\frac23}v_{s\bar{Q}}\end{pmatrix}&
\begin{pmatrix}\frac{10}{3}v_{nn}+\frac{10}{3}v_{ss}-4v_{ns}\\-\frac23v_{n\bar{Q}}-\frac23v_{s\bar{Q}}\end{pmatrix}&
\begin{pmatrix}-\frac{2}{3\sqrt{3}}v_{nn}+\frac{2}{3\sqrt{3}}v_{ss}\\-\frac{14}{3\sqrt{3}}\sqrt{\frac23}v_{n\bar{Q}}+\frac{14}{3\sqrt{3}}v_{s\bar{Q}}\end{pmatrix}&
2\sqrt{10}v_{n\bar{Q}}+2\sqrt{10}v_{s\bar{Q}} \\

\begin{pmatrix}-\frac{2}{9}\sqrt{2}v_{nn}-\frac{2}{9}\sqrt{2}v_{ss}\\-\frac{4}{9}\sqrt{2}v_{ns}\end{pmatrix}&
\begin{pmatrix}-\frac{2}{3\sqrt{3}}v_{nn}+\frac{2}{3\sqrt{3}}v_{ss}\\-\frac{14}{3\sqrt{3}}\sqrt{\frac23}v_{n\bar{Q}}+\frac{14}{3\sqrt{3}}v_{s\bar{Q}}\end{pmatrix}&
\begin{pmatrix}\frac{26}{9}v_{nn}+\frac{26}{9}v_{ss}-\frac{100}{9}v_{ns}\\+\frac{10}{3}v_{n\bar{Q}}+\frac{10}{3}v_{s\bar{Q}}\end{pmatrix}&
-\frac{2}{3}\sqrt{\frac{10}{3}}v_{n\bar{Q}}+\frac{2}{3}\sqrt{\frac{10}{3}}v_{s\bar{Q}}\\

\frac{16}{3}\sqrt{\frac53}v_{n\bar{Q}}-\frac{16}{3}\sqrt{\frac53}v_{s\bar{Q}} &
2\sqrt{10}v_{n\bar{Q}}+2\sqrt{10}v_{s\bar{Q}}&
-\frac{2}{3}\sqrt{\frac{10}{3}}v_{n\bar{Q}}+\frac{2}{3}\sqrt{\frac{10}{3}}v_{s\bar{Q}}&
\begin{pmatrix}\frac83v_{nn}+\frac83v_{ss}+\frac83v_{ns}\\-4v_{n\bar{Q}}-4v_{s\bar{Q}}\end{pmatrix}
\end{pmatrix}$
\\
$1(\frac12^{-})$&$\begin{pmatrix}
\begin{pmatrix}\frac{28}{9}v_{nn}+\frac{28}{9}v_{ss}+\frac{112}{9}v_{ns}\\+\frac{16}{3}v_{n\bar{Q}}+\frac{16}{3}v_{s\bar{Q}}\end{pmatrix}&
\begin{pmatrix}\frac{2}{3}\sqrt{\frac23}v_{nn}-\frac{2}{3}\sqrt{\frac23}v_{ss}\\+\frac{8}{3}\sqrt{\frac23}v_{n\bar{Q}}-\frac{8}{3}\sqrt{\frac23}v_{s\bar{Q}}\end{pmatrix}&
\begin{pmatrix}-\frac{2}{9}\sqrt{2}v_{nn}-\frac{2}{9}\sqrt{2}v_{ss}\\-\frac{4}{9}\sqrt{2}v_{ns}\end{pmatrix}&
-\frac{4}{3}\sqrt{\frac23}v_{n\bar{Q}}+\frac{4}{3}\sqrt{\frac23}v_{s\bar{Q}}\\

\begin{pmatrix}\frac{2}{3}\sqrt{\frac23}v_{nn}-\frac{2}{3}\sqrt{\frac23}v_{ss}\\+\frac{8}{3}\sqrt{\frac23}v_{n\bar{Q}}-\frac{8}{3}\sqrt{\frac23}v_{s\bar{Q}}\end{pmatrix}&
\begin{pmatrix}\frac{10}{3}v_{nn}+\frac{10}{3}v_{ss}-4v_{ns}\\+\frac43v_{n\bar{Q}}+\frac43v_{s\bar{Q}}\end{pmatrix}&
\begin{pmatrix}-\frac{2}{3\sqrt{3}}v_{nn}+\frac{2}{3\sqrt{3}}v_{ss}\\+\frac{28}{3\sqrt{3}}v_{n\bar{Q}}-\frac{28}{3\sqrt{3}}v_{s\bar{Q}}\end{pmatrix}&
-4v_{n\bar{Q}}-4v_{s\bar{Q}} \\

\begin{pmatrix}-\frac{2}{9}\sqrt{2}v_{nn}-\frac{2}{9}\sqrt{2}v_{ss}\\-\frac{4}{9}\sqrt{2}v_{ns}\end{pmatrix}&
\begin{pmatrix}-\frac{2}{3\sqrt{3}}v_{nn}+\frac{2}{3\sqrt{3}}v_{ss}\\+\frac{28}{3\sqrt{3}}v_{n\bar{Q}}-\frac{28}{3\sqrt{3}}v_{s\bar{Q}}\end{pmatrix}&
\begin{pmatrix}\frac{26}{9}v_{nn}+\frac{26}{9}v_{ss}-\frac{100}{9}v_{ns}\\-\frac{20}{3}v_{n\bar{Q}}-\frac{20}{3}v_{s\bar{Q}}\end{pmatrix}&
\frac{28}{3\sqrt3}v_{n\bar{Q}}-\frac{28}{3\sqrt3}v_{s\bar{Q}}\\

-\frac{4}{3}\sqrt{\frac23}v_{n\bar{Q}}+\frac{4}{3}\sqrt{\frac23}v_{s\bar{Q}} &
-4v_{n\bar{Q}}-4v_{s\bar{Q}}&
\frac{28}{3\sqrt3}v_{n\bar{Q}}-\frac{28}{3\sqrt3}v_{s\bar{Q}}&
\frac83v_{nn}+\frac83v_{ss}-\frac{16}{3}v_{ns}
\end{pmatrix}$\\
\cline{1-2}
$0(\frac52^{-})$&$-\frac43v_{nn}+\frac83v_{ss}+\frac{20}{3}v_{ns}-\frac43v_{n\bar{Q}}-\frac43v_{s\bar{Q}}$\\
$0(\frac32^{-})$&$\begin{pmatrix}
\begin{pmatrix}-\frac{14}{3}v_{nn}+\frac{8}{3}v_{ss}+\frac{14}{3}v_{ns}-\frac{7}{3}v_{n\bar{Q}}+v_{s\bar{Q}}\end{pmatrix}&
\begin{pmatrix}-\frac{10}{3}v_{nn}+\frac{10}{3}v_{ns}-\frac{5}{3}v_{n\bar{Q}}+\frac{5}{3}v_{s\bar{Q}}\end{pmatrix}&
\frac{11\sqrt{10}}{3}v_{n\bar{Q}}+\frac{\sqrt{10}}{3}v_{s\bar{Q}}\\

\begin{pmatrix}-\frac{10}{3}v_{nn}+\frac{10}{3}v_{ns}-\frac{5}{3}v_{n\bar{Q}}+\frac{5}{3}v_{s\bar{Q}}\end{pmatrix}&
\begin{pmatrix}-\frac{14}{3}v_{nn}+\frac{8}{3}v_{ss}-\frac{34}{3}v_{ns}+\frac{17}{3}v_{n\bar{Q}}+v_{s\bar{Q}}\end{pmatrix}&
\frac{\sqrt{10}}{3}v_{n\bar{Q}}-\frac{\sqrt{10}}{3}v_{s\bar{Q}} \\

\frac{11\sqrt{10}}{3}v_{n\bar{Q}}+\frac{\sqrt{10}}{3}v_{s\bar{Q}}&
\frac{\sqrt{10}}{3}v_{n\bar{Q}}-\frac{\sqrt{10}}{3}v_{s\bar{Q}}&
\begin{pmatrix}-\frac43v_{nn}+\frac83v_{ss}+\frac{10}{3}v_{ns}-10v_{n\bar{Q}}+2v_{s\bar{Q}}\end{pmatrix}
\end{pmatrix}$
\\
$0(\frac12^{-})$&$\begin{pmatrix}
\begin{pmatrix}-\frac{14}{3}v_{nn}+\frac{8}{3}v_{ss}+\frac{14}{3}\\v_{ns}+\frac{14}{3}v_{n\bar{Q}}-2v_{s\bar{Q}}\end{pmatrix}&
\begin{pmatrix}-\frac{10}{3}v_{nn}+\frac{10}{3}v_{ns}\\+\frac{10}{3}v_{n\bar{Q}}-\frac{10}{3}v_{s\bar{Q}}\end{pmatrix}&
-\frac43v_{n\bar{Q}}-\frac{20}{3}v_{s\bar{Q}}&
-\frac{4}{\sqrt3}v_{n\bar{Q}}+\frac{4}{\sqrt3}v_{s\bar{Q}}\\

\begin{pmatrix}-\frac{10}{3}v_{nn}+\frac{10}{3}v_{ns}\\+\frac{10}{3}v_{n\bar{Q}}-\frac{10}{3}v_{s\bar{Q}}\end{pmatrix}&
\begin{pmatrix}-\frac{14}{3}v_{nn}+\frac{8}{3}v_{ss}-\frac{34}{3}\\v_{ns}-\frac{34}{3}v_{n\bar{Q}}-2v_{s\bar{Q}}\end{pmatrix}&
\frac{16}{3}v_{n\bar{Q}}-\frac{16}{3}v_{s\bar{Q}}&
\frac{16}{\sqrt3}v_{n\bar{Q}}+\frac{8}{\sqrt3}v_{s\bar{Q}}\\

-\frac43v_{n\bar{Q}}-\frac{20}{3}v_{s\bar{Q}}&
\frac{16}{3}v_{n\bar{Q}}-\frac{16}{3}v_{s\bar{Q}}&
\begin{pmatrix}-\frac{19}{3}v_{nn}+\frac{11}{\sqrt3}v_{ss}+\frac{8}{3}v_{ns}\end{pmatrix}&
\begin{pmatrix}-\frac{5}{\sqrt3}v_{nn}-\frac{1}{\sqrt3}v_{ss}-\frac{4}{\sqrt3}v_{ns}\end{pmatrix}\\

-\frac{4}{\sqrt3}v_{n\bar{Q}}+\frac{4}{\sqrt3}v_{s\bar{Q}}&
\frac{16}{\sqrt3}v_{n\bar{Q}}+\frac{8}{\sqrt3}v_{s\bar{Q}}&
\begin{pmatrix}-\frac{5}{\sqrt3}v_{nn}-\frac{1}{\sqrt3}v_{ss}-\frac{4}{\sqrt3}v_{ns}\end{pmatrix}&
\begin{pmatrix}-3v_{nn}+3v_{ss}-16v_{ns}\end{pmatrix}
\end{pmatrix}$
\\
\bottomrule[0.5pt]
\bottomrule[1.5pt]
\end{tabular}
\end{lrbox}\scalebox{0.875}{\usebox{\tablebox}}
\end{table*}

\begin{table*}[t]
\centering \caption{
The overlaps of wave functions between a $nnns\bar{Q}$ or $nnss\bar{Q}$ pentaquark state and a particular baryon $\otimes$ meson state. The masses are all in units of MeV.
See the caption of Fig. \ref{fig-nnnsQ} for the meanings of ``$\diamond$" and ``$\star$".
}\label{eigenvector-nnnsC}
\begin{lrbox}{\tablebox}
\renewcommand\arraystretch{1.18}
\renewcommand\tabcolsep{1.4pt}
\begin{tabular}{cl|cc|cccccc|cl|cc|cccccc}
\bottomrule[1.5pt]
\bottomrule[0.5pt]
\multicolumn{2}{l|}{$nnns\bar{c}$}&\multicolumn{2}{c}{$nnn\bigotimes s\bar{c}$}&\multicolumn{6}{c}{$nns\bigotimes n\bar{c}$}&\multicolumn{2}{|l|}{$nnns\bar{b}$}&\multicolumn{2}{c}{$nnn\bigotimes s\bar{b}$}&\multicolumn{6}{c}{$nns\bigotimes n\bar{b}$}\\
$I(J^P)$&Mass&$\Delta \bar{D}_{s}^{*}$&$\Delta \bar{D}_{s}$&$\Sigma^{*}\bar{D}^{*}$&$\Sigma^{*}\bar{D}$&$\Sigma\bar{D}^{*}$&$\Sigma\bar{D}$&$\Lambda\bar{D}^{*}$&$\Lambda\bar{D}$
&$I(J^P)$&Mass&$\Delta B_{s}^{*}$&$\Delta B_{s}$&$\Sigma^{*}B^{*}$&$\Sigma^{*}B$&$\Sigma B^{*}$&$\Sigma B$&$\Lambda B^{*}$&$\Lambda B$\\
\bottomrule[0.5pt]
$\frac32(\frac{5}{2}^{-})$
&$3352\star$&1.000&&0.333&&&&&&$\frac32(\frac{5}{2}^{-})$&$6655\star$&1.000&&0.333&&&\\
$\frac32(\frac{3}{2}^{-})$
&3500&0.248&-0.291&0.508&-0.360&-0.088&&&&$\frac32(\frac{3}{2}^{-})$&6862&0.297&-0.257&0.496&-0.375&-0.081&\\
&$3343\star$&0.966&0.145&-0.191&-0.150&0.259&&&&&$6642\star$&0.906&0.380&-0.225&-0.187&0.193&\\
&$3177\star\diamond$&-0.072&0.946&0.042&0.264&0.270&&&&&6579&-0.302&0.889&-0.003&-0.216&-0.323&\\
$\frac32(\frac{1}{2}^{-})$
&3603&-0.442&&0.613&&-0.046&-0.043&&&$\frac32(\frac{1}{2}^{-})$&6895&-0.411&&0.619&&-0.044&-0.059\\
&3353&-0.580&&-0.100&&0.566&-0.049&&&&6657&-0.288&&-0.041&&0.607&-0.216\\
&3246&0.684&&0.149&&0.220&-0.467&&&&6601&0.865&&-0.152&&0.415&0.001\\
\Xcline{3-4}{0.5pt}\Xcline{13-14}{0.5pt}
\multicolumn{2}{l|}{}&$N \bar{D}_{s}^{*}$&$N\bar{D}_{s}$&&&&&&&
\multicolumn{2}{l|}{}&$N B_{s}^{*}$&$NB_{s}$&&&\\
\Xcline{3-4}{0.5pt}\Xcline{13-14}{0.5pt}
$\frac12(\frac{5}{2}^{-})$
&3405&&&0.666&&&&&&$\frac12(\frac{5}{2}^{-})$&6723&&&0.667\\
$\frac12(\frac{3}{2}^{-})$
&3376&0.159&&-0.580&-0.238&-0.069&&0.197&&$\frac12(\frac{3}{2}^{-})$&6702&0.072&&-0.507&-0.421&-0.032&&0.090\\
&3208&0.127&&-0.056&0.360&0.352&&0.427&&&6600&-0.201&&-0.298&0.454&-0.105&&-0.358\\
&3199&-0.266&&0.087&-0.437&0.4700&&0.035&&&6515&-0.263&&0.026&0.029&0.583&&0.282\\
&$3022\star\diamond$&0.942&&-0.072&0.110&0.275&&-0.236&&&$6324\star\diamond$&0.941&&0.077&0.072&0.271&&0.254\\
$\frac12(\frac{1}{2}^{-})$
&3309&0.316&0.108&0.486&&0.166&0.044&0.346&0.147&$\frac12(\frac{1}{2}^{-})$&6618&0.207&0.157&0.525&&0.111&0.072&0.265&0.243\\
&3172&0.274&-0.332&-0.234&&0.260&-0.113&0.328&-0.374&&6531&0.375&-0.271&-0.109&&0.269&-0.129&0.401&-0.334\\
&3082&-0.219&0.293&0.027&&0.377&0.446&0.165&0.175&&6473&0.053&0.261&0.029&&0.293&0.502&-0.175&0.228\\
&3000&-0.879&-0.208&-0.035&&0.263&-0.177&-0.139&-0.227&&6308&0.876&0.278&-0.063&&-0.004&0.343&0.107&0.173\\
&$2831\diamond$&-0.071&0.866&-0.054&&-0.093&0.379&-0.064&0.184&&$6253\diamond$&-0.217&0.869&0.063&&0.164&-0.340&0.071&0.160\\
\bottomrule[0.7pt]
\multicolumn{2}{l|}{$sssn\bar{c}$}&\multicolumn{2}{c}{$sss\bigotimes n\bar{c}$}&\multicolumn{4}{c}{$ssn\bigotimes s\bar{c}$}&&&\multicolumn{2}{l|}{$sssn\bar{b}$}&\multicolumn{2}{c}{$sss\bigotimes n\bar{b}$}&\multicolumn{4}{c}{$ssn\bigotimes s\bar{b}$}\\
$I(J^P)$&Mass&$\Omega \bar{D}^{*}$&$\Omega\bar{D}$&$\Xi^{*}\bar{D}_{s}^{*}$&$\Xi^{*}\bar{D}_{s}$&$\Xi\bar{D}_{s}^{*}$&$\Xi\bar{D}_{s}$
&&&$I(J^P)$&Mass&$\Omega B^{*}$&$\Omega B$&$\Xi^{*}B_{s}^{*}$&$\Xi^{*}B_{s}$&$\Xi B_{s}^{*}$&$\Xi B_{s}$\\
\bottomrule[0.5pt]
$\frac12(\frac{5}{2}^{-})$
&$3680\star$&1.000&&0.333&&&&&&$\frac12(\frac{5}{2}^{-})$&$6996\star$&1.000&&0.333&&\\
$\frac12(\frac{3}{2}^{-})$
&3724&0.627&-0.349&0.386&-0.345&0.068&&&
&$\frac12(\frac{3}{2}^{-})$&7079&0.571&-0.404&0.410&-0.331&0.061\\
&3655&0.772&0.396&-0.375&-0.041&0.226&&&
&&6976&0.733&0.642&0.290&0.179&-0.109\\
&3483&-0.102&0.849&-0.080&0.319&0.304&&&
&&6878&-0.368&0.653&0.209&-0.285&-0.364\\
$\frac12(\frac{1}{2}^{-})$
&3829&-0.648&&-0.551&&-0.025&-0.025&&
&$\frac12(\frac{1}{2}^{-})$&7114&-0.681&&-0.536&&-0.030&-0.042\\
&3601&-0.676&&-0.293&&0.427&0.150&&
&&6899&0.714&&0.339&&-0.311&-0.250\\
&3465&0.638&&-0.136&&-0.433&0.446&&
&&6813&0.163&&0.071&&0.522&-0.397\\
\bottomrule[1.0pt]
\multicolumn{2}{l|}{$nnss\bar{c}$}&\multicolumn{4}{c}{$nns\bigotimes s\bar{c}$}&\multicolumn{4}{c|}{$ssn\bigotimes n\bar{c}$}
&\multicolumn{2}{l|}{$nnss\bar{b}$}&\multicolumn{4}{c}{$nns\bigotimes s\bar{b}$}&\multicolumn{4}{c}{$ssn\bigotimes n\bar{b}$}\\
$I(J^P)$&Mass&
$\Sigma^{*} \bar{D}_{s}^{*}$&\multicolumn{1}{c}{$\Sigma^{*} \bar{D}_{s}$}&$\Sigma \bar{D}^{*}_{s}$&\multicolumn{1}{c|}{$\Sigma \bar{D}_{s}$}&$\Xi^{*} \bar{D}^{*}$&$\Xi^{*} \bar{D}$&$\Xi \bar{D}^{*}$&$\Xi \bar{D}$
&$I(J^P)$&Mass&$\Sigma^{*} B_{s}^{*}$&\multicolumn{1}{c}{$\Sigma^{*} B_{s}$}&$\Sigma B^{*}_{s}$&\multicolumn{1}{c|}{$\Sigma B_{s}$}&
$\Xi^{*} B^{*}$&$\Xi^{*} B$&$\Xi B^{*}$&$\Xi B$\\
\bottomrule[0.5pt]
$1(\frac52^{-})$
&3520&-0.577&\multicolumn{1}{c}{}&&\multicolumn{1}{c|}{}&-0.577&&&
&$1(\frac52^{-})$&6829&-0.577&\multicolumn{1}{c}{}&&\multicolumn{1}{c|}{}&-0.577&\\
$1(\frac32^{-})$
&3614&0.298&\multicolumn{1}{c}{-0.316}&-0.052&\multicolumn{1}{c|}{}&0.575&-0.360&0.122&
&$1(\frac32^{-})$&6972&0.344&\multicolumn{1}{c}{-0.290}&-0.049&\multicolumn{1}{c|}{}&0.542&-0.396&0.108\\
&3505&0.612&\multicolumn{1}{c}{0.051}&0.151&\multicolumn{1}{c|}{}&-0.369&-0.239&-0.272&
&&6813&-0.532&\multicolumn{1}{c}{-0.243}&-0.090&\multicolumn{1}{c|}{}&-0.410&-0.338&-0.185\\
&3368&0.006&\multicolumn{1}{c}{-0.330}&-0.367&\multicolumn{1}{c|}{}&0.074&0.266&-0.589&
&&6747&-0.244&\multicolumn{1}{c}{0.508}&-0.101&\multicolumn{1}{c|}{}&-0.060&0.366&-0.448\\
&3325&0.093&\multicolumn{1}{c}{-0.453}&0.533&\multicolumn{1}{c|}{}&-0.021&0.397&0.101&
&&6650&0.105&\multicolumn{1}{c}{-0.125}&0.651&\multicolumn{1}{c|}{}&-0.084&0.108&0.444\\
$1(\frac12^{-})$
&3716&0.497&\multicolumn{1}{c}{}&-0.030&\multicolumn{1}{c|}{-0.025}&-0.643&&-0.045&-0.051
&$1(\frac12^{-})$&7005&0.468&\multicolumn{1}{c}{}&-0.029&\multicolumn{1}{c|}{-0.036}&-0.661&&-0.051&-0.076\\
&3475&0.493&\multicolumn{1}{c}{}&0355&\multicolumn{1}{c|}{0.065}&0.320&&0.408&0.158
&&6772&-0.250&\multicolumn{1}{c}{}&0.245&\multicolumn{1}{c|}{-0.077}&-0.317&&-0.342&-0.287\\
&3350&-0.250&\multicolumn{1}{c}{}&0.393&\multicolumn{1}{c|}{-0.334}&0.187&&-0.331&0.441
&&6707&0.141&\multicolumn{1}{c}{}&-0.470&\multicolumn{1}{c|}{0.299}&-0.090&&0.412&-0.399\\
&3220&-0.061&\multicolumn{1}{c}{}&0.283&\multicolumn{1}{c|}{-0.548}&0.067&&-0.288&-0.441
&&6609&-0.106&\multicolumn{1}{c}{}&-0.295&\multicolumn{1}{c|}{-0.563}&0.099&&-0.268&-0.412\\
\Xcline{3-6}{0.5pt}\Xcline{13-16}{0.5pt}
\multicolumn{2}{l|}{}&$\Lambda \bar{D}^{*}_{s}$&\multicolumn{1}{c}{$\Lambda \bar{D}_{s}$}&&\multicolumn{1}{c|}{}&&&&&
\multicolumn{2}{l|}{}&$\Lambda B_{s}^{*}$&\multicolumn{1}{c}{$\Lambda B_{s}$}&&\multicolumn{1}{c|}{}&\\
\Xcline{3-6}{0.5pt}\Xcline{13-16}{0.5pt}
$0(\frac52^{-})$&3555&&\multicolumn{1}{c}{}&&\multicolumn{1}{c|}{}&0.817&&&&$0(\frac52^{-})$&6875&&\multicolumn{1}{c}{}&&\multicolumn{1}{c|}{}&-0.817\\
$0(\frac32^{-})$
&3524&-0.189&\multicolumn{1}{c}{}&&\multicolumn{1}{c|}{}&-0.706&-0.305&0.198&&$0(\frac32^{-})$&6854&0.082&\multicolumn{1}{c}{}&&\multicolumn{1}{c|}{}&-0.612&-0.527&0.086\\
&3351&-0.186&\multicolumn{1}{c}{}&&\multicolumn{1}{c|}{}&0.128&-0.674&-0.402&&&6743&-0.282&\multicolumn{1}{c}{}&&\multicolumn{1}{c|}{}&-0.060&0.366&-0.448\\
&$3216\diamond$&0.513&\multicolumn{1}{c}{}&&\multicolumn{1}{c|}{}&-0.021&0.397&0.101&&&$6527\diamond$&0.497&\multicolumn{1}{c}{}&&\multicolumn{1}{c|}{}&-0.121&0.115&-0.626\\
$0(\frac12^{-})$
&3451&-0.354&\multicolumn{1}{c}{-0.140}&&\multicolumn{1}{c|}{}&0.592&&-0.386&-0.148&$0(\frac12^{-})$&6759&-0.245&\multicolumn{1}{c}{-0.213}&&\multicolumn{1}{c|}{}&-0.638&&0.297&0.257\\
&3312&0.280&\multicolumn{1}{c}{-0.392}&&\multicolumn{1}{c|}{}&0.294&&0.473&-0.353&&6667&0.386&\multicolumn{1}{c}{-0.318}&&\multicolumn{1}{c|}{}&-0.138&&-0.517&0.360\\
&3213&0.609&\multicolumn{1}{c}{-0.062}&&\multicolumn{1}{c|}{}&-0.027&&-0.515&-0.161&&6513&0.567&\multicolumn{1}{c}{0.078}&&\multicolumn{1}{c|}{}&0.097&&0.520&0.244\\
&$3026\diamond$&0.095&\multicolumn{1}{c}{-0.489}&&\multicolumn{1}{c|}{}&0.080&&-0.033&0.641&&$6455\diamond$&0.230&\multicolumn{1}{c}{-0.514}&&\multicolumn{1}{c|}{}&0.094&&-0.114&0.572\\
\bottomrule[0.5pt]
\bottomrule[1.5pt]
\end{tabular}
\end{lrbox}\scalebox{0.88}{\usebox{\tablebox}}
\end{table*}

\begin{table*}[t]
\centering \caption{
The values of $k\cdot |c_{i}|^{2}$ for the $nnns\bar{Q}$ and $nnss\bar{Q}$  pentaquark states. The masses are all in units of MeV.
The kinetically forbidden decay channel is marked with ``$\times$''.
See the caption of Fig. \ref{fig-nnnsQ} for the meanings of ``$\diamond$" and ``$\star$".
One can roughly estimate the relative decay widths between different decay processes of different initial pentaquark states with this table if neglecting the $\gamma_i$ differences.
}\label{value-nnnsC}
\begin{lrbox}{\tablebox}
\renewcommand\arraystretch{1.18}
\renewcommand\tabcolsep{1.55pt}
\begin{tabular*}{182mm}{cl|cc|cccccc|cl|cc|cccccc}
\bottomrule[1.5pt]
\bottomrule[0.5pt]
\multicolumn{2}{l}{$nnns\bar{c}$}&\multicolumn{2}{c}{$nnn\bigotimes s\bar{c}$}&\multicolumn{6}{c|}{$nns\bigotimes n\bar{c}$}&\multicolumn{2}{l}{$nnns\bar{b}$}&\multicolumn{2}{c}{$nnn\bigotimes s\bar{b}$}&\multicolumn{6}{c}{$nns\bigotimes n\bar{b}$}\\
$I(J^P)$&Mass&$\Delta \bar{D}_{s}^{*}$&$\Delta \bar{D}_{s}$&$\Sigma^{*}\bar{D}^{*}$&$\Sigma^{*}\bar{D}$&$\Sigma\bar{D}^{*}$&$\Sigma\bar{D}$&$\Lambda\bar{D}^{*}$&$\Lambda\bar{D}$
&$I(J^P)$&\multicolumn{1}{c|}{Mass}&$\Delta B_{s}^{*}$&$\Delta B_{s}$&$\Sigma^{*}B^{*}$&$\Sigma^{*}B$&$\Sigma B^{*}$&$\Sigma B$&$\Lambda B^{*}$&$\Lambda B$\\
\bottomrule[0.5pt]
$\frac32(\frac{5}{2}^{-})$
&$3352\star$&111&&$\times$&&&&&&$\frac32(\frac{5}{2}^{-})$&$6655\star$&126&&$\times$&&&\\
$\frac32(\frac{3}{2}^{-})$
&3500&31&59&109&83&5&&&&$\frac32(\frac{3}{2}^{-})$&6862&60&50&145&95&6&\\
&$3343\star$&$\times$&10&$\times$&9&31&&&&&$6642\star$&$\times$&43&$\times$&$\times$&19&\\
&$3177\star\diamond$&$\times$&$\times$&$\times$&$\times$&$\times$&&&&&6579&$\times$&$\times$&$\times$&$\times$&37&\\
$\frac32(\frac{1}{2}^{-})$
&3603&127&&224&&2&2&&&$\frac32(\frac{1}{2}^{-})$&6895&123&&251&&2&3\\
&3353&39&&$\times$&&154&1&&&&6657&12&&$\times$&&196&29\\
&3246&$\times$&&$\times$&&12&56&&&&6601&$\times$&&$\times$&&70&$\times$\\
\Xcline{3-4}{0.5pt}\Xcline{13-14}{0.5pt}
\multicolumn{2}{l|}{}&$N \bar{D}_{s}^{*}$&$N\bar{D}_{s}$&&&&&&&
\multicolumn{2}{l|}{}&$N B_{s}^{*}$&$NB_{s}$&&&\\
\Xcline{3-4}{0.5pt}\Xcline{13-14}{0.5pt}
$\frac12(\frac{5}{2}^{-})$
&3405&&&61&&&&&&$\frac12(\frac{5}{2}^{-})$&6723&&&80&&&&\\
$\frac12(\frac{3}{2}^{-})$
&3376&17&&$\times$&25&3&&\multicolumn{1}{|c}{24}&&$\frac12(\frac{3}{2}^{-})$&6702&4&&$\times$&52&0.6&&\multicolumn{1}{|c}{6}\\
&3209&8&&$\times$&$\times$&11&&\multicolumn{1}{|c}{63}&&&6600&27&&$\times$&$\times$&5&&\multicolumn{1}{|c}{71}\\
&3199&32&&$\times$&$\times$&$\times$&&\multicolumn{1}{|c}{0.4}&&&6515&36&&$\times$&$\times$&$\times$&&\multicolumn{1}{|c}{30}\\
&$3022\star\diamond$&$\times$&&$\times$&$\times$&$\times$&&\multicolumn{1}{|c}{$\times$}&&&$6324\star\diamond$&$\times$&&$\times$&$\times$&$\times$&&\multicolumn{1}{|c}{$\times$}\\
$\frac12(\frac{1}{2}^{-})$
&3309&60&9&$\times$&&11&1&\multicolumn{1}{|c}{63}&15&$\frac12(\frac{1}{2}^{-})$&6618&29&18&$\times$&&6&3&\multicolumn{1}{|c}{41}&39\\
&3172&30&66&$\times$&&$\times$&5&\multicolumn{1}{|c}{28}&73&&6531&77&46&$\times$&&12&6&\multicolumn{1}{|c}{67}&57\\
&3082&10&42&$\times$&&$\times$&34&\multicolumn{1}{|c}{$\times$}&11&&6473&1&36&$\times$&&$\times$&11&\multicolumn{1}{|c}{8}&20\\
&3000&$\times$&15&$\times$&&$\times$&$\times$&\multicolumn{1}{|c}{$\times$}&7&&6308&$\times$&5&$\times$&&$\times$&$\times$&\multicolumn{1}{|c}{$\times$}&$\times$\\
&$2831\diamond$&$\times$&$\times$&$\times$&&$\times$&$\times$&\multicolumn{1}{|c}{$\times$}&$\times$&&$6253\diamond$&$\times$&$\times$&$\times$&&$\times$&$\times$&\multicolumn{1}{|c}{$\times$}&$\times$\\
\bottomrule[0.7pt]
\multicolumn{2}{l|}{$sssn\bar{c}$}&\multicolumn{2}{c}{$sss\bigotimes n\bar{c}$}&\multicolumn{4}{c}{$ssn\bigotimes s\bar{c}$}&&&\multicolumn{2}{l|}{$sssn\bar{b}$}&\multicolumn{2}{c}{$sss\bigotimes n\bar{b}$}&\multicolumn{4}{c}{$ssn\bigotimes s\bar{b}$}\\
$I(J^P)$&Mass&$\Omega \bar{D}^{*}$&$\Omega\bar{D}$&$\Xi^{*}\bar{D}_{s}^{*}$&$\Xi^{*}\bar{D}_{s}$&$\Xi\bar{D}_{s}^{*}$&$\Xi\bar{D}_{s}$
&&&$I(J^P)$&Mass&$\Omega B^{*}$&$\Omega B$&$\Xi^{*}B_{s}^{*}$&$\Xi^{*}B_{s}$&$\Xi B_{s}^{*}$&$\Xi B_{s}$\\
\bottomrule[0.5pt]
$\frac12(\frac{5}{2}^{-})$&$3680\star$&$\times$&&28&&&&&&$\frac12(\frac{5}{2}^{-})$&$6996\star$&$\times$&&38&&&\\
$\frac12(\frac{3}{2}^{-})$&3724&108&70&56&75&4&&&&$\frac12(\frac{3}{2}^{-})$&7079&149&94&95&73&3\\
&3655&$\times$&71&20&0.9&32&&&&&6976&$\times$&103&22&14&31\\
&3483&$\times$&$\times$&$\times$&$\times$&28&&&&&6878&$\times$&$\times$&$\times$&$\times$&75\\
$\frac12(\frac{1}{2}^{-})$&3829&219&&176&&0.5&0.6&&&$\frac12(\frac{1}{2}^{-})$&7114&255&&184&&0.8&2\\
&3601&$\times$&&$\times$&&98&16&&&&6899&$\times$&&$\times$&&59&44\\
&3465&$\times$&&$\times$&&47&108&&&&6813&$\times$&&$\times$&&115&84\\
\bottomrule[1.0pt]
\multicolumn{2}{l}{$nnss\bar{c}$}&\multicolumn{4}{c}{$nns\bigotimes s\bar{c}$}&\multicolumn{4}{c|}{$ssn\bigotimes n\bar{c}$}
&\multicolumn{2}{l}{$nnss\bar{b}$}&\multicolumn{4}{c}{$nns\bigotimes s\bar{b}$}&\multicolumn{4}{c}{$ssn\bigotimes n\bar{b}$}\\
$I(J^P)$&Mass&
$\Sigma^{*} \bar{D}_{s}^{*}$&\multicolumn{1}{c}{$\Sigma^{*} \bar{D}_{s}$}&$\Sigma \bar{D}^{*}_{s}$&\multicolumn{1}{c|}{$\Sigma \bar{D}_{s}$}&$\Xi^{*} \bar{D}^{*}$&$\Xi^{*} \bar{D}$&$\Xi \bar{D}^{*}$&$\Xi \bar{D}$
&$I(J^P)$&Mass&$\Sigma^{*} B_{s}^{*}$&\multicolumn{1}{c}{$\Sigma^{*} B_{s}$}&$\Sigma B^{*}_{s}$&\multicolumn{1}{c|}{$\Sigma B_{s}$}&
$\Xi^{*} B^{*}$&$\Xi^{*} B$&$\Xi B^{*}$&$\Xi B$\\
\bottomrule[0.5pt]
$1(\frac52^{-})$&3520&67&\multicolumn{1}{c}{}&&\multicolumn{1}{c|}{}&$\times$&&&&$1(\frac52^{-})$&6829&86&\multicolumn{1}{c}{}&&\multicolumn{1}{c|}{}&$\times$\\
$1(\frac32^{-})$
&3614&40&\multicolumn{1}{c}{67}&2&\multicolumn{1}{c|}{}&117&79&10&&$1(\frac32^{-})$&6972&74&\multicolumn{1}{c}{60}&2&\multicolumn{1}{c|}{}&156&98&10\\
&3505&47&\multicolumn{1}{c}{1}&13&\multicolumn{1}{c|}{}&$\times$&24&40&&&6813&50&\multicolumn{1}{c}{22}&5&\multicolumn{1}{c|}{}&$\times$&8&21\\
&3368&$\times$&\multicolumn{1}{c}{17}&42&\multicolumn{1}{c|}{}&$\times$&$\times$&90&&&6747&$\times$&\multicolumn{1}{c}{$\times$}&5&\multicolumn{1}{c|}{}&$\times$&$\times$&97\\
&3325&$\times$&\multicolumn{1}{c}{$\times$}&50&\multicolumn{1}{c|}{}&$\times$&$\times$&$\times$&&&6650&$\times$&\multicolumn{1}{c}{$\times$}&122&\multicolumn{1}{c|}{}&$\times$&$\times$&29\\
$1(\frac12^{-})$
&3716&152&\multicolumn{1}{c}{}&1&\multicolumn{1}{c|}{1}&230&&2&3&$1(\frac12^{-})$&7005&151&\multicolumn{1}{c}{}&1&\multicolumn{1}{c|}{1}&264&&2&6\\
&3475&$\times$&\multicolumn{1}{c}{}&65&\multicolumn{1}{c|}{3}&$\times$&&82&17&&6772&$\times$&\multicolumn{1}{c}{}&35&\multicolumn{1}{c|}{4}&$\times$&&63&52\\
&3350&$\times$&\multicolumn{1}{c}{}&41&\multicolumn{1}{c|}{60}&$\times$&&22&100&&6707&$\times$&\multicolumn{1}{c}{}&99&\multicolumn{1}{c|}{49}&$\times$&&65&79\\
&3220&$\times$&\multicolumn{1}{c}{}&$\times$&\multicolumn{1}{c|}{89}&$\times$&&$\times$&45&&6609&$\times$&\multicolumn{1}{c}{}&4&\multicolumn{1}{c|}{99}&$\times$&&$\times$&30\\
\Xcline{3-6}{0.5pt}\Xcline{13-16}{0.5pt}
\multicolumn{2}{l|}{}&$\Lambda \bar{D}^{*}_{s}$&\multicolumn{1}{c}{$\Lambda \bar{D}_{s}$}&&\multicolumn{1}{c|}{}&&&&&
\multicolumn{2}{l|}{}&$\Lambda B_{s}^{*}$&\multicolumn{1}{c}{$\Lambda B_{s}$}&&\multicolumn{1}{c|}{}&\\
\Xcline{3-6}{0.5pt}\Xcline{13-16}{0.5pt}
$0(\frac52^{-})$&3555&&\multicolumn{1}{c}{}&&\multicolumn{1}{c|}{}&101&&&&$0(\frac52^{-})$&6875&&\multicolumn{1}{c}{}&&\multicolumn{1}{c|}{}&69\\
$0(\frac32^{-})$
&3524&24&\multicolumn{1}{c}{}&&\multicolumn{1}{c|}{}&$\times$&168&22&&$0(\frac32^{-})$&6854&5&\multicolumn{1}{c}{}&&\multicolumn{1}{c|}{}&$\times$&89&5\\
&3351&15&\multicolumn{1}{c}{}&&\multicolumn{1}{c|}{}&$\times$&$\times$&33&&&6743&51&\multicolumn{1}{c}{}&&\multicolumn{1}{c|}{}&$\times$&$\times$&95\\
&$3216\diamond$&$\times$&\multicolumn{1}{c}{}&&\multicolumn{1}{c|}{}&$\times$&$\times$&$\times$&&&$6527\diamond$&$\times$&\multicolumn{1}{c}{}&&\multicolumn{1}{c|}{}&$\times$&$\times$&$\times$\\
$0(\frac12^{-})$
&3451&73&\multicolumn{1}{c}{15}&&\multicolumn{1}{c|}{}&$\times$&&67&14&$0(\frac12^{-})$&6759&40&\multicolumn{1}{c}{34}&&\multicolumn{1}{c|}{}&$\times$&&45&40\\
&3312&28&\multicolumn{1}{c}{90}&&\multicolumn{1}{c|}{}&$\times$&&$\times$&56&&6667&41&\multicolumn{1}{c}{61}&&\multicolumn{1}{c|}{}&$\times$&&65&51\\
&3213&$\times$&\multicolumn{1}{c}{2}&&\multicolumn{1}{c|}{}&$\times$&&$\times$&6&&6513&$\times$&\multicolumn{1}{c}{1}&&\multicolumn{1}{c|}{}&$\times$&&$\times$&$\times$\\
&$3026\diamond$&$\times$&\multicolumn{1}{c}{$\times$}&&\multicolumn{1}{c|}{}&$\times$&&$\times$&$\times$&&$6455\diamond$&$\times$&\multicolumn{1}{c}{$\times$}&&\multicolumn{1}{c|}{}&$\times$&&$\times$&$\times$\\
\bottomrule[0.5pt]
\bottomrule[1.5pt]
\end{tabular*}
\end{lrbox}\scalebox{0.88}{\usebox{\tablebox}}
\end{table*}



\begin{thebibliography}{300}
\bibitem{GellMann:1964nj}
M.~Gell-Mann,
A Schematic Model of Baryons and Mesons,
Phys. Lett. \textbf{8}, 214-215 (1964).
\bibitem{Zweig:1981pd}
G.~Zweig,
An SU(3) model for strong interaction symmetry and its breaking. Version 1,
CERN-TH-401.
\bibitem{Zweig:1964jf}
G.~Zweig,
An SU(3) model for strong interaction symmetry and its breaking. Version 2,
CERN-TH-412.
\bibitem{Aubert:1974js}
J.~J.~Aubert \textit{et al.} [E598],
Experimental Observation of a Heavy Particle $J$,
Phys. Rev. Lett. \textbf{33}, 1404-1406 (1974).
\bibitem{Augustin:1974xw}
J.~E.~Augustin \textit{et al.} [SLAC-SP-017],
Discovery of a Narrow Resonance in $e^+ e^-$ Annihilation,
Phys. Rev. Lett. \textbf{33}, 1406-1408 (1974).

\bibitem{Zyla:2020zbs}
P.A.~Zyla \textit{et al.} [Particle Data Group],
Review of Particle Physics,
PTEP \textbf{2020}, no.8, 083C01 (2020)
\bibitem{Eichten:1979ms}
E.~Eichten, K.~Gottfried, T.~Kinoshita, K.~D.~Lane and T.~M.~Yan,
Charmonium: Comparison with Experiment,
Phys. Rev. D \textbf{21}, 203 (1980).
\bibitem{Isgur:1978xj}
N.~Isgur and G.~Karl,
P Wave Baryons in the Quark Model,
Phys.\ Rev.\ D {\bf 18}, 4187 (1978).
\bibitem{Godfrey:1985xj}
S.~Godfrey and N.~Isgur,
Mesons in a Relativized Quark Model with Chromodynamics,
Phys.\ Rev.\ D {\bf 32}, 189 (1985).
\bibitem{Capstick:1986bm}
S.~Capstick and N.~Isgur,
Baryons in a Relativized Quark Model with Chromodynamics,
Phys.\ Rev.\ D {\bf 34}, 2809 (1986).
\bibitem{Aaij:2019vzc}
R.~Aaij {\it et al.} [LHCb Collaboration],
Observation of a narrow pentaquark state, $P_c(4312)^+$, and of two-peak structure of the $P_c(4450)^+$,
Phys.\ Rev.\ Lett.\  {\bf 122}, 222001 (2019).
\bibitem{Hosaka:2016pey}
A.~Hosaka, T.~Iijima, K.~Miyabayashi, Y.~Sakai and S.~Yasui,
Exotic hadrons with heavy flavors: X, Y, Z, and related states,
PTEP \textbf{2016}, 062C01 (2016).
\bibitem{Richard:2016eis}
J.~M.~Richard,
Exotic hadrons: review and perspectives,
Few Body Syst. \textbf{57}, 1185-1212 (2016).
\bibitem{Chen:2016qju}
H.~X.~Chen, W.~Chen, X.~Liu and S.~L.~Zhu,
The hidden-charm pentaquark and tetraquark states,
Phys. Rept. \textbf{639}, 1-121 (2016).
\bibitem{Guo:2017jvc}
F.~K.~Guo, C.~Hanhart, U.~G.~Mei\ss{}ner, Q.~Wang, Q.~Zhao and
B.~S.~Zou,
Rev. Mod. Phys. \textbf{90}, no.1, 015004 (2018)
\bibitem{Lebed:2016hpi}
R.~F.~Lebed, R.~E.~Mitchell and E.~S.~Swanson,
Prog. Part. Nucl. Phys. \textbf{93}, 143-194 (2017)
\bibitem{Esposito:2016noz}
A.~Esposito, A.~Pilloni and A.~D.~Polosa,
Phys. Rept. \textbf{668}, 1-97 (2017)
\bibitem{Ali:2017jda}
A.~Ali, J.~S.~Lange and S.~Stone,
Exotics: Heavy Pentaquarks and Tetraquarks,
Prog. Part. Nucl. Phys. \textbf{97}, 123-198 (2017).

\bibitem{Brambilla:2019esw}
N.~Brambilla, S.~Eidelman, C.~Hanhart, A.~Nefediev, C.~P.~Shen, C.~E.~Thomas, A.~Vairo and C.~Z.~Yuan,
The $XYZ$ states: experimental and theoretical status and perspectives,
Phys. Rept. \textbf{873}, 1-154 (2020).
\bibitem{Liu:2019zoy}
Y.~R.~Liu, H.~X.~Chen, W.~Chen, X.~Liu and S.~L.~Zhu,
Pentaquark and Tetraquark states,
Prog. Part. Nucl. Phys. \textbf{107}, 237-320 (2019).

\bibitem{Aaij:2020hon}
R.~Aaij \textit{et al.} [LHCb],
A model-independent study of resonant structure in $B^+\to D^+D^-K^+$ decays,
Phys. Rev. Lett. \textbf{125}, 242001 (2020).

\bibitem{LHCb:2020pxc}
R.~Aaij \textit{et al.} [LHCb],
Amplitude analysis of the $B^+\to D^+D^-K^+$ decay,
Phys. Rev. D \textbf{102} (2020), 112003.

\bibitem{Genovese:1997tm}
M.~Genovese, J.~M.~Richard, F.~Stancu and S.~Pepin,
Heavy flavor pentaquarks in a chiral constituent quark model,
Phys. Lett. B \textbf{425}, 171-176 (1998).
\bibitem{Richard:2018jkw}
J.~M.~Richard, A.~Valcarce and J.~Vijande,
Doubly-heavy baryons, tetraquarks, and related topics,
Bled Workshops Phys. \textbf{19}, 24 (2018).
\bibitem{Stewart:2004pd}
I.~W.~Stewart, M.~E.~Wessling and M.~B.~Wise,
Stable heavy pentaquark states,
Phys. Lett. B \textbf{590}, 185-189 (2004).
\bibitem{Sarac:2005fn}
Y.~Sarac, H.~Kim and S.~H.~Lee,
QCD sum rules for the anti-charmed pentaquark,
Phys. Rev. D \textbf{73}, 014009 (2006).

\bibitem{Lee:2005pn}
S.~H.~Lee, Y.~Kwon and Y.~Kwon,
Anti-charmed pentaquark from B decays,
Phys. Rev. Lett. \textbf{96}, 102001 (2006).
\bibitem{Aitala:1997ja}
E.~M.~Aitala \textit{et al.} [E791],
Search for the pentaquark via the $P_0(\bar{c} s)$ decay,
Phys. Rev. Lett. \textbf{81}, 44-48 (1998).

\bibitem{Aitala:1999ij}
E.~M.~Aitala \textit{et al.} [E791],
Search for the pentaquark via the $P_0(\bar{c} s)\rightarrow K^*_{0} K^- p$ decay,
Phys. Lett. B \textbf{448}, 303-310 (1999).


\bibitem{Aktas:2004qf}
A.~Aktas \textit{et al.} [H1],
Evidence for a narrow anti-charmed baryon state,
Phys. Lett. B \textbf{588}, 17 (2004).
\bibitem{Chekanov:2004qm}
S.~Chekanov \textit{et al.} [ZEUS],
Search for a narrow charmed baryonic state decaying to $D^{*\pm} p^{\mp}$ in $ep$ collisions at HERA,
Eur. Phys. J. C \textbf{38}, 29-41 (2004).
\bibitem{Link:2005ti}
J.~M.~Link \textit{et al.} [FOCUS],
Search for a strongly decaying neutral charmed pentaquark,
Phys. Lett. B \textbf{622}, 229-238 (2005).
\bibitem{Aubert:2006qu}
B.~Aubert \textit{et al.} [BaBar],
Search for the Charmed Pentaquark Candidate $\Theta_{c}(3100)^{0}$ in $e^{+}e^{-}$ annihilations at $\sqrt{s}=10.58$ GeV,
Phys. Rev. D \textbf{73}, 091101 (2006).
\bibitem{Aubert:2006qx}
B.~Aubert \textit{et al.} [BaBar],
Measurements of the Decays $B^0 \to \bar{D}^0 p \bar{p}$, $B^0 \to \bar{D}$* 0 $p \bar{p}$, $B^0 \to D^{-} p \bar{p} \pi^{+}$, and $B^0 \to D^{*-} p \bar{p} \pi^{+}$,
Phys. Rev. D \textbf{74}, 051101 (2006).
\bibitem{Aaij:2017jgf}
R.~Aaij \textit{et al.} [LHCb],
Search for weakly decaying $b$-flavored pentaquarks,
Phys. Rev. D \textbf{97}, 032010 (2018).


\bibitem{LHCb:2021ohr}
R.~Aaij \textit{et al.} [LHCb],
Observation of the suppressed $\Lambda_b^0\to D p K^-$ decay with $D\to K^+ \pi^-$ and measurement of its $C\!P$ asymmetry,
[arXiv:2109.02621 [hep-ex]].

\bibitem{Weng:2018mmf}
X.~Z.~Weng, X.~L.~Chen and W.~Z.~Deng,
Masses of doubly heavy-quark baryons in an extended chromomagnetic model,
Phys.\ Rev.\ D {\bf 97}, 054008 (2018).
\bibitem{Weng:2020jao}
X.~Z.~Weng, X.~L.~Chen, W.~Z.~Deng and S.~L.~Zhu,
Systematics of fully heavy tetraquarks,
Phys. Rev. D \textbf{103}, 034001 (2021).
\bibitem{Weng:2019ynva}
X.~Z.~Weng, X.~L.~Chen, W.~Z.~Deng and S.~L.~Zhu,
Hidden-charm pentaquarks and $P_c$ states,
Phys.\ Rev.\ D {\bf 100}, 016014 (2019).
\bibitem{Weng:2021ngd}
X.~Z.~Weng, W.~Z.~Deng and S.~L.~Zhu,
The triply heavy tetraquark states,
[arXiv:2109.05243].
\bibitem{Weng:2021hje}
X.~Z.~Weng, W.~Z.~Deng and S.~L.~Zhu,
Doubly heavy tetraquarks in an extended chromomagnetic model,
[arXiv:2108.07242].
\bibitem{Hogaasen:2013nca}
H.~H\o{}gaasen, E.~Kou, J.~M.~Richard and P.~Sorba,
Phys. Lett. B \textbf{732}, 97-100 (2014)
doi:10.1016/j.physletb.2014.03.027
[arXiv:1309.2049 [hep-ph]].
\bibitem{Karliner:2016zzc}
M.~Karliner, S.~Nussinov and J.~L.~Rosner,
Phys. Rev. D \textbf{95}, no.3, 034011 (2017)
doi:10.1103/PhysRevD.95.034011
[arXiv:1611.00348 [hep-ph]].
\bibitem{ParticleDataGroup:2020ssz}
P.~A.~Zyla \textit{et al.} [Particle Data Group],
\href{https://inspirehep.net/literature/1812251}{PTEP \textbf{2020},
no.8, 083C01 (2020)}
\bibitem{An:2020jix}
H.~T.~An, K.~Chen, Z.~W.~Liu and X.~Liu,
Fully heavy pentaquarks,
Phys. Rev. D \textbf{103}, 074006 (2021).

\bibitem{An:2021vwi}
H.~T.~An, K.~Chen, Z.~W.~Liu and X.~Liu,
Heavy flavor pentaquarks with four heavy quarks,
Phys. Rev. D \textbf{103} (2021) no.11, 114027.
\bibitem{Li:2018vhp}
S.~Y.~Li, Y.~R.~Liu, Y.~N.~Liu, Z.~G.~Si and J.~Wu,
Pentaquark states with the $QQQq\bar{q}$ configuration in a simple model,
Eur. Phys. J. C \textbf{79} (2019) no.1, 87.
\bibitem{Cheng:2019obk}
J.~B.~Cheng and Y.~R.~Liu,
$P_c(4457)^+$, $P_c(4440)^+$, and $P_c(4312)^+$: molecules or compact pentaquarks?,
Phys. Rev. D \textbf{100} (2019) no.5, 054002.
\bibitem{Cheng:2020nho}
J.~B.~Cheng, S.~Y.~Li, Y.~R.~Liu, Y.~N.~Liu, Z.~G.~Si and T.~Yao,
Spectrum and rearrangement decays of tetraquark states with four different flavors,
Phys. Rev. D \textbf{101}, 114017 (2020).
\bibitem{Wu:2018xdi}
J.~Wu, X.~Liu, Y.~R.~Liu and S.~L.~Zhu,
Systematic studies of charmonium-, bottomonium-, and $B_c$-like tetraquark states,
Phys. Rev. D \textbf{99} (2019) no.1, 014037.
\bibitem{Wu:2016vtq}
J.~Wu, Y.~R.~Liu, K.~Chen, X.~Liu and S.~L.~Zhu,
Heavy-flavored tetraquark states with the $QQ\bar{Q}\bar{Q}$ configuration,
Phys. Rev. D \textbf{97} (2018) no.9, 094015.
\bibitem{Jaffe:1976ig}
R.~L.~Jaffe,
Multi-Quark Hadrons. 1. The Phenomenology of $Q^{2}\bar{Q}^{2}$ Mesons,
Phys. Rev. D \textbf{15}, 267 (1977).
\bibitem{Strottman:1979qu}
D.~Strottman,
Multiquark baryons and the MIT Bag Model,
Phys. Rev. D \textbf{20}, 748-767 (1979).
\bibitem{Zhao:2014qva}
L.~Zhao, W.~Z.~Deng and S.~L.~Zhu,
Hidden-Charm Tetraquarks and Charged $Z_c$ States,
Phys. Rev. D \textbf{90}, 094031 (2014).
\bibitem{Wang:2015epa}
Z.~G.~Wang,
Analysis of $P_c(4380)$ and $P_c(4450)$ as pentaquark states in the diquark model with QCD sum rules,
Eur. Phys. J. C \textbf{76}, 70 (2016).


\bibitem{Jaffe:2003sg}
R.~L.~Jaffe and F.~Wilczek,
Diquarks and exotic spectroscopy,
Phys. Rev. Lett. \textbf{91}, 232003 (2003).
\bibitem{Nakano:2003qx}
T.~Nakano {\it et al.} [LEPS Collaboration],
Evidence for a narrow S = +1 baryon resonance in photoproduction from the neutron,
Phys.\ Rev.\ Lett.\  {\bf 91}, 012002 (2003).
\bibitem{Leibovich:2003tw}
A.~K.~Leibovich, Z.~Ligeti, I.~W.~Stewart and M.~B.~Wise,
Predictions for nonleptonic $\Lambda_{b}$ and $\Theta_{b}$ decays,
Phys. Lett. B \textbf{586}, 337-344 (2004).
\bibitem{Oh:1994np}
Y.~s.~Oh, B.~Y.~Park and D.~P.~Min,
Pentaquark exotic baryons in the Skyrme model,
Phys. Lett. B \textbf{331}, 362-370 (1994).
\bibitem{Park:2018oib}
W.~Park, S.~Cho and S.~H.~Lee,
Where is the stable pentaquark?,
Phys.\ Rev.\ D {\bf 99}, 094023 (2019).
\bibitem{Schael:2004nm}
S.~Schael \textit{et al.} [ALEPH],
Search for pentaquark states in Z decays,
Phys. Lett. B \textbf{599}, 1-16 (2004).
\bibitem{Litvintsev:2004yw}
D.~O.~Litvintsev [CDF],
Pentaquark searches at CDF,
Nucl. Phys. B Proc. Suppl. \textbf{142}, 374-377 (2005).
\bibitem{Gignoux:1987cn}
C.~Gignoux, B.~Silvestre-Brac and J.~M.~Richard,
Possibility of stable multiquark baryons,
Phys.\ Lett.\ B {\bf 193}, 323 (1987).

\bibitem{Lipkin:1987sk}
H.~J.~Lipkin,
New Possibilities for Exotic Hadrons: Anticharmed Strange Baryons,
Phys. Lett. B \textbf{195}, 484-488 (1987).

\bibitem{Yamagata-Sekihara:2018gah}
J.~Yamagata-Sekihara and T.~Sekihara,
$\bar{K}\bar{D} N$ molecular state as a ``$u u d s \bar{c}$ pentaquark" in a three-body calculation,
Phys. Rev. C \textbf{100}, 015203 (2019).

\end{thebibliography}
\end{document}